\RequirePackage{fix-cm}
\documentclass{svjour3}                     
\smartqed  
\usepackage{graphicx}
\usepackage{natbib}
\usepackage{color}
\usepackage{amssymb}
\usepackage{lineno}
\usepackage{txfonts}
\usepackage{amstext}
\usepackage{mathabx}
\usepackage{multirow}
\usepackage{ulem}
\usepackage{lscape}
\usepackage{txfonts}
\usepackage{subfigure}
\usepackage{float}

\usepackage{hyperref}

 \journalname{Space Science Reviews}

\newcommand{\jgr}{J. Geophys. Res. }
\newcommand{\grl}{Geophys. Res. Lett. }
\newcommand{\icarus}{Icarus }
\newcommand{\aap}{Astron. Astrophys. }
\newcommand{\apj}{Astrophys. J. }
\newcommand{\apjl}{Astrophys. J. Lett. }
\newcommand{\apjs}{Astrophys. J. Suppl. }
\newcommand{\aj}{Astron. J. }
\newcommand{\planss}{Planet. Space Sci. }

\newcommand{\ssr}{Space Sci. Rev. }
\newcommand{\mnras}{Mon. Not. R. Astron. Soc. }
\newcommand{\nat}{Nature }
\newcommand{\science}{Science }
\newcommand{\prb}{Phys. Rev. B }
\newcommand{\prl}{Phys. Rev. Lett. }

\newcommand{\pasp}{Publ. Astron. Soc. Pac }
\newcommand{\apss}{Astrophys. Space Sci. }

\newcommand{\pasa}{Publ. Astron. Soc. Aust. }
\newcommand{\procspie}{	Proc. SPIE }
\newcommand{\nar}{New Astron. Rev. }
\newcommand{\combust}{Combust. Flame }
\newcommand{\sovast}{Sov. Astron. }

\newcommand{\Kunit}{\,cm$^{2}\cdot$s$^{-1}$}

\newcommand{\fig}[1]{Fig.~\ref{#1}}

\newcommand{\tab}[1]{Table~\ref{#1}}

\newcommand{\degre}{\ensuremath{^\circ}}

\begin{document}

\title{The deep composition of Uranus and Neptune from in situ exploration and thermochemical modeling}

\titlerunning{The deep composition of Uranus and Neptune}        

\author{Thibault Cavali\'e \and
        Olivia Venot \and
        Yamila Miguel \and
        Leigh N. Fletcher \and
        Peter Wurz \and
        Olivier Mousis \and
        Roda Bounaceur \and
        Vincent Hue \and
        J\'er\'emy Leconte \and
        Michel Dobrijevic
}


\institute{Thibault Cavali\'e \at
              Laboratoire d'Astrophysique de Bordeaux, Univ. Bordeaux, CNRS\\
              B18N, all\'ee Geoffroy Saint-Hilaire, 33615 Pessac, France \\
              LESIA, Observatoire de Paris, PSL Research University, CNRS, Sorbonne Universit\'es, UPMC Univ. Paris 06, Univ. Paris Diderot, Sorbonne Paris Cit\'e\\
              F-92195 Meudon, France\\
              Tel.: +33-540003271\\
              \email{thibault.cavalie@u-bordeaux.fr}           
           \and
           Olivia Venot \at
              Laboratoire Interuniversitaire des Syst\`emes Atmosph\'eriques (LISA)\\
              UMR CNRS 7583, Universit\'e Paris-Est-Cr\'eteil, Universit\'e de Paris, Institut Pierre Simon Laplace, Cr\'eteil, France
           \and
           Yamila Miguel \at
              Leiden Observatory, University of Leiden\\
              Niels Bohrweg 2, 2333CA Leiden, The Netherlands
           \and
           Leigh Fletcher  \at
              School of Physics and Astronomy, University of Leicester\\
              University Road, Leicester, LE1 7RH, UK
           \and
           Peter Wurz \at
              Universit\"at Bern, Physikalisches Institut, Space Science and Planetology\\
              Bern, Switzerland
           \and
           Olivier Mousis \at
              Aix Marseille Université, CNRS, CNES, LAM\\
              Marseille, France
           \and
           Roda Bounaceur  \at
              Laboratoire R\'{e}actions et G\'{e}nie des Proc\'{e}d\'{e}s, LRGP UMP 7274 CNRS, Universit\'{e} de Lorraine\\
              1 rue Grandville, BP 20401, F-54001 Nancy, France
           \and
           Vincent Hue \at
              Southwest Research Institute\\
              San Antonio, Texas, USA
           \and
           J\'er\'emy Leconte \at
              Laboratoire d'astrophysique de Bordeaux, Univ. Bordeaux, CNRS\\
              B18N, all\'ee Geoffroy Saint-Hilaire, 33615 Pessac, France
           \and
           Michel Dobrijevic \at
              Laboratoire d'astrophysique de Bordeaux, Univ. Bordeaux, CNRS\\
              B18N, all\'ee Geoffroy Saint-Hilaire, 33615 Pessac, France
}

\date{Received: 14 October 2019 / Accepted: 18 April 2020}

\maketitle

\begin{abstract}
The distant ice giants of the Solar System, Uranus and Neptune, have only been visited by one space mission, Voyager 2. The current knowledge on their composition remains very limited despite some recent advances. A better characterization of their composition is however essential to constrain their formation and evolution, as a significant fraction of their mass is made of heavy elements, contrary to the gas giants Jupiter and Saturn. An in situ probe like Galileo would provide us with invaluable direct ground-truth composition measurements. However, some of the condensibles will remain out of the grasp of a shallow probe. While additional constraints could be obtained from a complementary orbiter, thermochemistry and diffusion modeling can further help us to increase the science return of an in situ probe.

\end{abstract}

\section{Introduction}
\label{intro}
In the early days of planetary sciences and space exploration, Uranus and Neptune seemed to be very much alike. They share relatively similar masses, radii and color, for example, suggesting these planets could be twins from their formation to their current state. However, even if these distant planets have only been visited once by a spacecraft, data acquired during the Voyager 2 flybys and more recently from ground-based and space-based facilities demonstrate that they are quite different. Their density differ by as much as 30\%, Uranus is almost in equilibrium with incoming solar radiation while Neptune emits more than it receives \citep{Pearl1990,Pearl1991}. Moreover, Uranus has a high obliquity causing an extreme seasonal forcing while Neptune's obliquity (and thus seasonal cycle) is probably more comparable to Saturn's one \citep{Moses2018}. Improved gravity field, shape and rotation rate data now seem to point to different internal structures and thermal evolution \citep{Nettelmann2013,Nettelmann2016,Helled2020}. 

As pointed out in e.g. \citet{Guillot2005}, \citet{Guillot2019}, \citet{Atreya2020}, \citet{Mousis2020}, constraining the deep elemental and isotopic composition of the ice giants is one of the keys to better understand their formation and evolution. Unfortunately, and despite some recent progress \citep{Sromovsky2008,Karkoschka2011,Irwin2018,Irwin2019a,Tollefson2019a}, deep abundance measurements in the ice giants remain scarce. The Galileo probe composition measurements in Jupiter's troposphere \citep{vonZahn1998,Niemann1998,Mahaffy2000,Atreya1999,Wong2004} have triggered a tremendous amount of studies on the planet's formation \citep[e.g.][]{Owen1999,Gautier2001,Lodders2004,Guillot2006,Mousis2012,Mousis2019a}, now favouring the core accretion scenario for these planets \citep{Pollack1996,Hubickyj2005} over the disk instability scenario \citep{Boss1997,Boss2002}. The formation and evolution of Uranus and Neptune, on the other hand, remains one of the most outstanding open question. Contemplating these major breakthroughs enabled by the Galileo probe measurements, now complemented by Juno observations \citep[e.g.][]{Bolton2017,Li2017,Li2020,Kaspi2018}, it seems obvious that the next great leap in understanding the formation and evolution of the Solar System will result from sending orbiters and probes to the ice giants. In addition, the expected advances in this field will undoubtedly have significant repercussions on our understanding of exoplanet formation and evolution, since a significant fraction of the currently detected exoplanets are in the Neptune size-class\footnote{\href{https://exoplanets.nasa.gov/}{https://exoplanets.nasa.gov/}, \href{http://exoplanet.eu}{http://exoplanet.eu}}.

Orbiter and probe missions to the ice giants that are currently under consideration \citep{Mousis2018,Simon2018,Simon2020} will provide us with invaluable measurements in many fields, including bulk composition. In Section \ref{Sect-compo} of this paper, we will review the current knowledge of ice giants composition, with a comparison to gas giants, and the foreseeable prospects offered by ground-based and space-based observatories in the next decade. We will then show in Section \ref{Sect-thermo} how thermochemical and diffusion modeling can help us further constrain the deep composition of ice giants in the absence of in situ composition measurements, and what the critical parameters of such models are. This will lead us to present in Section \ref{Sect-mass-spectro} the increased science return a descent probe making abundance measurements with a mass spectrometer in Uranus and/or Neptune would have if its results would be coupled to further thermochemical modeling, and to complementary remote sensing observations of the probe entry site for context, as well as the requirements on the instrument that such measurement/model coupling result in. Finally, we will review in Section \ref{Sect-link} how deep composition measurements constrain interior and planetary formation models.

\section{The composition of ice giants \label{Sect-compo}}
Thermochemical models attempt to provide fits to the observed composition of a planetary atmosphere, by assuming a temperature profile, a deep mixing profile, and a set of chemical reactions. The bulk composition can only be measured in situ. The abundances measured by probes like Galileo are expected to be representative of the elemental composition at any location on the planet, especially for noble gases. The only known exception for Galileo is H$_2$O, because the probe descended into a hotspot \citep{Orton1998}. In addition to in situ measurements, remote sensing techniques can provide hints on the deep composition of giant planets, but they generally provide us with lower limits for condensible species and uncertainties are generally too large to be constraining for formation models. In some cases however, remote sensing can probe deeper than a shallow probe and could give better limits on the deep volatile composition.
While the ultraviolet and mid-infrared can mostly reveal the stratospheric abundances of hydrocarbons, other wavelength ranges can be used to obtain more useful observations for the deep chemical abundance. For example, methane and hydrogen sulfide (H$_2$S) can be derived in the troposphere from the near-infrared reflectivity and, potentially, from remote sensing in the (sub)millimeter range, along with CO. Helium can be estimated from the far-infrared collision-induced continuum. These tropospheric species, which are largely pressure broadened, give us the strongest constraints on the deep composition.

In this Section, we will present the current knowledge on the upper tropospheric composition of the ice giants and a comparison with gas giants. We will conclude with the perspectives offered by current and future observatories that could be used prior to an ice giant probe arrival to derive the composition of these planets.

  \subsection{Observed elemental composition \label{Sect-Observed_compo}}
  The elemental abundances reviewed hereafter are summarized in \tab{tab-abondances} and compared to the solar and protosolar values. The present-day solar elemental abundances used in this paper are all taken from \citet{Grevesse2010,Grevesse2015}, \citet{Scott2015}, \citet{Amarsi2017} and \citet{Amarsi2018,Amarsi2019}, except for germanium and arsenic \citep{Lodders2010}. The protosolar elemental abundances are derived from the solar abundances following \citet{Grevesse2010}.
  
  The isotopic ratios are also very valuable in that they tell us the main reservoirs for the various elements. Isotopic measurements in hydrogen, noble gases, nitrogen, carbon, oxygen, etc. \citep[e.g.][]{Lellouch2001, Feuchtgruber2013,Mahaffy2000,Fouchet2000a,Fletcher2014} are therefore key in constraining protoplanetary disk physico-chemical conditions and planet formation models \citep[e.g.][]{Hersant2001,Owen2003,Mousis2014b}. However, isotopes have not yet been accounted for in thermochemical models. Recent progress in Titan photochemistry modeling \citep{Dobrijevic2016} will enable that in the future. They are presented in more details in \citet{Atreya2020} and \citet{Mousis2020} and will not be discussed further in this paper.

  \begin{table}
  \caption{Elemental abundances in the protosun and in giant planets. }             
  \label{tab-abondances}      
  \begin{center}          
  \begin{tabular}{ccccccc}
    \hline
    $Z$ & Element         & Protosun         & Jupiter/ & Saturn/  & Uranus/  &  Neptune/ \\
        &                 & (dex)            & Protosun & Protosun & Protosun & Protosun \\
    \hline
    2  & He$^\mathrm{a}$  & (10.99$\pm$0.01) & 0.80$\pm$0.02    & 0.69$\pm$0.19   & 0.92$\pm$0.20 & 0.90$\pm$0.17 \\
    6  & C$^\mathrm{b,c}$ & (8.49$\pm$0.02)  & 3.85$\pm$0.95    & 8.58$\pm$0.37   & 80$\pm$20     & 80$\pm$20 \\
    7  & N$^\mathrm{d}$   & (7.88$\pm$0.05)  & 4.38$\pm$1.69    & 3.76$\pm$0.44   & see text       & see text \\
    8  & O$^\mathrm{e}$   & (8.74$\pm$0.03)  & $>$0.45$\pm$0.15 &  & \\
    10 & Ne$^\mathrm{f}$  & (7.98$\pm$0.10)  & 0.13$\pm$0.02    &  & \\
    15 & P$^\mathrm{g}$   & (5.46$\pm$0.03)  & 3.74$\pm$0.24    & 12.8$\pm$0.9    &  & \\
    16 & S$^\mathrm{h}$   & (7.17$\pm$0.03)  & 3.01$\pm$0.72    & $\sim$9         & see text      & see text \\
    18 & Ar$^\mathrm{i}$  & (6.40$\pm$0.13)  & 3.23$\pm$0.65    &  & \\
    32 & Ge$^\mathrm{j}$  & (3.70$\pm$0.07)  & 0.058$\pm$0.008  & 0.046$\pm$0.046 \\
    33 & As$^\mathrm{k}$  & (2.37$\pm$0.04)  & 2.35$\pm$0.15    & 7.38$\pm$2.49   \\
    36 & Kr$^\mathrm{l}$  & (3.30$\pm$0.06)  & 2.33$\pm$0.44    &  & \\
    54 & Xe$^\mathrm{m}$  & (2.29$\pm$0.06)  & 2.28$\pm$0.46    &  & \\
    \hline
  \end{tabular}
  \end{center}
  \small{
  $^\mathrm{a}$ \citet{Grevesse2010} for the protosun, \citet{vonZahn1998} and \citet{Niemann1998} or Jupiter, \citet{Conrath2000} for Saturn, \citet{Conrath1987} for Uranus and \citet{Conrath1993} for Neptune.\\
  $^\mathrm{b}$ \citet{Amarsi2019} for the protosun, \citet{Wong2004} for Jupiter, \citet{Fletcher2009b} for Saturn, \citet{Sromovsky2008} for Uranus and \citet{Karkoschka2011} and \citet{Irwin2019b} for Neptune (at the equator for Uranus and Neptune). \\
  $^\mathrm{c}$ As the CH$_4$ equatorial abundance is non negligible compared to He below its condensation level in both planets, it is accounted for when computing the H$_2$ mole fraction.\\
  $^\mathrm{d}$ \citet{Grevesse2010} for the protosun, \citet{Wong2004} for Jupiter, \citet{Fletcher2011b} at the equator for Saturn. The recent Juno microwave measurement of \citet{Li2017} results in N/H$=$(2.76$\pm$0.30) times protosolar. For Uranus and Neptune, N/H is computed from S/H as an upper limit such that S/N$>$5$\times$solar \citet{Irwin2018,Irwin2019a}. \\
  $^\mathrm{e}$ \citet{Amarsi2018} for the protosun, lower limit from \citet{Wong2004} for Jupiter. \\
  $^\mathrm{f}$ \citet{Scott2015} for the protosun, \citet{Mahaffy2000} for Jupiter. \\
  $^\mathrm{g}$ \citet{Scott2015} for the protosun, \citet{Fletcher2009b} for Jupiter and Saturn. \\
  $^\mathrm{h}$ \citet{Scott2015} for the protosun, \citet{Wong2004} for Jupiter, estimate from \citet{Briggs1989} for Saturn. \\
  $^\mathrm{i}$ \citet{Scott2015} for the protosun, \citet{Mahaffy2000} at the equator for Jupiter. \\
  $^\mathrm{j}$ \citet{Grevesse2015} for the protosun, \citet{Giles2017} for Jupiter, \citet{Noll1991} for Saturn. \\
  $^\mathrm{k}$ \citet{Lodders2010} for the protosun, \citet{Giles2017} for Jupiter, \citet{Noll1991} for Saturn. \\
  $^\mathrm{l}$ \citet{Grevesse2015} for the protosun, \citet{Mahaffy2000} for Jupiter. \\
  $^\mathrm{m}$ \citet{Grevesse2015} for the protosun, \citet{Mahaffy2000} for Jupiter. \\
}
\end{table}

    \subsubsection{Helium and noble gases}
    Voyager 2 provided the first measurement of the helium abundance of the giant planets from infrared spectroscopy and radio occultation experiments \citep{Gautier1981,Conrath1984,Conrath1987,Conrath1991}. 

    In Jupiter, the Galileo probe refined the measurement to an helium-to-hydrogen ratio (He/H) of (7.85$\pm$0.16)$\times$10$^{-2}$ \citep{Niemann1998,vonZahn1998}. In Saturn, the initial He/H of \citet{Conrath1984} was revised to a higher value of (6.75$\pm$1.25)$\times$10$^{-2}$ by \citet{Conrath2000}. The He/H in Saturn remains uncertain and several attempts have been made recently to make new measurements.  Using Cassini instrumentation, \citet{Koskinen2018} and \citet{Waite2018} derived an He/H of (5.5$\pm$1.0)$\times$10$^{-2}$ and $\sim$8$\times$10$^{-2}$, respectively. Helium is therefore subsolar in both gas giant upper tropospheres, and this can be explained by the formation of helium droplets in metallic hydrogen \citep{Wilson2010}.

    The initial results at Uranus and Neptune helium indicated mole fractions of 0.152$\pm$0.033 \citep{Conrath1987} and 0.190$\pm$0.032 \citep{Conrath1991}, respectively. Accounting for an N$_2$ mole fraction of 0.003 in Neptune's atmosphere enabled \citet{Conrath1993} to revise their results to 0.15 for Neptune, bringing it in better agreement with the Uranus value. Later Infrared Space Observatory (ISO) observations by \citet{Burgdorf2003} seem to confirm and further refine the Neptune helium abundance to 0.149$^{+0.017}_{-0.022}$. The He/H in Uranus and Neptune would thus seem to be slightly subsolar with abundances of (8.88$\pm$2.00)$\times$10$^{-2}$ and (8.96$\pm$1.46)$\times$10$^{-2}$, respectively. However, the error bars remain too large (from subsolar to marginally supersolar) to constrain interior models accurately \citep{Guillot2005,Helled2011,Helled2018,Helled2020,Nettelmann2013}. Remote sensing can only provide tentative results and it is clear that only in situ measurements can provide us with a measurement accurate enough to constrain formation and evolution models. The goal of a probe is to reach an accuracy of 2\%~\citep{Mousis2018}, similar to Galileo.
    
    Noble gases beyond helium have only been measured in Jupiter by the Galileo probe. Argon, Xenon and Krypton were all found enriched by a factor of 2-4 with respect to the protosolar value. Only neon is found subsolar, because of dissolution in liquid helium deep in the atmosphere of Jupiter \citep{Roulston1995,Wilson2010}.

    \subsubsection{Carbon \label{Sect-compo-C}}
    Methane is the most abundant species after helium in all giant planets, and it is their main carbon reservoir. 

    In Jupiter, Galileo measured C/H$=$(1.19$\pm$0.29)$\times$10$^{-3}$ \citep{Wong2004}. At Saturn, \citet{Fletcher2009a} used Cassini to constrain C/H to (2.65$\pm$0.10)$\times$10$^{-3}$. 

    In the ice giants, methane condenses at $\sim$1 bar and must be measured below this level. Its mole fraction was initially measured to about 0.02 \citep{Lindal1987,Lindal1992,Baines1995} in both ice giants, i.e. more than an order of magnitude above its stratospheric abundance \citep{Lellouch2015}. More recent observations have, however, shown that the picture is more complicated than initially thought. \citet{Karkoschka2009,Karkoschka2011}, \citet{Sromovsky2008}, \citet{Sromovsky2011,Sromovsky2014},  and \citet{Irwin2019b} have used near-IR scans that sample both an H$_2$-collision induced opacity and a methane opacity to separate the effects of clouds and methane. From these spatially-resolved observations, they have shown that methane is more abundant at low latitudes than at the high latitudes sampled by the earlier observations. The equatorial mole fraction of methane is 0.04$\pm$0.01 decreasing towards the poles in the upper troposphere possibly because of tropospheric circulation \citep{Fletcher2020b}. This point will be briefly addressed in Section \ref{Sect-mixing}. In any case, methane is being measured at the CH$_4$-ice condensation point, and there is a possibility that there is additional internal stratification, as seen with jovian ammonia that is not well-mixed beneath the expected cloud-condensation level \citep[e.g.][]{Li2017}. The current measurements must therefore be seen as lower limits on the deep C/H in ice giants.

    \subsubsection{Sulphur and nitrogen}
    Sulphur and nitrogen should be mainly borne by H$_2$S and ammonia (NH$_3$) in the reducing part of the atmospheres of the giant planets, even if the $^{15}$N/$^{14}$N isotopic ratio in Jupiter and Saturn suggests nitrogen may have originally been delivered from N$_2$ \citep{Fouchet2000a,Fletcher2014,Mousis2014b}. Both nitrogen and sulphur should be enriched over the protosolar value.

    Both have been observed in Jupiter by Galileo with N/H$=$(3.32$\pm$1.27)$\times$10$^{-4}$ and S/H$=$(4.45$\pm$1.05)$\times$10$^{-5}$ \citep{Wong2004}. More recent microwave mapping observations of Juno indicate that NH$_3$ is not well-mixed in the jovian upper troposphere, at least above the 50-60 bar level \citep{Bolton2017,Li2017}, raising the question whether the Galileo measurement is representative of the nitrogen deep abundance. They find a deep NH$_3$ mole fraction of 362$\pm$33 ppm, i.e. N/H$=$(2.09$\pm$0.20)$\times$10$^{-3}$ only marginally consistent with the Galileo measurement done at 6.5\degre~north. In Saturn, \citet{Fletcher2011b} found N/H$=$2.85$\times$10$^{-4}$ at the equator from Cassini/VIMS, confirmed by Cassini/RADAR observations of \citet{Janssen2013} and \citet{Laraia2013}. However, its deep value remains quite uncertain due to meridional variability, similarly to the Jupiter case \citep{Li2017}. The detection of H$_2$S in Saturn remains uncertain \citep{Briggs1989}.

    In the ice giants, H$_2$S and NH$_3$ remained undetected for a long time despite repeated efforts. The reason is that both species are thought to form a cloud of ammonium hydrosulfide (NH$_4$SH) at around 30-50 bars from the NH$_3$(g) $+$ H$_2$S(g) $\rightarrow $ NH$_4$SH(s), only leaving traces of the most abundant species among the two up to their own condensation level \citep{deBoer1994}. The most abundant of the two would then condense in another cloud, at pressures between 5 and 10 bars. \citet{dePater1989a} and \citet{dePater1989b} found that NH$_3$ had to be $\sim$0.1-0.001 times solar in the probed part of the atmosphere to match their microwave spectra of the two planets. To explain this depletion, \citet{dePater1991} tentatively proposed an abundance of H$_2$S 10-30 times solar and an S/N at least 5 times solar in Uranus. Similar conclusion were reached for Neptune by \citet{deBoer1994,deBoer1996}. However, these abundances must all be understood as lower limits since none of these observations probed below the NH$_4$SH cloud base. Using near-infrared observations with the Gemini North telescope, \citet{Irwin2018} detected H$_2$S above the main cloud deck in Uranus, indicating that sulphur is more abundant than nitrogen and placing a lower limit on their ratio, with S/N $>$ 4.4-5.0 times the solar value (in agreement with \citealt{dePater1991}). Using a similar technique, \citet{Irwin2019a} derived a lower limit on H$_2$S in Neptune. Complementary broadband spectra obtained with the VLA and ALMA enabled \citet{Tollefson2019a,Tollefson2019b} to tentatively constrain S in Neptune to be 30 times protosolar and N to be protosolar.

    \subsubsection{Oxygen}
    Water, the main oxygen-bearing species in a giant planet interior, played a crucial role when giant planets formed. Water ice at the time of planetesimal formation provided a significant mass reservoir to build the planetary cores beyond the snowline, and the C/O ratio is a good diagnostic of the planet formation location \citep{Ali-Dib2014,Mousis2012,Mousis2014b,Oberg2011,Oberg2016}. 

    In addition, these ices played a fundamental role in that they trapped the other heavy elements. Depending on the pressure and temperature conditions at which the ices condensed, the heavy elements were either trapped on amorphous ices or in clathrates \citep{Bar-Nun1988,Owen1999,Lunine1985,Gautier2001,Gautier2005,Mousis2006}. If ices condensed in amorphous form, then the oxygen enrichment should be similar to the enrichment of other heavy element \citep{Owen2003,Owen2006}. On the other hand, the clathrate scenario requires a radically different oxygen abundance, i.e., $\sim$4 times more, to trap the heavy elements \citep{Mousis2014b,Mousis2018}. This is why constraining the deep oxygen abundance is so important to understand giant planet formation. 

    The Galileo probe entered a 5-$\mu$m hotspot and failed to reach the levels where water is uniformly mixed in Jupiter \citep{Atreya2003,Wong2004}. Juno is currently attempting to make this measurement from microwave radiometry during low-altitude perijove passes \citep{Matousek2007,Bolton2017}, now that the NH$_3$ distribution is established \citep{Li2017}. The first result obtained in the equatorial zone, where NH$_3$ is well-mixed up to its condensation level, indicates an O/H$=$2.7$^{+2.4}_{-1.7}$ times protosolar \citep{Li2020}. This result is key to better understanding the formation of Jupiter \citep{Helled2014a}, but will require additional measurements at other latitudes to assess whether this is the bulk abundance.

    In the other giants, the tropospheric temperatures are colder than in Jupiter, making water condensation happen deeper. While reaching the well-mixed layers in Saturn will be at the limit of the capabilities of recently proposed probe \citep{Mousis2014a,Mousis2016,Atkinson2016,Atkinson2018}, direct in situ measurement will remain highly improbable in the ice giants because water condenses at a pressure ranging between $\sim$200 and $\sim$1000\,bar, depending on the adopted temperature extrapolation model \citep{Atreya2005,Leconte2017}. Complementary measurements taken by a remote sensing instrumentation suite on future ice giant orbiters \citep[e.g.][]{Arridge2014} will therefore be needed for additional context and constraints.

    In the meantime, indirect measurements are the only possibility to constrain the deep oxygen abundance in these planets. We will detail these techniques and recent progress in Section \ref{Sect-thermo}.

    \subsubsection{Phosphorus and other heavy elements \label{Sect-compo-P}}
    Phosphorus, mainly carried by phosphine (PH$_3$), was observed with Cassini by \citet{Fletcher2009b} and the P/H ratio is (1.08$\pm$0.06)$\times$10$^{-6}$ in Jupiter and (3.70$\pm$0.23)$\times$10$^{-6}$ in Saturn. However, it still remains undetected in the ice giants \citep{Moreno2009,Teanby2019}. It may result from the destruction of this species by H$_2$O thermochemistry at depth, provided that the deep oxygen abundance is high enough in both planets \citep{Visscher2005}.

    Other heavy-element-bearing species have been observed in Jupiter and Saturn, like GeH$_4$ and AsH$_3$ \citep{Giles2017,Noll1991,Fletcher2011b}. As is supersolar in Jupiter, like most other heavy elements, but Ge is subsolar. This probably results from deep thermochemistry as Ge atoms are partly transferred from GeH$_4$ to GeS around the GeH$_4$ quench level \citep{Lodders1994}. A complication arises from the non uniform meridional abundances of these species. While GeH$_4$ and PH$_3$ peak at low latitudes and decreases poleward, as expected from models \citep{Wang2015}, AsH$_3$ is minimal at low latitudes and peaks at the poles \citep{Grassi2019}. Their deep abundance thus remains quite uncertain.

    \subsubsection{Summary}
    Most heavy element abundance measurements were made possible by sending an entry probe in Jupiter. This underlines the importance of sending such instrumentation to all giant planets in the Solar System to make comparable ground-truth measurements. If these were coupled to remote sensing from orbiting facilities, the direct measurement would help to break the degenerate effects of gaseous species on the planetary spectrum. 

    Besides the elements presented previously, Galileo enabled quantifying the abundances of noble gases such as neon, argon, krypton, and xenon \citep{Mahaffy2000}. All elements measured by the probe are 2-4 times solar (except oxygen for the reasons mentioned above). The Juno measurement of oxygen will complete this panorama, but preliminary results that pertain to Jupiter's equatorial zone are compatible with this picture \citep{Li2020}. 

    In Saturn, helium is subsolar probably because of helium rain, carbon and phosphorus are about 10 times solar, but nitrogen seems to be less enriched. The non uniformity of the meridional distribution of NH$_3$ \citep{Fletcher2011b}, similarly to Jupiter \citep{Bolton2017,Li2017}, complicates the derivation of the deep nitrogen abundance. The lack of measurements for other heavy elements, especially noble gases which should be uniform with altitude and latitude, makes it difficult to constrain Saturn formation models \citep[e.g.][]{Hersant2008}. Several probe proposals were developed in the recent years \citep{Atkinson2016,Atkinson2018,Mousis2014a,Mousis2016} but none was selected for flight so far.

    In Uranus and Neptune, the scarcity of heavy element abundance measurements is even more dramatic than in Saturn, as only carbon and, to some extent, sulphur have been measured, though the measurements of these condensible species bear large error bars and might be lower limits. The nominal abundance of methane at 1-2 bars and at low latitudes in both planets results in a C/H of 0.04$\pm$0.01, i.e., about 80 times protosolar, as expected from models \citep{Owen2003,Hersant2004}. Sulphur may be 20-30 times protosolar, slightly lower than predictions from those same models.

    This summary stresses the need for planetary probes at Saturn, and even more so at the ice giants.

  \subsection{Perspectives on ice giant elemental composition determination ahead of the 2040s} 
  If a probe-carrying mission is to be selected for Uranus and/or Neptune with a launch window in the 2029-2034 timeframe \citep{Simon2020}, such a mission will arrive in the 2040s. In this Section, we will attempt to list the progress on ice giant composition we can expect from existing and forthcoming ground-based and space-based observatories. In addition, these observations ahead of a mission arrival in the 2040s will enable temporal variation studies which will set the ground for the mission operations and help contextualize them further.

    \subsubsection{Radio}
    Radio wave observations probe the giant planet spectra where NH$_3$, H$_2$S and H$_2$O absorb. Single dish observations in the centimeter to decameter range remain difficult to calibrate accurately enough for the measurements to be constraining \citep{Courtin2015}. Interferometric observations of Saturn with LOFAR (Low Frequency Array, \citealt{Rottgering2003}) have not yet detected Saturn's emission unambiguously because of the low planetary flux combined with the rapidly varying background sky emission (D. Gautier, private communication, 2015). The implementation of the Square Kilometer Array (SKA) may enable achieving these long wavelength measurements to better constrain the deep NH$_3$ and H$_2$O abundances in the giant planets in the 2030s. 
    
    In the centimeter wavelengths, the e-VLA (expanded Very Large Array) remains the best radio observatory to date. A project to improve the capabilities in terms of spatial resolution and sensitivity, named the ng-VLA (next generation VLA), may enable to improve on the constraints on deep N, S and O in the ice giants \citep{dePater2018}. This project is aiming to start early science operations in the late 2020s and full science operations in the mid-2030s.
    
    However, it remains to be seen whether radio measurements can probe deep enough and reach the well-mixed layers with the required accuracy. Juno has shown for NH$_3$ in Jupiter that reaching the well-mixed region requires probing at tens of bars \citep{Bolton2017,Li2017}. Interpreting the radio emission uniquely remains a challenge because it is hard to separate the broad spectral effects of temperature and the gaseous opacity.

    \subsubsection{Millimeter and submillimeter}
    ALMA (Atacama Large Millimeter/submillimeter Array) and NOEMA (NOrthern Extended Millimeter Array) are currently the most sensitive millimeter and submillimeter interferometers. Both will still be operating in the 2020s and 2030s. 
    
    Aggregating broadband observations of these arrays with ng-VLA observations will help to improve our understanding of spatial distribution of H$_2$S and NH$_3$ (see \citealt{Tollefson2019b} for results using the current capabilities of these observatories) and of upper troposheric circulation \citep{Fletcher2020b} in the $\sim$1-50 bar pressure range. In addition, the determination of the meridional distribution of tropospheric CO in Uranus and Neptune from line spectroscopy will help to constrain further the deep oxygen abundance by coupling the observations to thermochemical modeling (see Section \ref{Sect-thermo}).

    \subsubsection{Near, mid- and far-infrared}
    In the near-IR, the techniques for separating the reflective aerosols from gaseous composition (specifically CH$_4$ and H$_2$S) have been established by ground-based observers using the largest astronomical facilities (e.g., Gemini, Keck, Very Large Telescope, etc.). These have demonstrated latitudinal variations of these volatiles, and provided lower limits on the potential bulk abundances of carbon and sulphur. Future near-infrared ground-based measurements with higher spatial resolutions (e.g., from the next generation of instrumentation on extremely large telescopes, such as the Extremely Large Telescope, Giant Magellan Telescope and Thirty Meter Telescope) might allow for further discrimination between aerosols and gaseous composition, but these may still be hampered by terrestrial atmospheric contamination. In the mid-infrared and far-infrared, measurements from ground- and airborne facilities (e.g., Stratospheric Observatory for Infrared Astronomy) could continue to determine stratospheric composition and thermal structure, but this may not be of use for the determination of bulk planetary composition (with the exception of deuterium-to-hydrogen ratio measurements, if possible in the far-infrared).
    
    In all of these cases, further progress could be made by being above the complicating effects of the terrestrial atmosphere. The James Webb Space Telescope (JWST, \citealt{Gardner2006}) carries instruments spanning the 1-30 micron range at exquisite spectral resolution and sensitivity that surpasses anything from the ground \citep{Norwood2016b,Norwood2016a}.  In the mid-infrared, the MIRI instrument will place new upper limits on the PH$_3$ and NH$_3$ content using bands near 5 and 10 microns that have never been observed before. MIRI will also constrain the collision-induced continuum in the far-infrared, which may enable separation of temperature, para-H2 and helium, via the same techniques as used on Voyager IRIS.  MIRI will also provide our first spatially-resolved glimpses of the stratospheric temperatures and chemistry \citep{Moses2018}.
    
    In the near-infrared, NIRSpec will enable more sensitive measurements of the H$_2$S and CH$_4$ abundances using the techniques honed on the ground.  Furthermore, they will provide access to fluorescent regions between 3.0-4.5 microns, where CO and CO$_2$ fluoresce \citep{Encrenaz2004,Fletcher2010}.  Along with sub-millimetre observations of CO, these provide another independent measurement of the CO abundance on the ice giants.  In addition, the JWST instruments will further refine the D/H ratio in CH$_4$ (and potentially other species), as a further constraint on planetary formation.  
    
    At longer wavelengths in the far-infrared and sub-millimetre, the proposed Origins Space Telescope (OST, \citealt{Leisawitz2018}) and the SPace Infrared telescope for Cosmology and Astrophysics (SPICA, \citealt{Roelfsema2018}) will both offer sensitive observations of the spectrum, potentially allowing new constraints on the shape of the hydrogen-helium continuum, and on the isotopic ratios within hydrogen (from far-IR HD features).  Depending on the final architecture of these missions, they may also provide new measurements of rotational lines of CO and CH$_4$.  Even with these new and sensitive instruments, the ice giants will likely be unresolved, such that no spatial variability in these gases will be measured.  For this, we have to be reliant on future orbital missions to the ice giants.
\\
These future observations concern several species that can be further used to constrain the deep abundance of some key elements by combining these observations with thermochemical modeling. This is the subject of the next Section.

\section{Thermochemical modeling of giant planet atmospheres \label{Sect-thermo}}
In this Section, we will first present the principle of inferring deep planet composition from thermochemical modeling. We will then review the models that dealt with giant planet thermochemistry through the quench level approximation and show the recent progress enabled by the development of more comprehensive thermochemical and diffusion models. Finally, we will detail the parameters these models rely on and what the prospects on improving their predictability is.

The deep hot troposphere of the giant planets is in thermochemical equilibrium. If applied to the upper troposphere and to the stratosphere, this equilibrium predicts extremely small abundances for many species that have nonetheless been detected \citep{Prinn1977,Fegley1985,Fegley1986,Fegley1994}, among which the methyl radical (CH$_3$ ; \citealt{Bezard1998,Bezard1999}, \citealt{Fouchet2018b}), stable hydrocarbons \citep{Gladstone1983,Fouchet2000b,Courtin1984,Orton2014b,Burgdorf2006,Meadows2008}, phosphine (PH$_3$ ; \citealp{Knacke1982,Bregman1975,Fletcher2009b}, carbon monoxide (CO ; \citealp{Beer1975,Bezard2002,Noll1986,Noll1991,Encrenaz2004,Marten1993,Marten2005}), carbon dioxide (CO$_2$ ; \citealt{Feuchtgruber1997}, \citealt{Burgdorf2006}), hydrogen cyanide (HCN ; \citealt{Lellouch1995}, \citealt{Bezard1997}, \citealt{Fouchet2018a}, \citealt{Marten1993}), carbon sulfide (CS ; \citealt{Lellouch1995,Moreno2017}). These species are generally observed in the stratosphere. They are produced from CH$_4$ photochemistry \citep{Moses2000a,Moses2005,Moses2018,Dobrijevic2010,Dobrijevic2011,Dobrijevic2020,Hue2015,Hue2016,Hue2018} or injected in the atmosphere from external sources \citep{Feuchtgruber1997,Moses2000b,Ollivier2000}, like interplanetary dust particles \citep{Landgraf2002,Moses2017}, large comet impacts \citep{Lellouch1995,Lellouch2005,Lellouch2006,Cavalie2008c,Cavalie2010,cavalie2012,Cavalie2013,Moreno2017}, and icy rings and satellites \citep{Connerney1984,Connerney1986,Prange2006,Hartogh2011,Waite2018,Perry2018,Cavalie2019}. However, others like CO and PH$_3$ are observed in the upper troposphere\footnote{CO can actually have an internal and an external component \citep{Bezard2002,Lellouch2005}.} with abundances that are tens of orders of magnitude above thermochemical equilibrium predictions. Their presence at these levels is caused by convective vertical mixing that quenches thermochemical equilibrium where the vertical transport timescale becomes shorter than the chemical timescale. 
  
Thermochemical and diffusion modeling can then be a powerful tool to infer the deep elemental composition of the giant planets from disequilibrium species, especially when the main carrier of an element does not reach the observable levels. The disequilibrium species abundances is used to track back their abundance at their respective quench level to then tie them back to the main element-carrier abundance.

In this Section, we will present the modeling principle of thermochemistry to constrain deep composition and show how it has been applied in the past decades, first using the quench level approximation, and then using more comprehensive chemical models. We will detail the parameters that are fundamental in getting accurate simulations and the prospects regarding future improvements.

  \subsection{Principle}
  Oxygen is mainly carried by water, but water condenses in the troposphere of the giant planets. While its condensation level occurs at $\sim$10 bar in Jupiter, it occurs at pressure ranging from $\sim$200 to $\sim$1000 bars in both Uranus and Neptune, according to temperature extrapolation models \citep{Leconte2017}. Only microwaves can probe that deep \citep{Janssen2005,dePater2016}, but limited calibration accuracy often prevents any direct constraint on the water abundance \citep{dePater1989a,dePater1989b,Courtin2015}. The idea then lies in measuring the upper tropospheric abundance of CO, which does not condense in giant planet atmospheres and is in disequilibrium because of efficient vertical mixing, and to tie it back to the deep water abundance with a chemistry and diffusion model. As CO is chemically linked to water, thermochemical and diffusion models have been used with this species to constrain the deep oxygen abundance ever since it was first detected in Jupiter by \citet{Beer1975}. 
  
  Other carbon bearing species can, in principle, be used similarly to constrain the deep water, like ethane (C$_2$H$_6$) \citep{Fegley1994}. Another example is phosphorus, which has been detected in PH$_3$ in Jupiter and Saturn, but neither in Uranus nor in Neptune \citep{Moreno2009}. This species can be destroyed by water if water is abundant enough. Its detection then results either from the relatively low water abundance or from its quenching at levels that are deeper than where it gets destroyed by water \citep{Fegley1994}. On the other hand, its absence can help to put additional constraints on the deep water abundance \citep{Visscher2005}.
  
  We come back to the example of carbon monoxide and water, as it is the most studied case to date. In the deep hot tropospheres of giant planets, CO and H$_2$O are in thermochemical equilibrium through the reaction 
  \begin{equation}
    \mathrm{H}_2\mathrm{O}+\mathrm{CH}_4=\mathrm{CO}+3\mathrm{H}_2. \label{equilibrium_reaction}
  \end{equation}
  Rearranging the equilibrium constant of the above equation enables to express the CO mole fraction as follows:
   \begin{equation}
    y_\mathrm{CO}=\frac{y_{\mathrm{CH}_4} y_{\mathrm{H}_2\mathrm{O}}}{y_{\mathrm{H}_2}^3 p^2} K_\mathrm{eq}
    \label{CO_abundance}
  \end{equation}
  where $p$ is the total pressure and $K_\mathrm{eq}$ is the equilibrium constant of reaction (\ref{equilibrium_reaction}). At higher and colder levels, the H$_2$O-CO equilibrium moves towards the reduced H$_2$O-CH$_4$ mixture and the conversion kinetics slows down. There is a level in the troposphere at which the temperature is low enough for the kinetics to become slower than the vertical mixing caused by convection. This is the level where the chemical lifetime of CO destruction $\tau_\mathrm{chem}$ equals the vertical mixing timescale $\tau_\mathrm{mix}$. Thermochemistry is quenched and the CO mole fraction fixed for all levels above this quench level. 
  
  There are two techniques that have been used to find the abundances of CO and water at the quench level: the quench level approximation and comprehensive thermochemical and diffusion modeling. In both cases, presented below, the determination of convective mixing is crucial.
  
  \subsection{Estimating convective mixing strength}
  The magnitude of vertical mixing caused by convection is key in fixing the level at which thermochemistry is quenched, and in turn in fixing upper tropospheric abundances of disequilibrium species: the stronger the mixing, the deeper the quench level.
  
\begin{figure}[t]
    \begin{center}
      \includegraphics[width=0.75\textwidth,keepaspectratio]{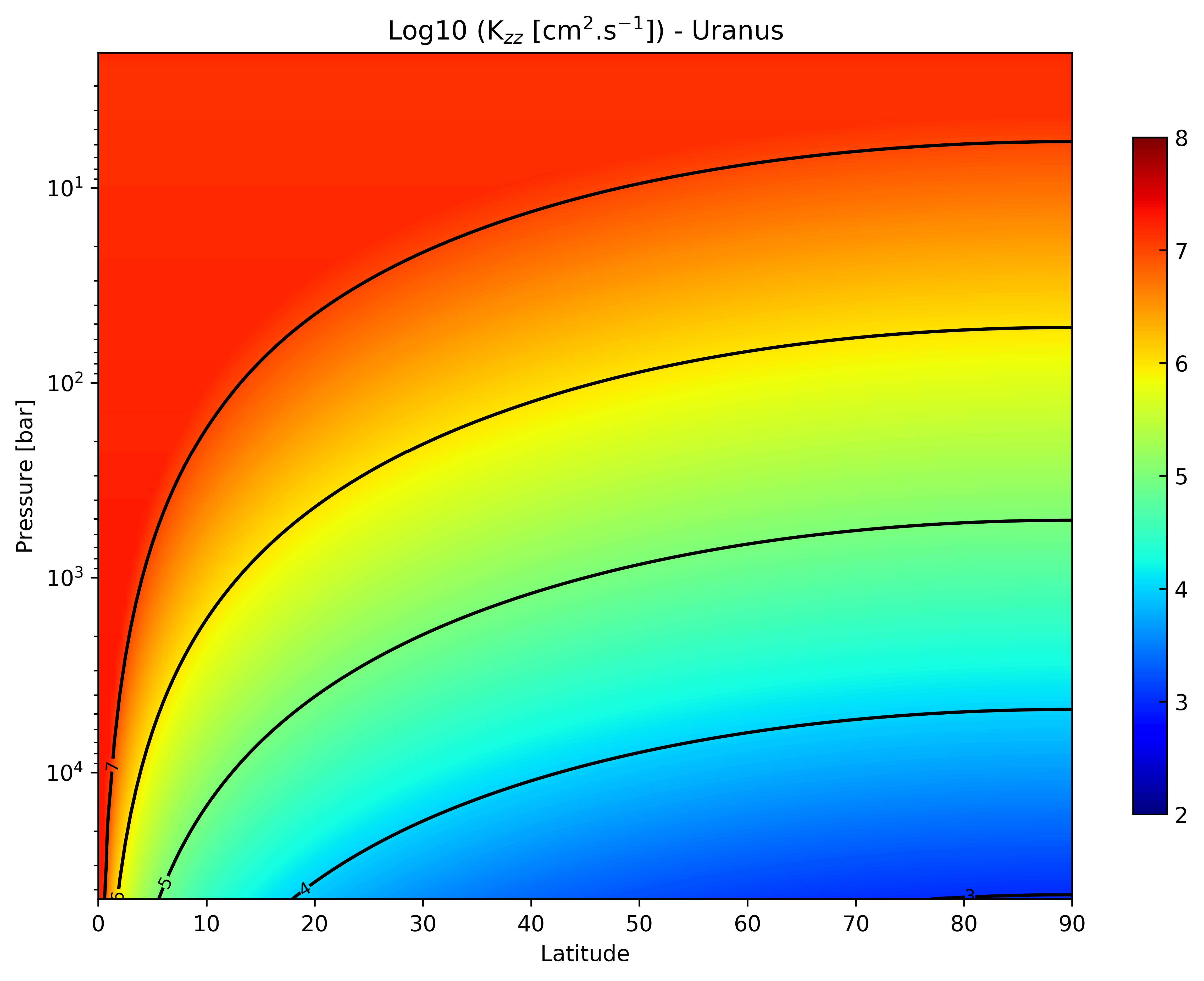}
      \includegraphics[width=0.75\textwidth,keepaspectratio]{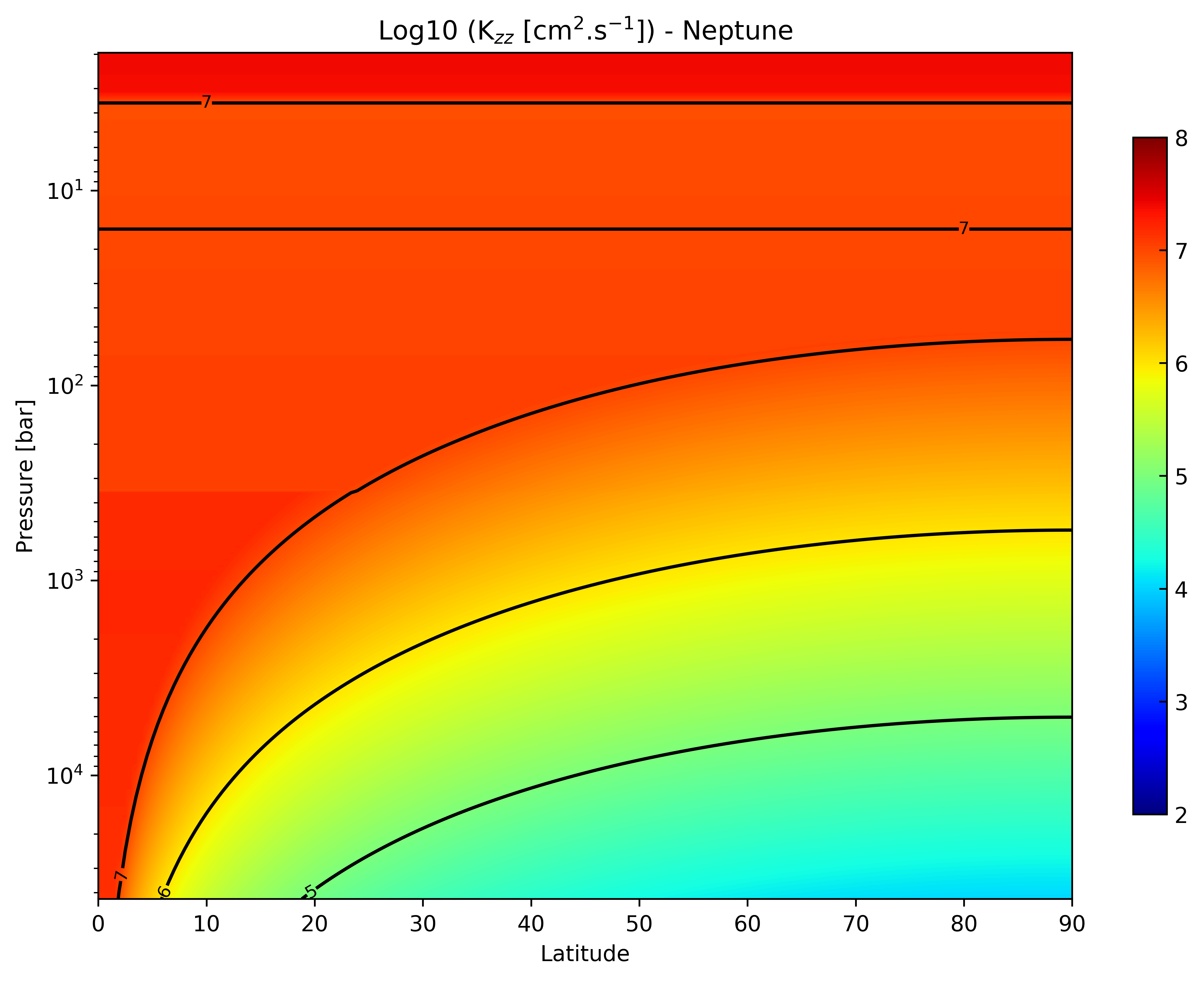}
    \end{center}
    \caption{Vertical mixing in the tropospheres of Uranus (top) and Neptune (bottom) as a function of pressure and latitude, using the prescription of \citet{Wang2015} and the temperature and abundance profiles of \citet{Venot2020}.}
    \label{fig-Kzz} 
  \end{figure}

  The vertical mixing timescale $\tau_\mathrm{mix}$ is given by
  \begin{equation}
    \tau_\mathrm{mix}=\frac{L^2}{K}, \label{eq-tau_mix}
  \end{equation} 
  where $K$ the vertical mixing coefficient and $L$ the length over which mixing occurs. The latter was taken as the atmospheric scale height $H$ in early studies. Convective mixing can be estimated from free-convection and mixing-length theories \citep{Stone1976,Gierasch1985} and modeled in 1D models by means of an eddy mixing coefficient. The scaling relationship
  \begin{equation}
    K\simeq\left(\frac{Fk_\mathrm{B}}{\rho m c_p}\right)^\frac{1}{3}H, \label{eq-Keddy}
  \end{equation}
  where $F$ is the internal heat flux of the planet, $k_\mathrm{B}$ is the Boltzmann constant, $\rho$ is the atmospheric mass density, $m$ is the atmospheric mean molecular mass, and $c_p$ is the atmospheric specific heat at constant pressure, applies in the absence of rapid rotation and a strong magnetic field. It is therefore only an approximation for giant planets. These estimates show that tropospheric $K$ is of the order of 10$^8$\,\Kunit, with a factor of 10 uncertainty, in the giant planets. \citet{Visscher2010} derived an altitude-latitude dependent expression for $K$ for fast rotating planets. They showed that $K$ decreased both with latitude and depth. The decrease with depth can however be neglected in thermochemical simulations because the variation is less than an order of magnitude between the top of the troposphere and the quench level. More recently, \citet{Wang2015} used rotating tank experiments to refine the scalings in the expression of $K$, and thus decrease the uncertainty on their estimation down to about 25\%. They also predicted that $K$ would be maximum at low latitudes and then decrease towards the high latitudes. They found that the decrease caused by depth and latitude was steeper for Saturn than for Jupiter. We illustrate the application of their prescription to Uranus and Neptune in \fig{fig-Kzz}. It essentially shows that disequilibrium species like CO, GeH$_4$ and PH$_3$ that are quenched where their abundance decreases with height should be more abundant in the upper troposphere at low latitudes. On the contrary, disequilibrium species like AsH$_3$ that are quenched where their abundance increases with height \citep{Fegley1994} should be more abundant at high latitude in the upper troposphere. This seems to be qualitatively in line with Juno/JIRAM observations of Jupiter \citep{Grassi2019}.
  
  \subsection{Quench level approximation}
  By decomposing the thermochemical equilibrium reaction (Equation \ref{equilibrium_reaction}) into the series of reactions that lead H$_2$O to be converted into CO (and vice versa), one can then try and identify the reaction which has the slowest kinetics, i.e. the rate-limiting reaction. The estimation of the rate-limiting reaction kinetics constrains the kinetics of the whole conversion scheme. By equating $\tau_\mathrm{chem}$ and $\tau_\mathrm{mix}$, it is then possible to derive the temperature at the quench level. Assuming a pressure-temperature relationship (e.g., dry or wet adiabat), it is then possible to compute $p$ in Equation \ref{CO_abundance}. The measured upper tropospheric mole fractions of CO and CH$_4$, which are the same as the one at the quench level, can eventually be used to solve the system and derive the deep value of $y_{\mathrm{H}_2\mathrm{O}}$.
  
  \citep{Prinn1977} first identified this rate-limiting reaction to be H$_2$ $+$ CH$_2$O $\rightleftharpoons$ CH$_3$ $+$ OH. By assuming a solar composition, they constrained vertical mixing to reproduce the CO detection of \citet{Beer1975}, thus using thermochemistry the other way around. Later work by \citet{Fegley1985,Fegley1988} and \citet{Fegley1994} further explored the deep composition of Jupiter and Saturn. \citet{Bezard2002} performed high spectral resolution observations in the 5$\mu$m window in the North Equatorial Belt of Jupiter to refine the planet's CO upper tropospheric abundance to 1.0$\pm$0.2 ppb. They applied the less ambitious kinetic scheme of \citet{Yung1988} for the CO-CH$_4$ conversion, in which the rate-limiting reaction is H $+$ H$_2$CO $+$ M $\rightleftharpoons$ CH$_3$O $+$ M. They also used the new method of \citet{Smith1998} to estimate the vertical scale for diffusion (in replacement of $H$ in Equation \ref{eq-tau_mix}). They derived a jovian deep oxygen abundance of 0.2 to 9 times the solar value.   
  
  The quench level approximation was later used in several studies \citep{Visscher2005,Cavalie2009,Luszcz-Cook2013} following the detections of CO in Saturn and Neptune by \citet{Noll1986} and \citet{Marten1993} to try and constrain the deep oxygen abundance in these planets.

  \subsection{1D kinetic and diffusion models}
  Another approach used to constrain the deep water abundance consists in using detailed kinetic and diffusion models that are able to reproduce accurately the chemical composition of hot atmospheric regions. The development of such models has been motivated by the discovery of hot giant exoplanets and the interpretation of their infrared spectra. Despite the high temperatures prevailing in their atmospheres, the regions probed by spectroscopic observations are not at thermochemical equilibrium. Disequilibrium processes are important and disturb the atmospheric composition. Thus, thermo-photochemical models have been developed specifically for the study of these peculiar atmospheres in which thermochemical equilibrium, mixing and photochemistry are at play \citep[e.g.][]{Moses2011, Visscher2011,Venot2012,Drummond2016,Tsai2017}.
  
  These models enable an accurate computation of the vertical profiles in the key pressure range where quenching occurs. They have been used for each solar system giant planet \citep{Visscher2010,Wang2016,Cavalie2014,Cavalie2017} to further constrain their deep oxygen abundances. \tab{tab-oxygen} summarizes the current status of model results regarding deep oxygen abundance in all giant planets.

  \begin{table}
  \caption{Deep oxygen abundance in giant planet deep atmospheres. }
  \label{tab-oxygen}      
  \begin{center}          
  \begin{tabular}{llll}
    \hline
           & CO mole fraction    & Deep O/H            & Reference \\
           & (upper troposphere) & ($\times$ protosun) &  \\
    \hline
    Jupiter & (1.0$\pm$0.2) ppb & 0.26-6.3 & \citet{Bezard2002}, \citet{Visscher2010} \\
    Saturn & $\sim$1 ppb        & 10-70 & \citet{Fouchet2017}, \citet{Wang2016} \\
    Uranus & $<$2.1 ppb         & $<$45 & \citet{Teanby2013}, \citet{Venot2020} \\
    Neptune & (0.20$\pm$0.05) ppm & 250 & \citet{Luszcz-Cook2013}, \\ 
            &                     &     & \citet{Moreno2011}, \citet{Venot2020} \\
    \hline
  \end{tabular}
  \end{center}
  \small{
  $^\mathrm{a}$ Oxygen abundances have been rescaled using the protosolar abundances of \tab{tab-abondances}. \\
}
\end{table}

  \subsection{Perspectives prior to an ice giant probe mission} 
  Thermochemical and diffusion models, like quench level models, still have to rely on several parameters that have to be assumed, i.e. the vertical mixing and the pressure-temperature profile. The main differences between their results then boil down to the differences in their chemical schemes. In this Section, we will review the progress we anticipate prior to the arrival of an ice giant probe in the 2040s regarding the determination of these input parameters.

    \subsubsection{Vertical mixing \label{Sect-mixing}}
    \citet{Visscher2010} showed that vertical mixing caused by convection in giant planet tropospheres depends on latitude and altitude, because of the planet rotation. \citet{Wang2015} further refined these calculations and concluded that the magnitude of this vertical mixing would decrease with latitude and depth. Its maximum is anticipated at the low latitudes. This means that the deepest quench levels, and therefore the highest abundances for species like CO and GeH$_4$, are expected to be observable at these same low latitudes. This is confirmed by recent Juno observations at Jupiter for GeH$_4$ \citep{Grassi2019}.
    
    However, the picture in giant planet upper tropospheres seems to be more complex than initially thought. In Jupiter, the abundance of NH$_3$ is far from the idealized well-mixed picture in the 1-50 bar range, with only a narrow band slightly north of the equator being uniformly mixed up to the NH$_3$ cloud \citep{Bolton2017,Li2017,dePater2019}. \citet{Guillot2019} proposed that this distribution is likely caused by the formation of NH$_3$-H$_2$O mesh balls in convective storms. Such an equatorial plume had already been identified by \citet{Fletcher2009b} in Jupiter's and Saturn's PH$_3$ distributions. In Neptune, \citet{Tollefson2019a,Tollefson2019b} showed that condensibles like H$_2$S and NH$_3$ were subject to tropospheric circulation and/or meteorology and that the circulation pattern extends down to the $\sim$30 bar level.
    
    Disk-resolved tropospheric observations with facilities like e.g., ALMA, e-VLA and JWST, and 3D general circulation model (GCM) are therefore required to better understand upper tropospheric circulation and chemistry \citep{Fletcher2020a,Fletcher2020b}. \citet{Venot2019,Venot2020} have proposed a reduced chemical scheme from their more complete 1D thermochemical model in view of their implementation in more complex 3D GCMs. Nailing down the latitude range where vertical mixing is most efficient in transporting disequilibrium species up to observable levels will be key in setting the entry latitude to target in priority with a shallow probe to increase its chances to access the well-mixed region of the explored atmosphere.

    \subsubsection{Temperature profile}
    One of the main unknown in giant planet tropospheres is the temperature-pressure field. It bears implication on circulation, kinetics, condensation layers, vertical mixing, etc. Except the Galileo probe measurements, which probed Jupiter down to the 22 bar level \citep{Seiff1998}, there is no such deep temperature measurement in any other giant planet. The fact that Galileo entered a 5$\mu$m hot spot further questions the representativeness of the measurements. In the other giants, there is a large uncertainty beneath the 2 bar level, which is the deepest level probed by occultation with Voyager 2 \citep{Lindal1985,Lindal1987,Lindal1990,Lindal1992}.
    Moreover, latitudinal variability remains unconstrained, even if the observed tropospheric distributions of several condensibles are hints of such variability \citep{Sromovsky2008,Karkoschka2011,Irwin2019b,Tollefson2019a,Molter2019}.
    
\begin{figure}[t]
      \begin{center}
        \includegraphics[width=\textwidth,keepaspectratio]{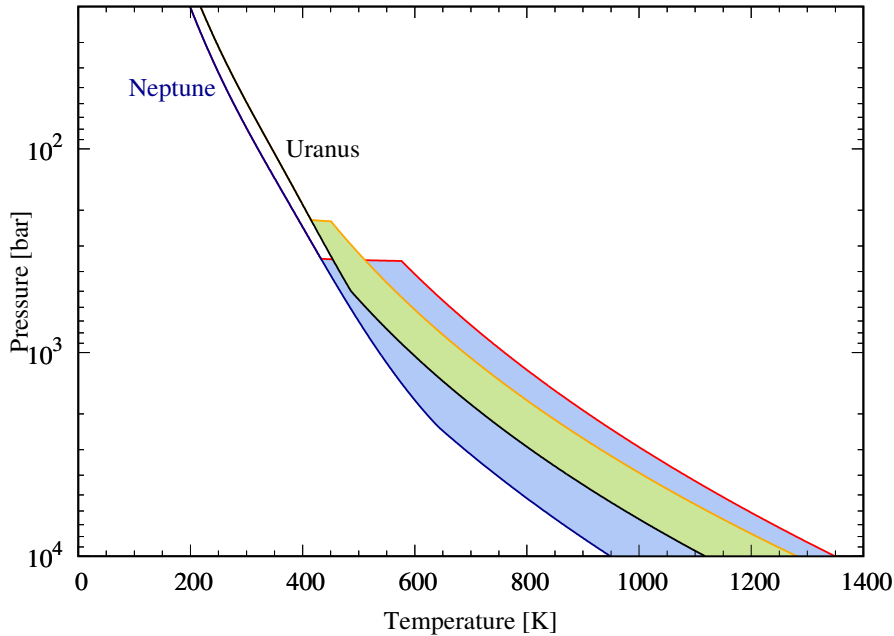}
      \end{center}
      \caption{Range of possible temperature profiles in the tropospheres of Uranus and Neptune, following \citet{Leconte2017} and \citet{Cavalie2017}. The coldest profiles, corresponding to the wet adiabat, are shown in black (Uranus) and dark blue (Neptune). The warmest profiles, corresponding to the ``3-layer profile'', are shown in orange (Uranus) and red (Neptune). The filled areas (green for Uranus and blue for Neptune) indicate the range of possible temperatures.}
      \label{fig-profiles} 
    \end{figure}

    Extrapolation to higher pressures are required for thermochemical computations and a dry or a wet adiabat has often been used \citep{Luszcz-Cook2013}. However, \citet{Guillot1995} first showed that Uranus and Neptune are in a situation where mean molecular weight gradients could inhibit convection at the condensation level of CH$_4$ and produce in a steep increase of the temperature. Later, \citet{Leconte2017} demonstrated that the effect of convection inhibition would be even more dramatic deeper, at the H$_2$O condensation level. The resulting profile would then be a ``3-layer profile'', starting from a wet adiabat in the uppermost levels, a radiative layer where the water vapor mixing ratio is between a fixed critical value and its maximum internal value, and a dry adiabat deeper down. The range of possible temperature profiles in Uranus and Neptune, between the wet adiabat (the coldest) and the convection inhibited one (the warmest), are shown in \fig{fig-profiles}. \citet{Cavalie2017} showed that the implications on the deep composition as derived from thermochemical modeling are significant. Therefore, any improvement in our knowledge of the tropospheric temperature is regarded as highly valuable.
    
    \subsubsection{Chemical scheme \label{Sect-chemistry}}
    The chemical scheme adopted in thermochemical calculation is obviously key on determining the quench level of the species of interest. \citet{Wang2016} compared the chemical schemes of \citet{Moses2011} and \citet{Venot2012} in applications to Jupiter and Saturn. They found that these two schemes resulted in differences of about an order of magnitude on the abundance of CO, all other parameters and deep composition being kept similar. \citet{Moses2014} already pointed out a significant difference in their carbon-oxygen chemistry, identifying a methanol (CH$_3$OH) conversion reaction as the main responsible. \citet{Venot2020} fully revised their CH$_3$OH chemistry, adopting recent experimental results of \citet{Burke2016}. The new scheme was validated over a wide range of temperature and pressure. The main changes concern the replacement of the reaction outlined by \citet{Moses2014} by a more detailed mechanism, in which pressure dependent reaction rates are adopted. Planets in which CO quenching occurs at high pressures are affected by the modifications.
    For Uranus and Neptune, the effect of this update is to lower the CO quenching level towards higher pressures, compared to the results obtained with the chemical scheme of \cite{Venot2012}. Consequently, to reproduce observational constraints of CH$_4$ and CO in the upper troposphere, a lower amount of H$_2$O is required in the deep tropospheric region where thermochemical equilibrium prevails. The O/H values found by \cite{Cavalie2017} using \citet{Venot2012}'s chemical scheme have been revised downwards. The O/H ratios necessary to reproduce current observations are $<$45 and 250 times protosolar value, for Uranus and Neptune respectively (\tab{tab-oxygen}).
    
    Chemical schemes currently used to model tropospheres of the ice giants contains only species made of C, H, O, and N. However, the recent detection of H$_2$S in Uranus and Neptune \citep{Irwin2018,Irwin2019a} make really necessary the addition of sulphur species. Such models would then require to account for cloud formation \citep{Atreya2005} as H$_2$S is involved in the formation of an NH$_4$SH cloud which consumes all NH$_3$ at these levels, and of an H$_2$S cloud above.
    Although not detected in ice giants, PH$_3$ might be present in these atmospheres also, as it is in Jupiter and Saturn. Alternately, its absence may serve as additional constraints for the deep oxygen abundance \citep{Visscher2005}. The addition of phosphorous species in chemical schemes is one of the next necessary step concerning the improvement of chemical schemes used to study ice giant atmospheres.
    
    As we said in Sect. \ref{Sect-mixing}, the heterogeneity of the troposphere, as seen in disk-resolved tropospheric observations, makes necessary the development of GCMs including a detailed chemistry. Full chemical schemes are too heavy ($\sim$100 species and $\sim$2000 reactions) to be incorporated in 3D models, as it would result to unreasonable computational time. The solution is to include a reduced chemical scheme, valid for a limited number of species of interest. In this purpose, reduced schemes have already been proposed by \cite{Venot2019,Venot2020} for H, C, N and O species. Such reduced schemes must be regularly updated, e.g. to account for sulphur and phosphorus species.

    \subsubsection{Summary}
    \citet{Cavalie2017} have shown the range of O/H values one can derive for Uranus and Neptune given the current limited knowledge of several key parameters in thermochemical modeling. Future progress in deep composition derivation from thermochemical modeling of the tropospheres of the ice giants require improvements to be made on the knowledge of the parameters this kind of models rely on. A better understanding of the 3D dynamics and chemistry to better constrain the disk variability of vertical mixing and temperature, both crucial in fixing quench levels, will involve a combination of disk-resolved observations, chemical and general circulation modeling work. Chemical networks will need to be extended to other key element bearing species and will have to include phase change processes for condensible species. Reaction rates for which either the temperature validity range or the accuracy are insufficient will need to be identified and improved \citep[see e.g.][]{Dobrijevic2010}.

\section{Thermochemical modeling in support of an ice giant atmospheric probe mass spectrometer \label{Sect-mass-spectro}}
  In this Section, we will briefly remind the baseline objectives of an ice giant mass spectrometer. We will then present the synergistic coupling of mass spectrometry with thermochemical modeling, and the requirements on the instrument such coupling drives. We will finally show how increasing the probe penetration depth could improve the science return of the probe mission. More details on the possible mass spectrometer can be found in \citet{Vorburger2020}.

  \subsection{Baseline ice giant probe mass spectrometer}
  In the current baseline scenario proposed for ice giant atmospheric probes (e.g. \citealt{Mousis2018} and \citealt{Vorburger2020}, and references therein), inherited from recent Saturn probe proposals \citep{Mousis2014a,Mousis2016,Atkinson2016,Atkinson2018}, the nominally targeted depth is the 10-bar level. The mass spectrometer proposed for the Hera mission to Saturn and that is now considered for an ice giant probe mission consisted of several units, among which a time-of-flight mass spectrometer (TOF-MS) which has a nominal mass resolution of $\sim$1000 used for neutral gas composition, and a tunable laser spectrometer used for selected isotopic ratio measurements \citep{Mousis2016,Wurz2012}. The TOF-MS will be 1000 times more sensitive than the Ion and Neutral Mass Spectrometer of the Cassini mission.
  
  Reaching the 10-bar level with such an instrument will ensure accurate measurements of helium (within 2\%) and the other noble gases (within 10\%) that are expected to be well-mixed in both altitude and latitude. If the entry latitude is close to the equator, where methane is most abundant (see Section \ref{Sect-compo-C}), the probe may also measure a carbon abundance representative of the deep C/H value. However Juno has shown with NH$_3$ that the well-mixed region for condensible species can occur much below than the cloud base of that species \citep{Bolton2017,Li2017}. 
  
  It will also measure the abundance of sulphur above the NH$_4$SH cloud, and thus the minimum S/N. However, N/H and S/H will remain out of reach, as the NH$_4$SH cloud deck is expected at 40 bars or so. Oxygen will also remain out of reach for a direct measurement, as water condenses as deep as a few hundred bars already in the ice giants \citep{Atreya2005,Cavalie2017}.

  \subsection{Synergistic coupling of in situ mass spectrometry and thermochemical modeling in ice giants}
  During its descent in the upper troposphere of an ice giant, the probe mass spectrometer will be sensitive to several gases (beyond helium, nobles gases, and methane) of key importance to constrain the deep composition of the ice giant from thermochemical modeling, provided that more ambitious mass resolution requirements are fulfilled. 
  
  The first species of interest is CO especially in Uranus, where its tropospheric component has not yet been unambiguously identified \citep{Encrenaz2004,Cavalie2014}. Combining mass spectrometry determination of the CO abundance within 10\%, accurate temperature-pressure measurements of the Atmospheric Structure Instrument \citep{Ferri2019}, and thermochemical modeling as detailed in Section \ref{Sect-thermo}, it will be possible to constrain the deep O/H of the ice giants more accurately than possible before. One limitation though regarding the deep O/H derivation is the single entry point of the probe which will result in a single temperature-pressure profile. Any variability over the planet, that is likely to occur, will remain out of reach to the probe. One key will then consist in picking the probe entry point such that we get a profile which is as much as possible representative for the whole planet by trajectory design and by knowing  what places to avoid (e.g., avoid Great Dark Spots).
  
  But directly measuring the abundance of CO bears several implications for the mass spectrometer. First, carbon dioxide (CO$_2$) needs also to be measured accurately as well as its fragmentation into CO inside the instrument. As CO$_2$ has the same mass as propane (C$_3$H$_8$), a mass resolution $m/\Delta m$ $>$600 is already required. Moreover, the instrument must be able to mass-separate CO from dinitrogen (N$_2$) and ethylene (C$_2$H$_4$). These species all reside at mass 28 on a mass spectrum. To separate them, a mass resolution $m/\Delta m$ $>$3000 is required at comparable abundance of CO and N$_2$.
  
  As already stated in \ref{Sect-compo-P}, additional constraints on the deep O/H can be obtained from measuring the abundance of PH$_3$ by solving the following thermochemical equation:
    \begin{equation}
      4\mathrm{PH}_3 + 6\mathrm{H}_2\mathrm{O} = \mathrm{P}_4\mathrm{O}_6 + 12\mathrm{H}_2.
  \end{equation}
  This would, in turn, require a mass resolution $m/\Delta m$ $>$4000, or a suitable chemical pre-separation \citep{Vorburger2020}, to separate PH$_3$ from H$_2$S, another mass-34 species detected in both ice giants. In the same spirit, ethane (C$_2$H$_6$) and acetylene (C$_2$H$_2$) can also be used as an additional constraint in the carbon-oxygen thermochemistry \citep{Fegley1985,Fegley1994}.
  
\begin{figure}[t]
    \begin{center}
      \includegraphics[width=\textwidth,keepaspectratio]{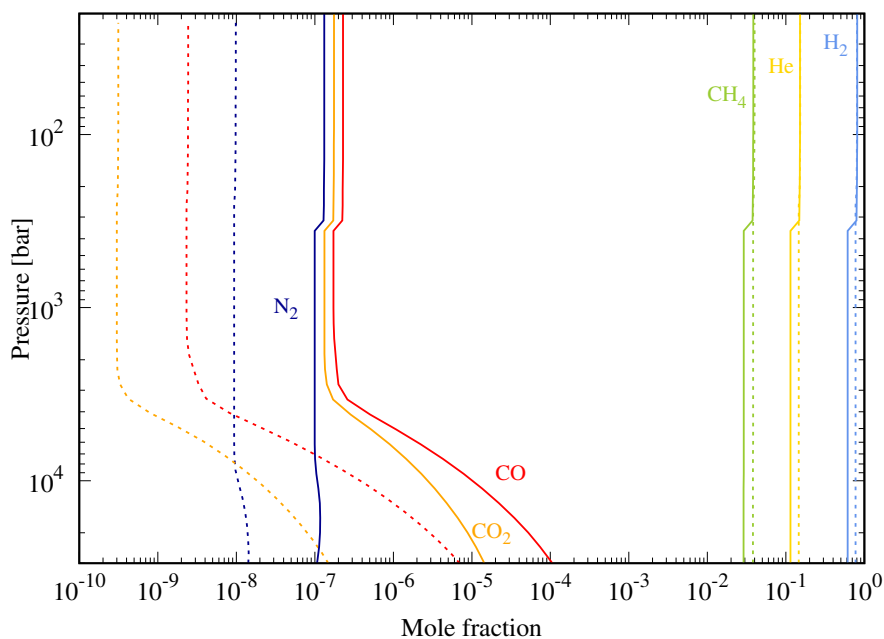}
    \end{center}
    \caption{Vertical profiles of CO (red), CO$_2$ (orange) and N$_2$ (dark blue) for Uranus (dashed lines) and Neptune (solid lines). Other main species like H$_2$, He and CH$_4$ are also shown with the corresponding layout.}
    \label{fig-carbon-nitrogen} 
  \end{figure}

  The direct benefit of such a high mass resolution would be a measurement of the N$_2$ abundance. In the same way CO is used to constrain the deep H$_2$O, N$_2$ can be used in thermochemical modeling to reproduce its upper tropospheric abundance and constrain the deep NH$_3$ abundance and thus the deep N/H, without the need for the probe to go beneath the NH$_4$SH cloud deck. \fig{fig-carbon-nitrogen} shows the vertical profiles of CO and N$_2$ for Uranus and Neptune using the model described in \citet{Venot2020} and assuming the deep N/H of \tab{tab-abondances}. It shows that N$_2$ could be present in both planets with abundances comparable or even higher than CO. Having the deep N/H established this way, it would then be possible to derive the deep S/H from the combined reconstruction of the deep NH$_3$ and H$_2$S abundance profiles below the NH$_4$SH cloud deck and current H$_2$S observations above its own cloud \citep{Irwin2018,Irwin2019a}. The current limitation of a descent probe in ice giants to measure directly N/H and S/H because of end-of-operations at 10 bars, i.e. before reaching the NH$_4$SH cloud deck at 40 bars or so, would thus be waived. 
  
  \subsection{The question of depth}
  It is obvious that even more robust N/H and S/H values could be directly measured by an ice giant probe mass spectrometer, provided that it would reach below the NH$_4$SH cloud. However, such a depth goal bears implications on several technical aspects.
  
  The descent would take longer to reach this level rather than the 10-bar level. The relay spacecraft would thus have to fly slower above the entry point to keep the radio link with the probe. For an orbiter, this would imply a higher orbit. However, placing the relay spacecraft further away from the probe would degrade the data rate. The situation on the data rate side is even more challenging as the atmospheric opacity increases exponentially with depth, especially beyond 15 bars, even though the situation is less critical now that it has been established that the main absorber in the altitude range will be H$_2$S rather than NH$_3$. To overcome this problem, two possibilities are being discussed: a second relay spacecraft could be sent or the communication system could use optical laser instead of radio frequencies.

  \subsection{The question of the probe entry latitude}
  To measure abundances of major species that are representative of their deep values, a probe should target an entry site where the material is uniformly mixed. There is already observational evidence that the high latitude may be depleted, at least in the upper troposphere, in CH$_4$ and H$_2$S \citep{Sromovsky2014,Irwin2019b,Tollefson2019a}. This, in turn, implies targeting latitudes where tropospheric mixing is maximum, i.e. the low latitudes in the ice giants according to \fig{fig-Kzz}. For disequilibrium species, which are quenched in layers where their abundance increase with depth (e.g., CO and PH$_3$), to be more likely detected by a mass spectrometer, low latitudes should also be favored. It should be noted however that there are some disequilibrium species (e.g., AsH$_3$) for which high latitudes should be more favorable.
  \\
  Now that we have reviewed how the bulk composition of the ice giants can be constrained from the combination of in situ measurements and thermochemical modeling (possibly supplemented by remote sensing observations), we will review how it can help us better understand the interior of these planets and the processes that led to their formation.

\section{Link between deep composition, interior models, and planet formation \label{Sect-link}}

Because the atmospheres and interiors of the giant planets are intimately linked and there is no probe that can go very deep into either planet, a proper understanding of Uranus and Neptune's atmospheres is crucial to characterise their interiors.  The atmospheric thermal profiles and deep compositions put constraints and impact directly on the interior model calculations \citep{Guillot2005,Guillot2015,Helled2018}. 

 The internal structure of Uranus and Neptune is estimated using interior models that fit the observational data for mass, radius, luminosity, atmospheric temperature, atmospheric abundances and gravity data. With only one mission (Voyager 2) visiting these planets so far, the gravity data that was obtained by remote sensing is much more limited than what we have for Jupiter \citep{Bolton2017,Iess2018} and Saturn \citep{Iess2019}. In \tab{interior-data}, we show the parameters used for interior model calculations for Uranus and Neptune with the exception of the atmospheric abundances, already shown in \tab{tab-abondances}. The data for Jupiter and Saturn are shown for comparison.  

\begin{table}
\begin{center}
\begin{tabular}{c c c c c}        
\hline                
Parameter & Jupiter & Saturn & Uranus & Neptune \\   
\hline                       
Mass/$10^{24}$ (kg) & $1898.187 \pm 0.088^\mathrm{a}$ & 568.336 $\pm$ 0.026$^\mathrm{b}$ & 86.8127 $\pm$ 0.0040$^\mathrm{c}$& 102.4126 $\pm$ 0.0048$^\mathrm{d}$\\
Equatorial radius (km) & 71492 $\pm$ 4$^\mathrm{e}$ & 60268 $\pm$ 4$^\mathrm{e}$ & 25559 $\pm$ 4$^\mathrm{e}$ & 24764 $\pm$ 15$^\mathrm{e}$ \\
Temperature$_{1bar}$ (K) & 165 $\pm$ 4$^\mathrm{f}$ & 135 $\pm$ 5$^\mathrm{f}$ & 76 $\pm$ 2$^\mathrm{f}$ & 72 $\pm$ 2$^\mathrm{f}$ \\
Intrinsic flux ($J~s^{-1}~m^{-2}$) & 5.44 $\pm$ 0.43$^\mathrm{g}$ & 2.01 $\pm$ 0.14$^\mathrm{g}$ & 0.042$^{+0.047}_{-0.042}{}^\mathrm{g}$ &  0.433 $\pm$ 0.046$^\mathrm{g}$ \\
J$_2$/$10^{6}$ & 14696.572 $\pm$ 0.0046$^\mathrm{h}$ & 16290.573 $\pm$ 0.0093$^\mathrm{i}$ & 3516 $\pm$ 3.2$^\mathrm{j}$ & 3408.4 $\pm$ 3404.5$^d$\\
J$_3$/$10^{6}$ & $-$0.042 $\pm$ 0.0033$^\mathrm{h}$ & 0.059 $\pm$ 0.0076$^\mathrm{i}$ & $-$ & $-$ \\
J$_4$/$10^{6}$ & $-$586.609 $\pm$ 0.0013$^\mathrm{h}$ & $-$935.314 $\pm$ 0.0123$^\mathrm{i}$ & $-$35.4 $\pm$ 34.1$^j$ & $-$33.4 $\pm$ 32.9$^d$ \\ 
J$_5$/$10^{6}$ &  $-$0.069 $\pm$ 0.0026$^\mathrm{h}$ & $-$0.224 $\pm$ 0.018$^\mathrm{i}$ &$-$ & $-$ \\ 
J$_6$/$10^{6}$ & 34.198 $\pm$ 0.003$^\mathrm{h}$ & 86.340 $\pm$ 0.029$^\mathrm{i}$ & $-$ & $-$ \\ 
J$_7$/$10^{6}$ & 0.124 $\pm$ 0.0056$^\mathrm{h}$ & $-$ & $-$ & $-$ \\ 
J$_8$/$10^{6}$ & $-$2.426 $\pm$ 0.0083$^\mathrm{h}$ & $-$14.624 $\pm$ 0.0683$^\mathrm{i}$ & $-$ & $-$ \\ 
J$_9$/$10^{6}$ & $-$0.106 $\pm$ 0.0146$^\mathrm{h}$ & $-$ & $-$ & $-$ \\ 
J$_{10}$/$10^{6}$ & 0.172 $\pm$ 0.023$^\mathrm{h}$ & 4.672 $\pm$ 0.14$^\mathrm{i}$ & $-$ & $-$ \\ 
J$_{12}$/$10^{6}$ & $-$ & $-$0.997 $\pm$ 0.224$^\mathrm{i}$ & $-$ & $-$ \\ 
\hline                                   
\end{tabular}
\label{interior-data}
\end{center}
  \small{
  $^\mathrm{a}$ Jacobson 2003 - published in the JPL website: https://ssd.jpl.nasa.gov/?planet\_phys\_par \\
  $^\mathrm{b}$ \citet{Jacobson2006}\\
  $^\mathrm{c}$\citet{Jacobson2014}\\
  $^\mathrm{d}$\citet{Jacobson2009}\\
  $^\mathrm{e}$\citet{Archinal2018}\\
  $^\mathrm{f}$\citet{Lindal1992}, note that \citet{Seiff1998} derived 166.1 K for Jupiter\\
  $^\mathrm{g}$\citet{Pearl1991}\\
  $^\mathrm{h}$\citet{Iess2018}\\
  $^\mathrm{i}$\citet{Iess2019}\\
  $^\mathrm{j}$\citet{Lindal1981}, \citet{Helled2013} derive slightly different values
  }
\end{table}

The information in \tab{interior-data} is combined with interior models to calculate the mass of heavy elements and their distribution in the interior, investigating all possible interior structures for Uranus and Neptune (see Section \ref{structure}). Given that one of the most accepted theories for the formation of these planets requires that a core forms first and the gas is accreted later on (see Section \ref{formation}), the constraints obtained from the interior models are crucial to understand the history of these planets.  

Uranus and Neptune are usually referred to as twin planets, but in reality they have many differences. When looking at their masses and radii we notice that Neptune is denser than Uranus, by approximately 30$\%$. The reason for this difference is not clear, but it was suggested that giant impacts during their formation and evolution might have affected their structure \citep{Podolak2012}. Uranus has a much higher obliquity when compared with Neptune and all the other giants, that is also explained with a giant impact during its formation, and that may cause differences in the atmospheres between the two ice giants \citep{Safronov1966}. In addition, \tab{interior-data} shows that the intrinsic flux of these two planets is quite different. While Neptune emits more energy than it receives from the Sun, Uranus has an emitted flux an order of magnitude lower than its neighbour.  This implies that while Neptune is still cooling, Uranus is almost in equilibrium with the solar irradiation, which implies differences in the energy transport in their interiors and points towards different evolution for these two planets.

Regarding the link between the atmosphere and interior, one of the most important constraints needed for interior models are the atmospheric abundances, which have been extensively discussed in the previous Sections. Uranus and Neptune are different from Jupiter and Saturn because they are not merely dominated by hydrogen and helium, and may be highly enriched in heavy elements. While H and He are consistent with the protosolar abundances, C has an enrichment of 80$\pm$20 compared to the protosun \citep{Atreya2018}, but this may as well be a lower limit only. 

Another relevant parameter used in interior models is the temperature at 1 bar, that sets the upper boundary for these calculations.  This parameter is obtained from stellar and ring occultations, that also provides determination of the shape of the planets. However, we have to note that this data are limited to low-pressure values, approximately 0.1\,bar and even lower pressures \citep{French1998}, and this can bring uncertainties in the radius used to model these planets \citep{Helled2010}.  In addition to this, the thermal profile inferred to reach the 1 bar level is highly degenerate (it depends on many unknown parameters such as the refractivity which depends on the mean molecular weight and the temperature at each pressure level). Therefore, the temperature inferred corresponds to one possible solution, but there might be other possibilities \citep{Guillot1995,Sromovsky2011}. 

The magnetic field is another observable quantity that provides constraints to understand the boundary between the deep atmosphere and the interior. Observations suggest that there is a convective and electrically conductive region that extends down to 20\% of the radius \citep{Stanley2004,Stanley2006,Redmer2011}. This is directly linked with the dynamics of Uranus and Neptune's atmospheres, with zonal winds that extend down to approximately 1000 km below the clouds \citep{Kaspi2013} and putting constraints on the interior models and linking it with the deep atmosphere. 

\subsection{Formation theories}\label{formation}
The most accepted scenario to explain the formation of the giant planets is the core accretion model, where the planets grow first their cores and then, once they reach a critical core mass, start accreting gas and forming their gaseous envelopes \citep{Pollack1996}.  There are different theories to explain how the core was first formed, that can be either by accreting planetesimals, bodies of some km in size \citep[e.g.][]{Alibert2005}, or by pebbles of some mm to cm in size \citep[e.g.][]{Lambrechts2014}. Regarding their gaseous envelope, once the critical core mass is reached, the giant planets start accreting gas in a runaway fashion, and one of the long standing questions in the case of Uranus and Neptune is how to stop such gas accretion to prevent them of accreting a massive gaseous envelope and becoming gas giants. One of the ideas to solve this problem suggests that, in a planetesimal-driven scenario, the planets formed in a region with a smaller density of solids when compared to where Jupiter and Saturn were formed. Their cores therefore grew slowly enough for the protoplanetary disk to be almost dissipated by the time the protoplanets started the gas accretion phase. This is why they are sometimes referred as ``failed giants'' \citep{Pollack1996,Helled2014b}. Other ideas require fine tuning of the models to prevent the planets of entering the gas accretion mode \citep{Frelikh2017}.

The other theory to explain the formation of these planets is the disk instability.  According to this scenario, clumps formed in the protosolar disk due to gravitational instabilities that gave rise to the giant planets. Uranus and Neptune could have been formed in this scenario if there was substantial gaseous mass loss in the disk caused by tidal stripping or photo-evaporation (see \citealt{Helled2014c} and references therein). 

Given the different possible scenarios and competing theories, interior model calculations are crucial to disentangle these competing scenarios, and thus better understand the formation and evolution of these planets.

\subsection{Internal Structure of Uranus and Neptune}\label{structure}
\begin{figure}
 \begin{center}
\includegraphics[angle=0,width=\textwidth]{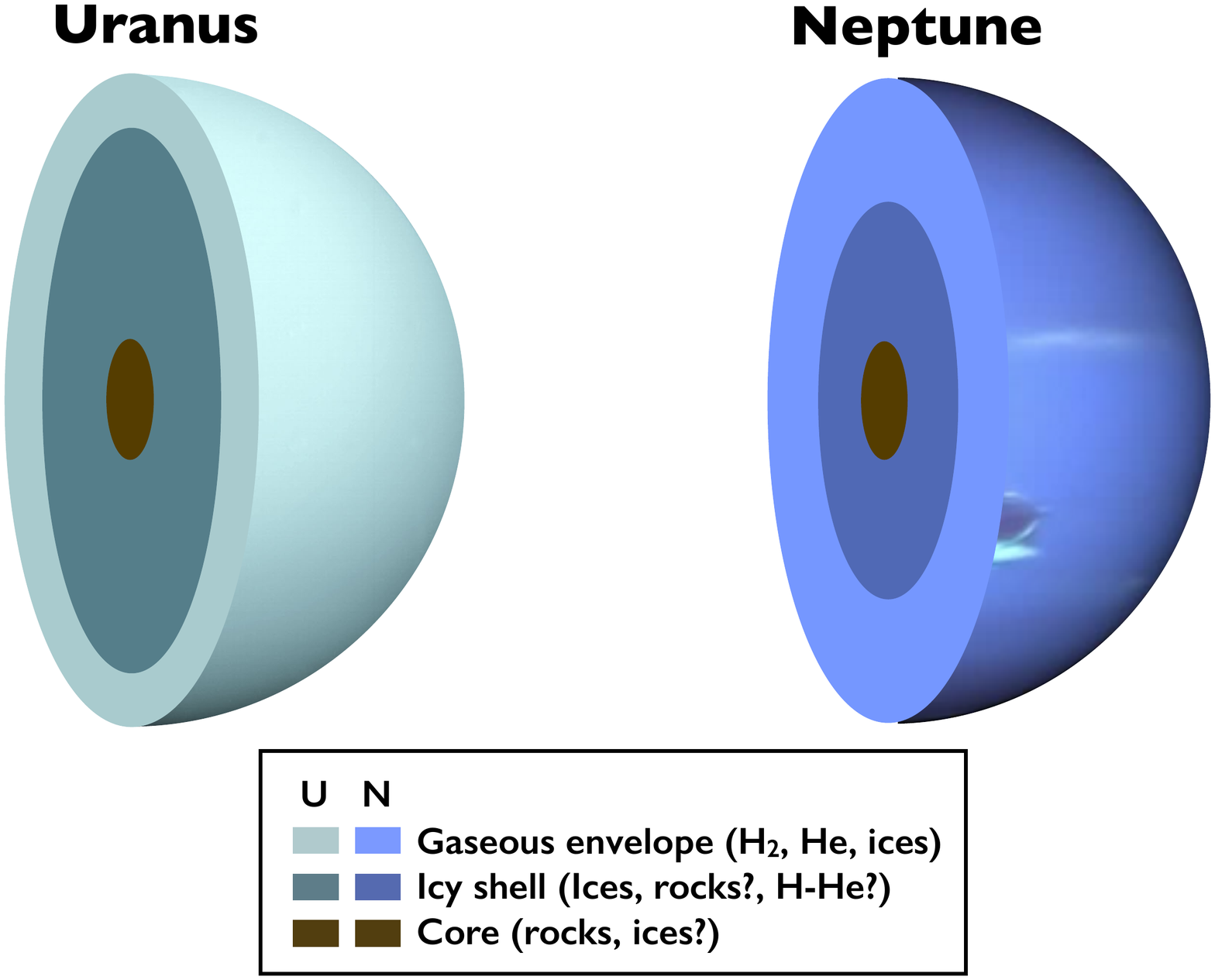}
 \end{center}
 \caption{Schematic view of Uranus and Neptune's interior structures.}
   \label{interior-figure}
\end{figure}

Interior models are constructed assuming hydrostatic, thermodynamic, mass and energy conservation, solving the following set of differential equations: 
\begin{equation}
\frac{\partial P}{\partial r} = -\rho~g
\end{equation}
\begin{equation}
\frac{\partial T}{\partial r} = \frac{\partial P}{\partial r} \frac{T}{P}\nabla_T
\end{equation}
\begin{equation}
\frac{\partial m}{\partial r} = 4\pi r^2 \rho
\end{equation}
\begin{equation}
\frac{\partial L}{\partial r} = 4\pi r^2 \rho \Bigg(\dot{\epsilon} - T \frac{\partial S}{\partial t} \Bigg)
\end{equation}
with $P$ the pressure, $r$ the radius, $\rho$ the density, $g$ the gravitational acceleration, $T$ the temperature, $m$ the mass, $L$ the planet luminosity and $S$ its entropy.  

\begin{figure}
 \begin{center}
\includegraphics[angle=0,width=\textwidth]{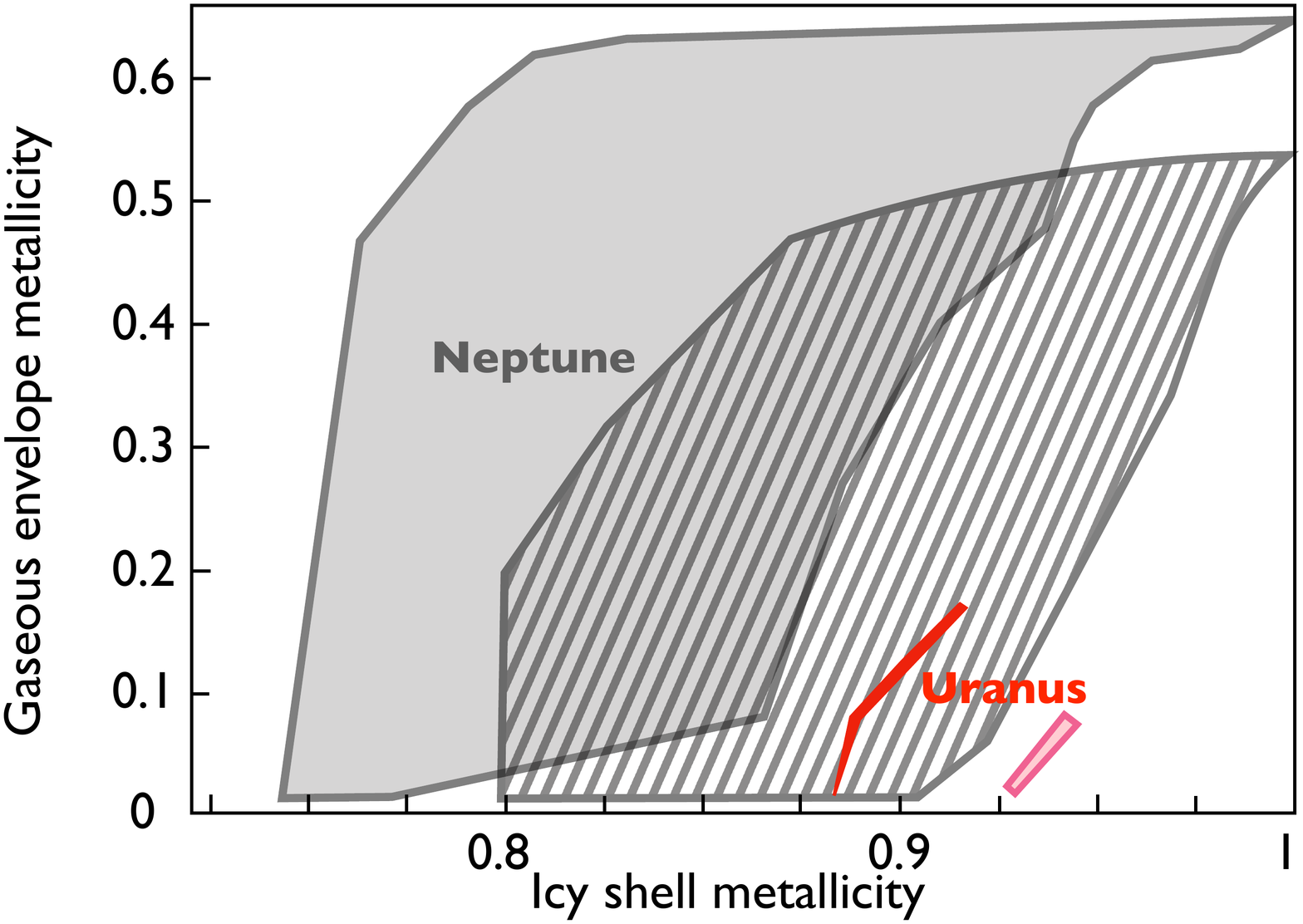}
 \end{center}
 \caption{Heavy elements mass fraction in the icy shell vs. the gaseous envelope. Structure models solutions for Uranus models are shown in red and for Neptune in grey (dashed). Models with a modified shape and rotation data for Uranus (pink) and Neptune (solid grey) are also shown. Adapted from \citet{Nettelmann2013}.}
   \label{interior2}
\end{figure}

Given the poor gravity constraints for these planets (see \tab{interior-data}), one of the major obstacles found when modeling their interiors and constraining the ice-to-rock ratio is the significant degeneracies in their potential composition \citep{Podolak1991, Hubbard1995, Baraffe2014}. Some of the structure models for Uranus and Neptune use three fully adiabatic layers (a rocky core, an icy shell and a gaseous envelope) and \textit{ab initio} equations of state (EOS) \citep{Nettelmann2013}. Nevertheless, other methods using no pre-stablished assumption regarding the structure or equations of state \citep[e.g.][]{Marley1995, Helled2011} also proved to be useful.  All these approaches find that the heavy element concentration increases towards the planetary centre, as shown by \fig{interior-figure}. Note that \fig{interior-figure} is a schematic representation where there are sharp boundaries between the different layers, but a more realistic idea is to consider a gradient of heavy elements and change in composition towards the interior \citep{Helled2018} (see also Section \ref{challenges}). More specific values for the metallicities in the gaseous envelope and the icy shell can be found in \fig{interior2}, which shows results found by \citet{Nettelmann2013}. As seen in \fig{interior2}, there are still big uncertainties in the internal structure of these planets. Some of the uncertainties are related to the fact that the core mass, the ice-to-rock ratio, the equations of state of mixtures of materials, the pressure of separation between the different layers, the depth of the winds and extent of differential rotation and the extent of compositional gradients, are highly unknown for these planets. Because the observational data are crucial to tackle these degeneracies, a more accurate determination of the gravity field and a proper characterization of the atmospheres of Uranus and Neptune are needed to get a better knowledge of their interior structures.

\subsection{Remaining questions and challenges for the future}\label{challenges}
Despite the substantial progress in the modeling of planetary interiors in the last decades, there are still several unsolved questions regarding the nature of Uranus and Neptune. One of the most important parameters when modeling the interior of these planets is EOS. In the last decade, there has been great progress in this area, with new EOS published for hydrogen and helium \citep{Militzer2013,Becker2014, Chabrier2019} and also in heavier material such as water \citep{Nettelmann2008,Mazevet2019}. Nevertheless, disagreement between the different EOS still cause differences in the internal structure of these planets (e.g. \citealt{Miguel2016} for Jupiter), and better constraints on EOS not only of individual elements but also in mixtures, together with higher pressure experiments could deeply improve interior structure models. Another important aspect of the interior modeling is the energy transport mechanism. The source of the different cooling rates of the planets is still unsolved. Better modeling, especially with potentially non-adiabatic models and a more realistic distribution of heavy elements in the interior, could help unveiling this story \citep{Helled2018,Vazan2020}. Last but not least, we need to understand the bulk composition of these planets: are they really formed by ices or do they have a substantial amount of rocks in their interiors? And how are these heavy elements distributed? These are questions that are far from being solved. 

When thinking of formation mechanisms, there are still several key questions that remain open: Where in the primitive nebula were Uranus and Neptune formed? Was pebble accretion or planetesimal accretion the primary mechanism that formed their cores? What are the mechanisms at play regarding gas accretion? What is the enrichment of the gaseous envelope and the radial distribution of heavy elements during the planet formation and subsequent evolution?
Understanding the connection between the atmosphere, interior and link with formation of these planets is still incomplete and one of the big challenges in planetary science for the future. New studies on the deposition of heavy elements in the forming giant planet \citep{Valletta2019} and recent results in exoplanet studies indicate that measurements of the envelope metallicities are relevant diagnostics of the bulk metallicity \citep{Thorngren2019}. Measurements from the Earth, but more importantly, at least for the gravity data and bulk composition, future space missions to Uranus and Neptune carrying in situ probes will provide constraints to reduce the degeneracies in calculations towards a better understanding on the atmosphere-interior connection, on the internal structure and ultimately the history of these worlds.

\section{Conclusion \label{Sect-conclu}}
An entry probe is the only means to measure the deep abundance of a number of species of key importance, notably the noble gases. These can put significant constraints on formation of Uranus and Neptune \citep{Mousis2018}. The difficulty with those cold distant worlds lies in the condensation of some key species, like CH$_4$, and to a more critical extent, H$_2$S, NH$_3$, and H$_2$O, which render their direct in situ measurement complicated, or even impossible.

Designing a probe that would reach the 40-50 bar level and return data to measure not only He/H (and other noble gases) and C/H, but also N/H and S/H, will be very challenging in the current timeframe (possible launch dates range from 2029 to 2034, \citealt{Simon2020}). The coupling of high resolution mass spectrometry ($m/\Delta m$ $>$4000) with accurate temperature-pressure measurements with thermochemical modeling at 10 bar is thus an interesting combination to infer the deep elemental abundances of condensible species not reachable by a shallow probe, like H$_2$O, NH$_3$ and H$_2$S, in the ice giants. 

The results of such an entry probe, combined with a better knowledge of gravity moments and magnetic field obtained from an orbiter, will undoubtedly result in major breakthroughs in our understanding of the formation and evolution of the ice giants of our Solar System, Uranus and Neptune.

\begin{acknowledgements}
T. Cavali\'e, O. Venot, and O. Mousis acknowledge support from CNES and the Programme National de Plan\'etologie (PNP) of CNRS/INSU. 
\end{acknowledgements}

\bibliographystyle{spbasic}      

\begin{thebibliography}{245}
\providecommand{\natexlab}[1]{#1}
\providecommand{\url}[1]{{#1}}
\providecommand{\urlprefix}{URL }
\expandafter\ifx\csname urlstyle\endcsname\relax
  \providecommand{\doi}[1]{DOI~\discretionary{}{}{}#1}\else
  \providecommand{\doi}{DOI~\discretionary{}{}{}\begingroup
  \urlstyle{rm}\Url}\fi
\providecommand{\eprint}[2][]{\url{#2}}

\bibitem[{{Ali-Dib} et~al.(2014){Ali-Dib}, {Mousis}, {Petit}, and
  {Lunine}}]{Ali-Dib2014}
{Ali-Dib} M, {Mousis} O, {Petit} JM, {Lunine} JI (2014) {The Measured
  Compositions of Uranus and Neptune from their Formation on the CO Ice Line}.
  \apj 793:9, \doi{10.1088/0004-637X/793/1/9}

\bibitem[{{Alibert} et~al.(2005){Alibert}, {Mordasini}, {Benz}, and
  {Winisdoerffer}}]{Alibert2005}
{Alibert} Y, {Mordasini} C, {Benz} W, {Winisdoerffer} C (2005) {Models of giant
  planet formation with migration and disc evolution}. \aap 434(1):343--353,
  \doi{10.1051/0004-6361:20042032}

\bibitem[{{Amarsi} and {Asplund}(2017)}]{Amarsi2017}
{Amarsi} AM, {Asplund} M (2017) {The solar silicon abundance based on 3D
  non-LTE calculations}. \mnras 464(1):264--273, \doi{10.1093/mnras/stw2445}

\bibitem[{{Amarsi} et~al.(2018){Amarsi}, {Barklem}, {Asplund}, {Collet}, and
  {Zatsarinny}}]{Amarsi2018}
{Amarsi} AM, {Barklem} PS, {Asplund} M, {Collet} R, {Zatsarinny} O (2018)
  {Inelastic O+H collisions and the O I 777 nm solar centre-to-limb variation}.
  \aap 616:A89, \doi{10.1051/0004-6361/201832770}

\bibitem[{{Amarsi} et~al.(2019){Amarsi}, {Barklem}, {Collet}, {Grevesse}, and
  {Asplund}}]{Amarsi2019}
{Amarsi} AM, {Barklem} PS, {Collet} R, {Grevesse} N, {Asplund} M (2019) {3D
  non-LTE line formation of neutral carbon in the Sun}. \aap 624:A111,
  \doi{10.1051/0004-6361/201833603}

\bibitem[{{Archinal} et~al.(2018){Archinal}, {Acton}, {A'Hearn}, {Conrad},
  {Consolmagno}, {Duxbury}, {Hestroffer}, {Hilton}, {Kirk}, {Klioner},
  {McCarthy}, {Meech}, {Oberst}, {Ping}, {Seidelmann}, {Tholen}, {Thomas}, and
  {Williams}}]{Archinal2018}
{Archinal} BA, {Acton} CH, {A'Hearn} MF, {Conrad} A, {Consolmagno} GJ,
  {Duxbury} T, {Hestroffer} D, {Hilton} JL, {Kirk} RL, {Klioner} SA, {McCarthy}
  D, {Meech} K, {Oberst} J, {Ping} J, {Seidelmann} PK, {Tholen} DJ, {Thomas}
  PC, {Williams} IP (2018) {Report of the IAU Working Group on Cartographic
  Coordinates and Rotational Elements: 2015}. Celest Mech Dyn Astr 130:22,
  \doi{10.1007/s10569-017-9805-5}

\bibitem[{{Arridge} et~al.(2014){Arridge}, {Achilleos}, {Agarwal}, {Agnor},
  {Ambrosi}, {Andr{\'e}}, {Badman}, {Baines}, {Banfield}, {Barth{\'e}l{\'e}my},
  {Bisi}, {Blum}, {Bocanegra-Bahamon}, {Bonfond}, {Bracken}, {Brandt},
  {Briand}, {Briois}, {Brooks}, {Castillo-Rogez}, {Cavali{\'e}}, {Christophe},
  {Coates}, {Collinson}, {Cooper}, {Costa-Sitja}, {Courtin}, {Daglis}, {de
  Pater}, {Desai}, {Dirkx}, {Dougherty}, {Ebert}, {Filacchione}, {Fletcher},
  {Fortney}, {Gerth}, {Grassi}, {Grodent}, {Gr{\"u}n}, {Gustin}, {Hedman},
  {Helled}, {Henri}, {Hess}, {Hillier}, {Hofstadter}, {Holme}, {Horanyi},
  {Hospodarsky}, {Hsu}, {Irwin}, {Jackman}, {Karatekin}, {Kempf}, {Khalisi},
  {Konstantinidis}, {Kr{\"u}ger}, {Kurth}, {Labrianidis}, {Lainey}, {Lamy},
  {Laneuville}, {Lucchesi}, {Luntzer}, {MacArthur}, {Maier}, {Masters},
  {McKenna-Lawlor}, {Melin}, {Milillo}, {Moragas-Klostermeyer}, {Morschhauser},
  {Moses}, {Mousis}, {Nettelmann}, {Neubauer}, {Nordheim}, {Noyelles}, {Orton},
  {Owens}, {Peron}, {Plainaki}, {Postberg}, {Rambaux}, {Retherford}, {Reynaud},
  {Roussos}, {Russell}, {Rymer}, {Sallantin}, {S{\'a}nchez-Lavega}, {Santolik},
  {Saur}, {Sayanagi}, {Schenk}, {Schubert}, {Sergis}, {Sittler}, {Smith},
  {Spahn}, {Srama}, {Stallard}, {Sterken}, {Sternovsky}, {Tiscareno}, {Tobie},
  {Tosi}, {Trieloff}, {Turrini}, {Turtle}, {Vinatier}, {Wilson}, and
  {Zarka}}]{Arridge2014}
{Arridge} CS, {Achilleos} N, {Agarwal} J, {Agnor} CB, {Ambrosi} R, {Andr{\'e}}
  N, {Badman} SV, {Baines} K, {Banfield} D, {Barth{\'e}l{\'e}my} M, {Bisi} MM,
  {Blum} J, {Bocanegra-Bahamon} T, {Bonfond} B, {Bracken} C, {Brandt} P,
  {Briand} C, {Briois} C, {Brooks} S, {Castillo-Rogez} J, {Cavali{\'e}} T,
  {Christophe} B, {Coates} AJ, {Collinson} G, {Cooper} JF, {Costa-Sitja} M,
  {Courtin} R, {Daglis} IA, {de Pater} I, {Desai} M, {Dirkx} D, {Dougherty} MK,
  {Ebert} RW, {Filacchione} G, {Fletcher} LN, {Fortney} J, {Gerth} I, {Grassi}
  D, {Grodent} D, {Gr{\"u}n} E, {Gustin} J, {Hedman} M, {Helled} R, {Henri} P,
  {Hess} S, {Hillier} JK, {Hofstadter} MH, {Holme} R, {Horanyi} M,
  {Hospodarsky} G, {Hsu} S, {Irwin} P, {Jackman} CM, {Karatekin} O, {Kempf} S,
  {Khalisi} E, {Konstantinidis} K, {Kr{\"u}ger} H, {Kurth} WS, {Labrianidis} C,
  {Lainey} V, {Lamy} LL, {Laneuville} M, {Lucchesi} D, {Luntzer} A, {MacArthur}
  J, {Maier} A, {Masters} A, {McKenna-Lawlor} S, {Melin} H, {Milillo} A,
  {Moragas-Klostermeyer} G, {Morschhauser} A, {Moses} JI, {Mousis} O,
  {Nettelmann} N, {Neubauer} FM, {Nordheim} T, {Noyelles} B, {Orton} GS,
  {Owens} M, {Peron} R, {Plainaki} C, {Postberg} F, {Rambaux} N, {Retherford}
  K, {Reynaud} S, {Roussos} E, {Russell} CT, {Rymer} AM, {Sallantin} R,
  {S{\'a}nchez-Lavega} A, {Santolik} O, {Saur} J, {Sayanagi} KM, {Schenk} P,
  {Schubert} J, {Sergis} N, {Sittler} EC, {Smith} A, {Spahn} F, {Srama} R,
  {Stallard} T, {Sterken} V, {Sternovsky} Z, {Tiscareno} M, {Tobie} G, {Tosi}
  F, {Trieloff} M, {Turrini} D, {Turtle} EP, {Vinatier} S, {Wilson} R, {Zarka}
  P (2014) {The science case for an orbital mission to Uranus: Exploring the
  origins and evolution of ice giant planets}. \planss 104:122--140,
  \doi{10.1016/j.pss.2014.08.009}

\bibitem[{{Atkinson} et~al.(2016){Atkinson}, {Simon}, {Banfield}, {Atreya},
  {Blacksberg}, {Brinckerhoff}, {Colaprete}, {Coustenis}, {Fletcher},
  {Guillot}, {Hofstadter}, {Lunine}, {Mahaffy}, {Marley}, {Mousis}, {Spilker},
  {Trainer}, and {Webster}}]{Atkinson2016}
{Atkinson} DH, {Simon} AA, {Banfield} D, {Atreya} SK, {Blacksberg} J,
  {Brinckerhoff} W, {Colaprete} A, {Coustenis} A, {Fletcher} L, {Guillot} T,
  {Hofstadter} M, {Lunine} JI, {Mahaffy} P, {Marley} MS, {Mousis} O, {Spilker}
  TR, {Trainer} MG, {Webster} C (2016) {Exploring Saturn - The Saturn PRobe
  Interior and aTmosphere Explorer (SPRITE) Mission}. In: AAS/Division for
  Planetary Sciences Meeting Abstracts, AAS/Division for Planetary Sciences
  Meeting Abstracts, vol~48, p \#123.29

\bibitem[{{Atkinson} et~al.(2018){Atkinson}, {Simon}, {Banfield}, {Atreya},
  {Blacksberg}, {Brinckerhoff}, {Colaprete}, {Coustenis}, {Fletcher},
  {Guillot}, {Hofstadter}, {Lunine}, {Mahaffy}, {Marley}, {Mousis}, {Spilker},
  {Trainer}, and {Webster}}]{Atkinson2018}
{Atkinson} DH, {Simon} A, {Banfield} D, {Atreya} S, {Blacksberg} J,
  {Brinckerhoff} W, {Colaprete} A, {Coustenis} A, {Fletcher} L, {Guillot} T,
  {Hofstadter} M, {Lunine} J, {Mahaffy} P, {Marley} M, {Mousis} O, {Spilker} T,
  {Trainer} M, {Webster} C (2018) {SPRITE (Saturn PRobe Interior and aTmosphere
  Explorer): A Saturn Entry Probe Mission Concept}. In: European Planetary
  Science Congress, pp EPSC2018--65

\bibitem[{{Atreya} and {Wong}(2005)}]{Atreya2005}
{Atreya} SK, {Wong} AS (2005) {Coupled Clouds and Chemistry of the Giant
  Planets{\textemdash} A Case for Multiprobes}. \ssr 116(1-2):121--136,
  \doi{10.1007/s11214-005-1951-5}

\bibitem[{{Atreya} et~al.(1999){Atreya}, {Wong}, {Owen}, {Mahaffy}, {Niemann},
  {de Pater}, {Drossart}, and {Encrenaz}}]{Atreya1999}
{Atreya} SK, {Wong} MH, {Owen} TC, {Mahaffy} PR, {Niemann} HB, {de Pater} I,
  {Drossart} P, {Encrenaz} T (1999) {A comparison of the atmospheres of Jupiter
  and Saturn: deep atmospheric composition, cloud structure, vertical mixing,
  and origin}. \planss 47:1243--1262, \doi{10.1016/S0032-0633(99)00047-1}

\bibitem[{{Atreya} et~al.(2003){Atreya}, {Mahaffy}, {Niemann}, {Wong}, and
  {Owen}}]{Atreya2003}
{Atreya} SK, {Mahaffy} PR, {Niemann} HB, {Wong} MH, {Owen} TC (2003)
  {Composition and origin of the atmosphere of Jupiter - an update, and
  implications for the extrasolar giant planets}. Planetary and Space Science
  51(2):105--112, \doi{10.1016/S0032-0633(02)00144-7}

\bibitem[{Atreya et~al.(2018)Atreya, Crida, Guillot, Lunine, Madhusudhan, and
  Mousis}]{Atreya2018}
Atreya SK, Crida A, Guillot T, Lunine JI, Madhusudhan N, Mousis O (2018) The
  Origin and Evolution of Saturn, with Exoplanet Perspective, Cambridge
  University Press, p 5?43. Cambridge Planetary Science,
  \doi{10.1017/9781316227220.002}

\bibitem[{{Atreya} et~al.(2020){Atreya}, {Hofstadter}, {In}, {Mousis}, {Reh},
  and {Wong}}]{Atreya2020}
{Atreya} SK, {Hofstadter} MH, {In} JH, {Mousis} O, {Reh} K, {Wong} MH (2020)
  {Deep Atmosphere Composition, Structure, Origin, and Exploration, with
  Particular Focus on Critical in situ Science at the Icy Giants}. \ssr
  216(1):18, \doi{10.1007/s11214-020-0640-8}

\bibitem[{{Baines} et~al.(1995){Baines}, {Mickelson}, {Larson}, and
  {Ferguson}}]{Baines1995}
{Baines} KH, {Mickelson} ME, {Larson} LE, {Ferguson} DW (1995) {The abundances
  of methane and ortho/para hydrogen on Uranus and Neptune: Implications of New
  Laboratory 4-0 H2 quadrupole line parameters}. \icarus 114:328--340,
  \doi{10.1006/icar.1995.1065}

\bibitem[{{Bar-Nun} et~al.(1988){Bar-Nun}, {Kleinfeld}, and
  {Kochavi}}]{Bar-Nun1988}
{Bar-Nun} A, {Kleinfeld} I, {Kochavi} E (1988) {Trapping of gas mixtures by
  amorphous water ice}. \prb 38:7749--7754, \doi{10.1103/PhysRevB.38.7749}

\bibitem[{{Baraffe} et~al.(2014){Baraffe}, {Chabrier}, {Fortney}, and
  {Sotin}}]{Baraffe2014}
{Baraffe} I, {Chabrier} G, {Fortney} J, {Sotin} C (2014) {Planetary Internal
  Structures}. In: {Beuther} H, {Klessen} RS, {Dullemond} CP, {Henning} T (eds)
  Protostars and Planets VI, p 763,
  \doi{10.2458/azu_uapress_9780816531240-ch033}

\bibitem[{{Becker} et~al.(2014){Becker}, {Lorenzen}, {Fortney}, {Nettelmann},
  {Sch{\"o}ttler}, and {Redmer}}]{Becker2014}
{Becker} A, {Lorenzen} W, {Fortney} JJ, {Nettelmann} N, {Sch{\"o}ttler} M,
  {Redmer} R (2014) {Ab Initio Equations of State for Hydrogen (H-REOS.3) and
  Helium (He-REOS.3) and their Implications for the Interior of Brown Dwarfs}.
  \apjs 215(2):21, \doi{10.1088/0067-0049/215/2/21}

\bibitem[{{Beer}(1975)}]{Beer1975}
{Beer} R (1975) {Detection of carbon monoxide in Jupiter}. \apjl
  200:L167--L169, \doi{10.1086/181923}

\bibitem[{{B{\'e}zard} et~al.(1997){B{\'e}zard}, {Griffith}, {Kelly}, {Lacy},
  {Greathouse}, and {Orton}}]{Bezard1997}
{B{\'e}zard} B, {Griffith} CA, {Kelly} DM, {Lacy} JH, {Greathouse} T, {Orton} G
  (1997) {Thermal Infrared Imaging Spectroscopy of Shoemaker-Levy 9 Impact
  Sites: Temperature and HCN Retrievals}. \icarus 125(1):94--120,
  \doi{10.1006/icar.1996.5610}

\bibitem[{{B{\'e}zard} et~al.(1998){B{\'e}zard}, {Feuchtgruber}, {Moses}, and
  {Encrenaz}}]{Bezard1998}
{B{\'e}zard} B, {Feuchtgruber} H, {Moses} JI, {Encrenaz} T (1998) {Detection of
  methyl radicals (CH\_3) on Saturn}. \aap 334:L41--L44

\bibitem[{{B{\'e}zard} et~al.(1999){B{\'e}zard}, {Romani}, {Feuchtgruber}, and
  {Encrenaz}}]{Bezard1999}
{B{\'e}zard} B, {Romani} PN, {Feuchtgruber} H, {Encrenaz} T (1999) {Detection
  of the Methyl Radical on Neptune}. \apj 515(2):868--872, \doi{10.1086/307070}

\bibitem[{{B{\'e}zard} et~al.(2002){B{\'e}zard}, {Lellouch}, {Strobel},
  {Maillard}, and {Drossart}}]{Bezard2002}
{B{\'e}zard} B, {Lellouch} E, {Strobel} D, {Maillard} JP, {Drossart} P (2002)
  {Carbon Monoxide on Jupiter: Evidence for Both Internal and External
  Sources}. \icarus 159:95--111, \doi{10.1006/icar.2002.6917}

\bibitem[{{Bolton} et~al.(2017){Bolton}, {Adriani}, {Adumitroaie}, {Allison},
  {Anderson}, {Atreya}, {Bloxham}, {Brown}, {Connerney}, {DeJong}, {Folkner},
  {Gautier}, {Grassi}, {Gulkis}, {Guillot}, {Hansen}, {Hubbard}, {Iess},
  {Ingersoll}, {Janssen}, {Jorgensen}, {Kaspi}, {Levin}, {Li}, {Lunine},
  {Miguel}, {Mura}, {Orton}, {Owen}, {Ravine}, {Smith}, {Steffes}, {Stone},
  {Stevenson}, {Thorne}, {Waite}, {Durante}, {Ebert}, {Greathouse}, {Hue},
  {Parisi}, {Szalay}, and {Wilson}}]{Bolton2017}
{Bolton} SJ, {Adriani} A, {Adumitroaie} V, {Allison} M, {Anderson} J, {Atreya}
  S, {Bloxham} J, {Brown} S, {Connerney} JEP, {DeJong} E, {Folkner} W,
  {Gautier} D, {Grassi} D, {Gulkis} S, {Guillot} T, {Hansen} C, {Hubbard} WB,
  {Iess} L, {Ingersoll} A, {Janssen} M, {Jorgensen} J, {Kaspi} Y, {Levin} SM,
  {Li} C, {Lunine} J, {Miguel} Y, {Mura} A, {Orton} G, {Owen} T, {Ravine} M,
  {Smith} E, {Steffes} P, {Stone} E, {Stevenson} D, {Thorne} R, {Waite} J,
  {Durante} D, {Ebert} RW, {Greathouse} TK, {Hue} V, {Parisi} M, {Szalay} JR,
  {Wilson} R (2017) {Jupiter's interior and deep atmosphere: The initial
  pole-to-pole passes with the Juno spacecraft}. Science 356(6340):821--825,
  \doi{10.1126/science.aal2108}

\bibitem[{{Boss}(1997)}]{Boss1997}
{Boss} AP (1997) {Giant planet formation by gravitational instability.} Science
  276:1836--1839, \doi{10.1126/science.276.5320.1836}

\bibitem[{{Boss}(2002)}]{Boss2002}
{Boss} AP (2002) {Formation of gas and ice giant planets}. Earth and Planetary
  Science Letters 202(3-4):513--523, \doi{10.1016/S0012-821X(02)00808-7}

\bibitem[{{Bregman} et~al.(1975){Bregman}, {Lester}, and {Rank}}]{Bregman1975}
{Bregman} JD, {Lester} DF, {Rank} DM (1975) {Observations of the
  {\ensuremath{\nu}}$_{2}$ band of PH$_{3}$ in the atmosphere of Saturn.} \apjl
  202:L55, \doi{10.1086/181979}

\bibitem[{{Briggs} and {Sackett}(1989)}]{Briggs1989}
{Briggs} FH, {Sackett} PD (1989) {Radio observations of Saturn as a probe of
  its atmosphere and cloud structure}. \icarus 80(1):77--103,
  \doi{10.1016/0019-1035(89)90162-0}

\bibitem[{{Burgdorf} et~al.(2003){Burgdorf}, {Orton}, {Davis}, {Sidher},
  {Feuchtgruber}, {Griffin}, and {Swinyard}}]{Burgdorf2003}
{Burgdorf} M, {Orton} GS, {Davis} GR, {Sidher} SD, {Feuchtgruber} H, {Griffin}
  MJ, {Swinyard} BM (2003) {Neptune's far-infrared spectrum from the ISO
  long-wavelength and short-wavelength spectrometers}. \icarus 164:244--253,
  \doi{10.1016/S0019-1035(03)00138-6}

\bibitem[{{Burgdorf} et~al.(2006){Burgdorf}, {Orton}, {van Cleve}, {Meadows},
  and {Houck}}]{Burgdorf2006}
{Burgdorf} M, {Orton} G, {van Cleve} J, {Meadows} V, {Houck} J (2006)
  {Detection of new hydrocarbons in Uranus' atmosphere by infrared
  spectroscopy}. \icarus 184:634--637, \doi{10.1016/j.icarus.2006.06.006}

\bibitem[{{Burke} et~al.(2016){Burke}, {Metcalfe}, {Burke}, {Heufer}, {Dagaut},
  and {Curran}}]{Burke2016}
{Burke} U, {Metcalfe} WK, {Burke} SM, {Heufer} KA, {Dagaut} P, {Curran} HJ
  (2016) {A detailed chemical kinetic modeling, ignition delay time and
  jet-stirred reactor study of methanol oxidation}. \combust 165:125--136

\bibitem[{{Cavali{\'e}} et~al.(2008){Cavali{\'e}}, {Billebaud}, {Biver},
  {Dobrijevic}, {Lellouch}, {Brillet}, {Lecacheux}, {Hjalmarson}, {Sandqvist},
  {Frisk}, {Olberg}, {Bergin}, and {The Odin Team}}]{Cavalie2008c}
{Cavali{\'e}} T, {Billebaud} F, {Biver} N, {Dobrijevic} M, {Lellouch} E,
  {Brillet} J, {Lecacheux} A, {Hjalmarson} {\AA}, {Sandqvist} A, {Frisk} U,
  {Olberg} M, {Bergin} EA, {The Odin Team} (2008) {Observation of water vapor
  in the stratosphere of Jupiter with the Odin space telescope}. \planss
  56:1573--1584, \doi{10.1016/j.pss.2008.04.013}

\bibitem[{{Cavali{\'e}} et~al.(2009){Cavali{\'e}}, {Billebaud}, {Dobrijevic},
  {Fouchet}, {Lellouch}, {Encrenaz}, {Brillet}, {Moriarty-Schieven},
  {Wouterloot}, and {Hartogh}}]{Cavalie2009}
{Cavali{\'e}} T, {Billebaud} F, {Dobrijevic} M, {Fouchet} T, {Lellouch} E,
  {Encrenaz} T, {Brillet} J, {Moriarty-Schieven} GH, {Wouterloot} JGA,
  {Hartogh} P (2009) {First observation of CO at 345 GHz in the atmosphere of
  Saturn with the JCMT: New constraints on its origin}. \icarus 203:531--540,
  \doi{10.1016/j.icarus.2009.05.024}

\bibitem[{{Cavali{\'e}} et~al.(2010){Cavali{\'e}}, {Hartogh}, {Billebaud},
  {Dobrijevic}, {Fouchet}, {Lellouch}, {Encrenaz}, {Brillet}, and
  {Moriarty-Schieven}}]{Cavalie2010}
{Cavali{\'e}} T, {Hartogh} P, {Billebaud} F, {Dobrijevic} M, {Fouchet} T,
  {Lellouch} E, {Encrenaz} T, {Brillet} J, {Moriarty-Schieven} GH (2010) {A
  cometary origin for CO in the stratosphere of Saturn?} \aap 510:A88

\bibitem[{{Cavali{\'e}} et~al.(2012){Cavali{\'e}}, {Biver}, {Hartogh},
  {Dobrijevic}, {Billebaud}, {Lellouch}, {Sandqvist}, {Brillet}, {Lecacheux},
  {Hjalmarson}, {Frisk}, {Olberg}, and {Odin Team}}]{cavalie2012}
{Cavali{\'e}} T, {Biver} N, {Hartogh} P, {Dobrijevic} M, {Billebaud} F,
  {Lellouch} E, {Sandqvist} A, {Brillet} J, {Lecacheux} A, {Hjalmarson} {\AA},
  {Frisk} U, {Olberg} M, {Odin Team} (2012) {Odin space telescope monitoring of
  water vapor in the stratosphere of Jupiter}. \planss 61:3--14,
  \doi{10.1016/j.pss.2011.04.001}

\bibitem[{{Cavali{\'e}} et~al.(2013){Cavali{\'e}}, {Feuchtgruber}, {Lellouch},
  {de Val-Borro}, {Jarchow}, {Moreno}, {Hartogh}, {Orton}, {Greathouse},
  {Billebaud}, {Dobrijevic}, {Lara}, {Gonz{\'a}lez}, and
  {Sagawa}}]{Cavalie2013}
{Cavali{\'e}} T, {Feuchtgruber} H, {Lellouch} E, {de Val-Borro} M, {Jarchow} C,
  {Moreno} R, {Hartogh} P, {Orton} G, {Greathouse} TK, {Billebaud} F,
  {Dobrijevic} M, {Lara} LM, {Gonz{\'a}lez} A, {Sagawa} H (2013) {Spatial
  distribution of water in the stratosphere of Jupiter from Herschel HIFI and
  PACS observations}. \aap 553:A21, \doi{10.1051/0004-6361/201220797}

\bibitem[{{Cavali{\'e}} et~al.(2014){Cavali{\'e}}, {Moreno}, {Lellouch},
  {Hartogh}, {Venot}, {Orton}, {Jarchow}, {Encrenaz}, {Selsis}, {Hersant}, and
  {Fletcher}}]{Cavalie2014}
{Cavali{\'e}} T, {Moreno} R, {Lellouch} E, {Hartogh} P, {Venot} O, {Orton} GS,
  {Jarchow} C, {Encrenaz} T, {Selsis} F, {Hersant} F, {Fletcher} LN (2014) {The
  first submillimeter observation of CO in the stratosphere of Uranus}. \aap
  562:A33, \doi{10.1051/0004-6361/201322297}

\bibitem[{{Cavali{\'e}} et~al.(2017){Cavali{\'e}}, {Venot}, {Selsis},
  {Hersant}, {Hartogh}, and {Leconte}}]{Cavalie2017}
{Cavali{\'e}} T, {Venot} O, {Selsis} F, {Hersant} F, {Hartogh} P, {Leconte} J
  (2017) {Thermochemistry and vertical mixing in the tropospheres of Uranus and
  Neptune. How convection inhibition can affect the derivation of deep oxygen
  abundances.} \icarus 291:1--16

\bibitem[{{Cavali{\'e}} et~al.(2019){Cavali{\'e}}, {Hue}, {Hartogh}, {Moreno},
  {Lellouch}, {Feuchtgruber}, {Jarchow}, {Cassidy}, {Fletcher}, {Billebaud},
  {Dobrijevic}, {Rezac}, {Orton}, {Rengel}, {Fouchet}, and
  {Guerlet}}]{Cavalie2019}
{Cavali{\'e}} T, {Hue} V, {Hartogh} P, {Moreno} R, {Lellouch} E, {Feuchtgruber}
  H, {Jarchow} C, {Cassidy} T, {Fletcher} LN, {Billebaud} F, {Dobrijevic} M,
  {Rezac} L, {Orton} GS, {Rengel} M, {Fouchet} T, {Guerlet} S (2019) {Herschel
  map of Saturn's stratospheric water, delivered by the plumes of Enceladus}.
  \aap 630:A87, \doi{10.1051/0004-6361/201935954}

\bibitem[{{Chabrier} et~al.(2019){Chabrier}, {Mazevet}, and
  {Soubiran}}]{Chabrier2019}
{Chabrier} G, {Mazevet} S, {Soubiran} F (2019) {A New Equation of State for
  Dense Hydrogen-Helium Mixtures}. \apj 872(1):51,
  \doi{10.3847/1538-4357/aaf99f}

\bibitem[{{Connerney}(1986)}]{Connerney1986}
{Connerney} JEP (1986) {Magnetic connection for Saturn's rings and atmosphere}.
  \grl 13:773--776, \doi{10.1029/GL013i008p00773}

\bibitem[{{Connerney} and {Waite}(1984)}]{Connerney1984}
{Connerney} JEP, {Waite} JH (1984) {New model of Saturn's ionosphere with an
  influx of water from the rings}. \nat 312(5990):136--138,
  \doi{10.1038/312136a0}

\bibitem[{{Conrath} et~al.(1987){Conrath}, {Hanel}, {Gautier}, {Marten}, and
  {Lindal}}]{Conrath1987}
{Conrath} B, {Hanel} R, {Gautier} D, {Marten} A, {Lindal} G (1987) {The helium
  abundance of Uranus from Voyager measurements}. \jgr 92:15003--15010,
  \doi{10.1029/JA092iA13p15003}

\bibitem[{{Conrath} and {Gautier}(2000)}]{Conrath2000}
{Conrath} BJ, {Gautier} D (2000) {Saturn Helium Abundance: A Reanalysis of
  Voyager Measurements}. \icarus 144:124--134, \doi{10.1006/icar.1999.6265}

\bibitem[{{Conrath} et~al.(1984){Conrath}, {Gautier}, {Hanel}, and
  {Hornstein}}]{Conrath1984}
{Conrath} BJ, {Gautier} D, {Hanel} RA, {Hornstein} JS (1984) {The helium
  abundance of Saturn from Voyager measurements}. The Astrophysical Journal
  282:807--815, \doi{10.1086/162267}

\bibitem[{{Conrath} et~al.(1991){Conrath}, {Gautier}, {Lindal}, {Samuelson},
  and {Shaffer}}]{Conrath1991}
{Conrath} BJ, {Gautier} D, {Lindal} GF, {Samuelson} RE, {Shaffer} WA (1991)
  {The helium abundance of Neptune from Voyager measurements}. \jgr 96:18907

\bibitem[{{Conrath} et~al.(1993){Conrath}, {Gautier}, {Owen}, and
  {Samuelson}}]{Conrath1993}
{Conrath} BJ, {Gautier} D, {Owen} TC, {Samuelson} RE (1993) {Constraints on N
  $_{2}$ in Neptune's Atmosphere from Voyager Measurements}. \icarus
  101(1):168--171, \doi{10.1006/icar.1993.1014}

\bibitem[{{Courtin} et~al.(1984){Courtin}, {Gautier}, {Marten}, {Bezard}, and
  {Hanel}}]{Courtin1984}
{Courtin} R, {Gautier} D, {Marten} A, {Bezard} B, {Hanel} R (1984) {The
  composition of Saturn's atmosphere at northern temperate latitudes from
  Voyager IRIS spectra - NH3, PH3, C2H2, C2H6, CH3D, CH4, and the Saturnian D/H
  isotopic ratio}. \apj 287:899--916, \doi{10.1086/162748}

\bibitem[{{Courtin} et~al.(2015){Courtin}, {Pandey-Pommier}, {Gautier},
  {Zarka}, {Hofstadter}, {Hersant}, and {Girard}}]{Courtin2015}
{Courtin} R, {Pandey-Pommier} M, {Gautier} D, {Zarka} P, {Hofstadter} M,
  {Hersant} F, {Girard} J (2015) {Metric Observations of Saturn with the Giant
  Metrewave Radio Telescope}. In: SF2A-2015: Proceedings of the Annual meeting
  of the French Society of Astronomy and Astrophysics, pp 241--245

\bibitem[{{de Pater} and {Richmond}(1989)}]{dePater1989a}
{de Pater} I, {Richmond} M (1989) {Neptune's microwave spectrum from 1 MM to 20
  CM}. \icarus 80:1--13, \doi{10.1016/0019-1035(89)90158-9}

\bibitem[{{de Pater} et~al.(1989){de Pater}, {Romani}, and
  {Atreya}}]{dePater1989b}
{de Pater} I, {Romani} PN, {Atreya} SK (1989) {Uranus deep atmosphere
  revealed}. \icarus 82(2):288--313, \doi{10.1016/0019-1035(89)90040-7}

\bibitem[{{de Pater} et~al.(1991){de Pater}, {Romani}, and
  {Atreya}}]{dePater1991}
{de Pater} I, {Romani} PN, {Atreya} SK (1991) {Possible microwave absorption by
  H2S gas in Uranus' and Neptune's atmospheres}. \icarus 91:220--233,
  \doi{10.1016/0019-1035(91)90020-T}

\bibitem[{{de Pater} et~al.(2016){de Pater}, {Sault}, {Butler}, {DeBoer}, and
  {Wong}}]{dePater2016}
{de Pater} I, {Sault} RJ, {Butler} B, {DeBoer} D, {Wong} MH (2016) {Peering
  through Jupiter{\textquoteright}s clouds with radio spectral imaging}.
  Science 352(6290):1198--1201, \doi{10.1126/science.aaf2210}

\bibitem[{{de Pater} et~al.(2018){de Pater}, {Butler}, {Sault}, {Moullet},
  {Moeckel}, {Tollefson}, {de Kleer}, {Gurwell}, and {Milam}}]{dePater2018}
{de Pater} I, {Butler} B, {Sault} RJ, {Moullet} A, {Moeckel} C, {Tollefson} J,
  {de Kleer} K, {Gurwell} M, {Milam} S (2018) {Potential for Solar System
  Science with the ngVLA}. In: {Murphy} E (ed) Science with a Next Generation
  Very Large Array, Astronomical Society of the Pacific Conference Series, vol
  517, p~49

\bibitem[{{de Pater} et~al.(2019){de Pater}, {Sault}, {Wong}, {Fletcher},
  {DeBoer}, and {Butler}}]{dePater2019}
{de Pater} I, {Sault} RJ, {Wong} MH, {Fletcher} LN, {DeBoer} D, {Butler} B
  (2019) {Jupiter's ammonia distribution derived from VLA maps at 3-37 GHz}.
  \icarus 322:168--191, \doi{10.1016/j.icarus.2018.11.024}

\bibitem[{{DeBoer} and {Steffes}(1994)}]{deBoer1994}
{DeBoer} DR, {Steffes} PG (1994) {Laboratory measurements of the microwave
  properties of H2S under simulated Jovian conditions with an application to
  Neptune}. \icarus 109:352--366, \doi{10.1006/icar.1994.1099}

\bibitem[{{DeBoer} and {Steffes}(1996)}]{deBoer1996}
{DeBoer} DR, {Steffes} PG (1996) {Estimates of the Tropospheric Vertical
  Structure of Neptune Based on Microwave Radiative Transfer Studies}. \icarus
  123:324--335, \doi{10.1006/icar.1996.0161}

\bibitem[{{Dobrijevic} et~al.(2010){Dobrijevic}, {Cavali{\'e}}, {H{\'e}brard},
  {Billebaud}, {Hersant}, and {Selsis}}]{Dobrijevic2010}
{Dobrijevic} M, {Cavali{\'e}} T, {H{\'e}brard} E, {Billebaud} F, {Hersant} F,
  {Selsis} F (2010) {Key reactions in the photochemistry of hydrocarbons in
  Neptune's stratosphere}. \planss 58:1555--1566,
  \doi{10.1016/j.pss.2010.07.024}

\bibitem[{{Dobrijevic} et~al.(2011){Dobrijevic}, {Cavali{\'e}}, and
  {Billebaud}}]{Dobrijevic2011}
{Dobrijevic} M, {Cavali{\'e}} T, {Billebaud} F (2011) {A methodology to
  construct a reduced chemical scheme for 2D-3D photochemical models:
  Application to Saturn}. \icarus 214:275--285,
  \doi{10.1016/j.icarus.2011.04.027}

\bibitem[{{Dobrijevic} et~al.(2016){Dobrijevic}, {Loison}, {Hickson}, and
  {Gronoff}}]{Dobrijevic2016}
{Dobrijevic} M, {Loison} JC, {Hickson} KM, {Gronoff} G (2016) {1D-coupled
  photochemical model of neutrals, cations and anions in the atmosphere of
  Titan}. \icarus 268:313--339, \doi{10.1016/j.icarus.2015.12.045}

\bibitem[{{Dobrijevic} et~al.(2020){Dobrijevic}, {Loison}, {Hue},
  {Cavali{\'e}}, and {Hickson}}]{Dobrijevic2020}
{Dobrijevic} M, {Loison} JC, {Hue} V, {Cavali{\'e}} T, {Hickson} KM (2020) {1D
  photochemical model of the ionosphere and the stratosphere of Neptune}.
  \icarus 335:113375, \doi{10.1016/j.icarus.2019.07.009}

\bibitem[{{Drummond} et~al.(2016){Drummond}, {Tremblin}, {Baraffe}, {Amundsen},
  {Mayne}, {Venot}, and {Goyal}}]{Drummond2016}
{Drummond} B, {Tremblin} P, {Baraffe} I, {Amundsen} DS, {Mayne} NJ, {Venot} O,
  {Goyal} J (2016) {The effects of consistent chemical kinetics calculations on
  the pressure-temperature profiles and emission spectra of hot Jupiters}. \aap
  594:A69, \doi{10.1051/0004-6361/201628799}

\bibitem[{{Encrenaz} et~al.(2004){Encrenaz}, {Lellouch}, {Drossart},
  {Feuchtgruber}, {Orton}, and {Atreya}}]{Encrenaz2004}
{Encrenaz} T, {Lellouch} E, {Drossart} P, {Feuchtgruber} H, {Orton} GS,
  {Atreya} SK (2004) {First detection of CO in Uranus}. \aap 413:L5--L9,
  \doi{10.1051/0004-6361:20034637}

\bibitem[{{Fegley} and {Prinn}(1988)}]{Fegley1988}
{Fegley} B, {Prinn} RG (1988) {Chemical constraints on the water and total
  oxygen abundances in the deep atmosphere of Jupiter}. The Astrophysical
  Journal 324:621--625, \doi{10.1086/165922}

\bibitem[{{Fegley} and {Prinn}(1985)}]{Fegley1985}
{Fegley} J B, {Prinn} RG (1985) {Equilibrium and nonequilibrium chemistry of
  Saturn's atmosphere - Implications for the observability of PH3, N2, CO, and
  GeH4.} \apj 299:1067--1078, \doi{10.1086/163775}

\bibitem[{{Fegley} and {Prinn}(1986)}]{Fegley1986}
{Fegley} J B, {Prinn} RG (1986) {Chemical Models of the Deep Atmosphere of
  Uranus}. \apj 307:852, \doi{10.1086/164472}

\bibitem[{{Fegley} and {Lodders}(1994)}]{Fegley1994}
{Fegley} J Bruce, {Lodders} K (1994) {Chemical Models of the Deep Atmospheres
  of Jupiter and Saturn}. \icarus 110(1):117--154, \doi{10.1006/icar.1994.1111}

\bibitem[{{Ferri} and {colleagues}(2019)}]{Ferri2019}
{Ferri} F, {colleagues} (2019) {The atmospheric structure of the ice giants
  planets from in situ measurements}. \ssr this issue

\bibitem[{{Feuchtgruber} et~al.(1997){Feuchtgruber}, {Lellouch}, {de Graauw},
  {B{\'e}zard}, {Encrenaz}, and {Griffin}}]{Feuchtgruber1997}
{Feuchtgruber} H, {Lellouch} E, {de Graauw} T, {B{\'e}zard} B, {Encrenaz} T,
  {Griffin} M (1997) {External supply of oxygen to the atmospheres of the giant
  planets}. \nat 389:159--162, \doi{10.1038/38236}

\bibitem[{{Feuchtgruber} et~al.(2013){Feuchtgruber}, {Lellouch}, {Orton}, {de
  Graauw}, {Vandenbussche}, {Swinyard}, {Moreno}, {Jarchow}, {Billebaud},
  {Cavali{\'e}}, {Sidher}, and {Hartogh}}]{Feuchtgruber2013}
{Feuchtgruber} H, {Lellouch} E, {Orton} G, {de Graauw} T, {Vandenbussche} B,
  {Swinyard} B, {Moreno} R, {Jarchow} C, {Billebaud} F, {Cavali{\'e}} T,
  {Sidher} S, {Hartogh} P (2013) {The D/H ratio in the atmospheres of Uranus
  and Neptune from Herschel-PACS observations}. \aap 551:A126,
  \doi{10.1051/0004-6361/201220857}

\bibitem[{{Fletcher} et~al.(2009{\natexlab{a}}){Fletcher}, {Orton}, {Teanby},
  and {Irwin}}]{Fletcher2009b}
{Fletcher} LN, {Orton} GS, {Teanby} NA, {Irwin} PGJ (2009{\natexlab{a}})
  {Phosphine on Jupiter and Saturn from Cassini/CIRS}. \icarus 202:543--564,
  \doi{10.1016/j.icarus.2009.03.023}

\bibitem[{{Fletcher} et~al.(2009{\natexlab{b}}){Fletcher}, {Orton}, {Teanby},
  {Irwin}, and {Bjoraker}}]{Fletcher2009a}
{Fletcher} LN, {Orton} GS, {Teanby} NA, {Irwin} PGJ, {Bjoraker} GL
  (2009{\natexlab{b}}) {Methane and its isotopologues on Saturn from
  Cassini/CIRS observations}. \icarus 199:351--367,
  \doi{10.1016/j.icarus.2008.09.019}

\bibitem[{{Fletcher} et~al.(2010){Fletcher}, {Drossart}, {Burgdorf}, {Orton},
  and {Encrenaz}}]{Fletcher2010}
{Fletcher} LN, {Drossart} P, {Burgdorf} M, {Orton} GS, {Encrenaz} T (2010)
  {Neptune's atmospheric composition from AKARI infrared spectroscopy}. \aap
  514:A17, \doi{10.1051/0004-6361/200913358}

\bibitem[{{Fletcher} et~al.(2011){Fletcher}, {Baines}, {Momary}, {Showman},
  {Irwin}, {Orton}, {Roos-Serote}, and {Merlet}}]{Fletcher2011b}
{Fletcher} LN, {Baines} KH, {Momary} TW, {Showman} AP, {Irwin} PGJ, {Orton} GS,
  {Roos-Serote} M, {Merlet} C (2011) {Saturn's tropospheric composition and
  clouds from Cassini/VIMS 4.6-5.1 {\ensuremath{\mu}}m nightside spectroscopy}.
  \icarus 214(2):510--533, \doi{10.1016/j.icarus.2011.06.006}

\bibitem[{{Fletcher} et~al.(2014){Fletcher}, {Greathouse}, {Orton}, {Irwin},
  {Mousis}, {Sinclair}, and {Giles}}]{Fletcher2014}
{Fletcher} LN, {Greathouse} TK, {Orton} GS, {Irwin} PGJ, {Mousis} O, {Sinclair}
  JA, {Giles} RS (2014) {The origin of nitrogen on Jupiter and Saturn from the
  15N/14N ratio}. \icarus 238:170--190, \doi{10.1016/j.icarus.2014.05.007}

\bibitem[{{Fletcher} et~al.(2020{\natexlab{a}}){Fletcher}, {de Pater}, {Orton},
  {Hofstadter}, {Irwin}, {Roman}, and {Toledo}}]{Fletcher2020b}
{Fletcher} LN, {de Pater} I, {Orton} GS, {Hofstadter} MD, {Irwin} PGJ, {Roman}
  MT, {Toledo} D (2020{\natexlab{a}}) {Ice Giant Circulation Patterns:
  Implications for Atmospheric Probes}. \ssr 216(2):21,
  \doi{10.1007/s11214-020-00646-1}

\bibitem[{{Fletcher} et~al.(2020{\natexlab{b}}){Fletcher}, {Kaspi}, {Guillot},
  and {Showman}}]{Fletcher2020a}
{Fletcher} LN, {Kaspi} Y, {Guillot} T, {Showman} AP (2020{\natexlab{b}}) {How
  Well Do We Understand the Belt/Zone Circulation of Giant Planet Atmospheres?}
  \ssr 216(2):30, \doi{10.1007/s11214-019-0631-9}

\bibitem[{{Fouchet} et~al.(2000{\natexlab{a}}){Fouchet}, {Lellouch},
  {B{\'e}zard}, {Encrenaz}, {Drossart}, {Feuchtgruber}, and {de
  Graauw}}]{Fouchet2000a}
{Fouchet} T, {Lellouch} E, {B{\'e}zard} B, {Encrenaz} T, {Drossart} P,
  {Feuchtgruber} H, {de Graauw} T (2000{\natexlab{a}}) {ISO-SWS Observations of
  Jupiter: Measurement of the Ammonia Tropospheric Profile and of the $^{15}$N/
  $^{14}$N Isotopic Ratio}. \icarus 143(2):223--243,
  \doi{10.1006/icar.1999.6255}

\bibitem[{{Fouchet} et~al.(2000{\natexlab{b}}){Fouchet}, {Lellouch},
  {B{\'e}zard}, {Feuchtgruber}, {Drossart}, and {Encrenaz}}]{Fouchet2000b}
{Fouchet} T, {Lellouch} E, {B{\'e}zard} B, {Feuchtgruber} H, {Drossart} P,
  {Encrenaz} T (2000{\natexlab{b}}) {Jupiter's hydrocarbons observed with
  ISO-SWS: vertical profiles of C\_2H\_6 and C\_2H\_2, detection of
  CH\_3C\_2H}. \aap 355:L13--L17

\bibitem[{{Fouchet} et~al.(2017){Fouchet}, {Lellouch}, {Cavali{\'e}}, and
  {B{\'e}zard}}]{Fouchet2017}
{Fouchet} T, {Lellouch} E, {Cavali{\'e}} T, {B{\'e}zard} B (2017) {First
  determination of the tropospheric CO abundance in Saturn}. In: AAS/Division
  for Planetary Sciences Meeting Abstracts \#49, AAS/Division for Planetary
  Sciences Meeting Abstracts, p 209.05

\bibitem[{{Fouchet} et~al.(2018{\natexlab{a}}){Fouchet}, {Greathouse},
  {B{\'e}zard}, {Richter}, and {Moses}}]{Fouchet2018b}
{Fouchet} T, {Greathouse} T, {B{\'e}zard} B, {Richter} M, {Moses} J
  (2018{\natexlab{a}}) {First Measurements of Methyl Radical (CH$_{3}$) in
  Jupiter's atmosphere using TEXES/IRTF and EXES/SOFIA}. In: AAS/Division for
  Planetary Sciences Meeting Abstracts, p 507.07

\bibitem[{{Fouchet} et~al.(2018{\natexlab{b}}){Fouchet}, {Moreno},
  {Cavali{\'e}}, and {Lellouch}}]{Fouchet2018a}
{Fouchet} T, {Moreno} R, {Cavali{\'e}} T, {Lellouch} E (2018{\natexlab{b}})
  {Detection of HCN in Saturn's atmosphere}, {Central Bureau Electronic
  Telegram No. 4535}

\bibitem[{{Frelikh} and {Murray-Clay}(2017)}]{Frelikh2017}
{Frelikh} R, {Murray-Clay} RA (2017) {The Formation of Uranus and Neptune:
  Fine-tuning in Core Accretion}. \aj 154(3):98, \doi{10.3847/1538-3881/aa81c7}

\bibitem[{{French} et~al.(1998){French}, {McGhee}, and {Sicardy}}]{French1998}
{French} RG, {McGhee} CA, {Sicardy} B (1998) {Neptune's Stratospheric Winds
  from Three Central Flash Occultations}. \icarus 136(1):27--49,
  \doi{10.1006/icar.1998.6001}

\bibitem[{{Gardner} et~al.(2006){Gardner}, {Mather}, {Clampin}, {Doyon},
  {Greenhouse}, {Hammel}, {Hutchings}, {Jakobsen}, {Lilly}, {Long}, {Lunine},
  {McCaughrean}, {Mountain}, {Nella}, {Rieke}, {Rieke}, {Rix}, {Smith},
  {Sonneborn}, {Stiavelli}, {Stockman}, {Windhorst}, and
  {Wright}}]{Gardner2006}
{Gardner} JP, {Mather} JC, {Clampin} M, {Doyon} R, {Greenhouse} MA, {Hammel}
  HB, {Hutchings} JB, {Jakobsen} P, {Lilly} SJ, {Long} KS, {Lunine} JI,
  {McCaughrean} MJ, {Mountain} M, {Nella} J, {Rieke} GH, {Rieke} MJ, {Rix} HW,
  {Smith} EP, {Sonneborn} G, {Stiavelli} M, {Stockman} HS, {Windhorst} RA,
  {Wright} GS (2006) {The James Webb Space Telescope}. \ssr 123(4):485--606,
  \doi{10.1007/s11214-006-8315-7}

\bibitem[{{Gautier} and {Hersant}(2005)}]{Gautier2005}
{Gautier} D, {Hersant} F (2005) {Formation and Composition of Planetesimals}.
  \ssr 116:25--52, \doi{10.1007/s11214-005-1946-2}

\bibitem[{{Gautier} et~al.(1981){Gautier}, {Conrath}, {Flasar}, {Hanel},
  {Kunde}, {Chedin}, and {Scott}}]{Gautier1981}
{Gautier} D, {Conrath} B, {Flasar} M, {Hanel} R, {Kunde} V, {Chedin} A, {Scott}
  N (1981) {The helium abundance of Jupiter from Voyager}. Journal of
  Geophysical Research 86(A10):8713--8720, \doi{10.1029/JA086iA10p08713}

\bibitem[{{Gautier} et~al.(2001){Gautier}, {Hersant}, {Mousis}, and
  {Lunine}}]{Gautier2001}
{Gautier} D, {Hersant} F, {Mousis} O, {Lunine} JI (2001) {Enrichments in
  Volatiles in Jupiter: A New Interpretation of the Galileo Measurements}.
  \apjl 550:L227--L230, \doi{10.1086/319648}

\bibitem[{{Gierasch} and {Conrath}(1985)}]{Gierasch1985}
{Gierasch} PJ, {Conrath} BJ (1985) {Energy conversion processes in the outer
  planets}. In: {Hunt} GE (ed) Recent Advances in Planetary Meteorology, pp
  121--146

\bibitem[{{Giles} et~al.(2017){Giles}, {Fletcher}, and {Irwin}}]{Giles2017}
{Giles} RS, {Fletcher} LN, {Irwin} PGJ (2017) {Latitudinal variability in
  Jupiter's tropospheric disequilibrium species: GeH$_{4}$, AsH$_{3}$ and
  PH$_{3}$}. \icarus 289:254--269, \doi{10.1016/j.icarus.2016.10.023}

\bibitem[{{Gladstone} and {Yung}(1983)}]{Gladstone1983}
{Gladstone} GR, {Yung} YL (1983) {An analysis of the reflection spectrum of
  Jupiter from 1500 A to 1740 A}. \apj 266:415--424, \doi{10.1086/160789}

\bibitem[{{Grassi} et~al.(2019){Grassi}, {Adriani}, {Mura}, {Bolton},
  {Plainaki}, and {the JIRAM Juno team}}]{Grassi2019}
{Grassi} D, {Adriani} A, {Mura} A, {Bolton} S, {Plainaki} G C~{Sindoni}, {the
  JIRAM Juno team} (2019) {On the content of minor species in the upper Jupiter
  troposphere as inferred from JIRAM Juno data}. In: EPSC Abstracts, vol~13, pp
  EPSC--DPS2019--239--1

\bibitem[{{Grevesse} et~al.(2010){Grevesse}, {Asplund}, {Sauval}, and
  {Scott}}]{Grevesse2010}
{Grevesse} N, {Asplund} M, {Sauval} AJ, {Scott} P (2010) {The chemical
  composition of the Sun}. \apss 328(1-2):179--183,
  \doi{10.1007/s10509-010-0288-z}

\bibitem[{{Grevesse} et~al.(2015){Grevesse}, {Scott}, {Asplund}, and
  {Sauval}}]{Grevesse2015}
{Grevesse} N, {Scott} P, {Asplund} M, {Sauval} AJ (2015) {The elemental
  composition of the Sun. III. The heavy elements Cu to Th}. \aap 573:A27,
  \doi{10.1051/0004-6361/201424111}

\bibitem[{{Guillot}(1995)}]{Guillot1995}
{Guillot} T (1995) {Condensation of Methane, Ammonia, and Water and the
  Inhibition of Convection in Giant Planets}. Science 269:1697--1699,
  \doi{10.1126/science.7569896}

\bibitem[{{Guillot}(2005)}]{Guillot2005}
{Guillot} T (2005) {The Interiors of Giant Planets: Models and Outstanding
  Questions}. Annual Review of Earth and Planetary Sciences 33:493--530,
  \doi{10.1146/annurev.earth.32.101802.120325}, \eprint{astro-ph/0502068}

\bibitem[{{Guillot} and {Gautier}(2015)}]{Guillot2015}
{Guillot} T, {Gautier} D (2015) 10.16 - giant planets. In: Schubert G (ed)
  Treatise on Geophysics (Second Edition), second edition edn, Elsevier,
  Oxford, pp 529 -- 557,
  \doi{https://doi.org/10.1016/B978-0-444-53802-4.00176-7},
  \urlprefix\url{http://www.sciencedirect.com/science/article/pii/B9780444538024001767}

\bibitem[{{Guillot} and {Hueso}(2006)}]{Guillot2006}
{Guillot} T, {Hueso} R (2006) {The composition of Jupiter: sign of a
  (relatively) late formation in a chemically evolved protosolar disc}. \mnras
  367(1):L47--L51, \doi{10.1111/j.1745-3933.2006.00137.x}

\bibitem[{{Guillot} et~al.(2019){Guillot}, {Stevenson}, {Li}, {Atreya},
  {Ingersoll}, and S.}]{Guillot2019}
{Guillot} T, {Stevenson} DJ, {Li} C, {Atreya} S, {Ingersoll} A, S B (2019)
  {Storms and the distribution of ammonia in Jupiter's atmosphere}. In: EPSC
  Abstracts, vol~13, pp EPSC--DPS2019--1142--1

\bibitem[{{Hartogh} et~al.(2011){Hartogh}, {Lellouch}, {Moreno},
  {Bockel{\'e}e-Morvan}, {Biver}, {Cassidy}, {Rengel}, {Jarchow},
  {Cavali{\'e}}, {Crovisier}, {Helmich}, and {Kidger}}]{Hartogh2011}
{Hartogh} P, {Lellouch} E, {Moreno} R, {Bockel{\'e}e-Morvan} D, {Biver} N,
  {Cassidy} T, {Rengel} M, {Jarchow} C, {Cavali{\'e}} T, {Crovisier} J,
  {Helmich} FP, {Kidger} M (2011) {Direct detection of the Enceladus water
  torus with Herschel}. \aap 532:L2, \doi{10.1051/0004-6361/201117377}

\bibitem[{{Helled} and {Bodenheimer}(2014)}]{Helled2014c}
{Helled} R, {Bodenheimer} P (2014) {The Formation of Uranus and Neptune:
  Challenges and Implications for Intermediate-mass Exoplanets}. \apj
  789(1):69, \doi{10.1088/0004-637X/789/1/69}

\bibitem[{{Helled} and {Guillot}(2013)}]{Helled2013}
{Helled} R, {Guillot} T (2013) {Interior Models of Saturn: Including the
  Uncertainties in Shape and Rotation}. \apj 767(2):113,
  \doi{10.1088/0004-637X/767/2/113}

\bibitem[{{Helled} and {Guillot}(2018)}]{Helled2018}
{Helled} R, {Guillot} T (2018) {Internal Structure of Giant and Icy Planets:
  Importance of Heavy Elements and Mixing}, Springer International Publishing
  AG, part of Springer Nature, p~44. \doi{10.1007/978-3-319-55333-7_44}

\bibitem[{{Helled} and {Lunine}(2014)}]{Helled2014a}
{Helled} R, {Lunine} J (2014) {Measuring Jupiter's water abundance by Juno: the
  link between interior and formation models}. \mnras 441:2273--2279,
  \doi{10.1093/mnras/stu516}

\bibitem[{{Helled} et~al.(2010){Helled}, {Anderson}, and
  {Schubert}}]{Helled2010}
{Helled} R, {Anderson} JD, {Schubert} G (2010) {Uranus and Neptune: Shape and
  rotation}. \icarus 210(1):446--454, \doi{10.1016/j.icarus.2010.06.037}

\bibitem[{{Helled} et~al.(2011){Helled}, {Anderson}, {Podolak}, and
  {Schubert}}]{Helled2011}
{Helled} R, {Anderson} JD, {Podolak} M, {Schubert} G (2011) {Interior Models of
  Uranus and Neptune}. \apj 726:15, \doi{10.1088/0004-637X/726/1/15}

\bibitem[{{Helled} et~al.(2014){Helled}, {Bodenheimer}, {Podolak}, {Boley},
  {Meru}, {Nayakshin}, {Fortney}, {Mayer}, {Alibert}, and {Boss}}]{Helled2014b}
{Helled} R, {Bodenheimer} P, {Podolak} M, {Boley} A, {Meru} F, {Nayakshin} S,
  {Fortney} JJ, {Mayer} L, {Alibert} Y, {Boss} AP (2014) {Giant Planet
  Formation, Evolution, and Internal Structure}. In: {Beuther} H, {Klessen} RS,
  {Dullemond} CP, {Henning} T (eds) Protostars and Planets VI, p 643,
  \doi{10.2458/azu_uapress_9780816531240-ch028}

\bibitem[{{Helled} et~al.(2020){Helled}, {Nettelmann}, and
  {Guillot}}]{Helled2020}
{Helled} R, {Nettelmann} N, {Guillot} T (2020) {Uranus and Neptune: Origin,
  Evolution and Internal Structure}. \ssr 216(3):38,
  \doi{10.1007/s11214-020-00660-3}

\bibitem[{{Hersant} et~al.(2001){Hersant}, {Gautier}, and
  {Hur{\'e}}}]{Hersant2001}
{Hersant} F, {Gautier} D, {Hur{\'e}} JM (2001) {A Two-dimensional Model for the
  Primordial Nebula Constrained by D/H Measurements in the Solar System:
  Implications for the Formation of Giant Planets}. \apj 554:391--407,
  \doi{10.1086/321355}

\bibitem[{{Hersant} et~al.(2004){Hersant}, {Gautier}, and
  {Lunine}}]{Hersant2004}
{Hersant} F, {Gautier} D, {Lunine} JI (2004) {Enrichment in volatiles in the
  giant planets of the Solar System}. \planss 52:623--641,
  \doi{10.1016/j.pss.2003.12.011}

\bibitem[{{Hersant} et~al.(2008){Hersant}, {Gautier}, {Tobie}, and
  {Lunine}}]{Hersant2008}
{Hersant} F, {Gautier} D, {Tobie} G, {Lunine} JI (2008) {Interpretation of the
  carbon abundance in Saturn measured by Cassini}. \planss 56:1103--1111,
  \doi{10.1016/j.pss.2008.02.007}

\bibitem[{{Hubbard} et~al.(1995){Hubbard}, {Podolak}, and
  {Stevenson}}]{Hubbard1995}
{Hubbard} WB, {Podolak} M, {Stevenson} DJ (1995) {The interior of Neptune.} In:
  Neptune and Triton, pp 109--138

\bibitem[{{Hubickyj} et~al.(2005){Hubickyj}, {Bodenheimer}, and
  {Lissauer}}]{Hubickyj2005}
{Hubickyj} O, {Bodenheimer} P, {Lissauer} JJ (2005) {Accretion of the gaseous
  envelope of Jupiter around a 5 10 Earth-mass core}. \icarus 179(2):415--431,
  \doi{10.1016/j.icarus.2005.06.021}

\bibitem[{{Hue} et~al.(2015){Hue}, {Cavali{\'e}}, {Dobrijevic}, {Hersant}, and
  {Greathouse}}]{Hue2015}
{Hue} V, {Cavali{\'e}} T, {Dobrijevic} M, {Hersant} F, {Greathouse} TK (2015)
  {2D photochemical modeling of Saturn's stratosphere. Part I: Seasonal
  variation of atmospheric composition without meridional transport}. \icarus
  257:163--184, \doi{10.1016/j.icarus.2015.04.001}

\bibitem[{{Hue} et~al.(2016){Hue}, {Greathouse}, {Cavali{\'e}}, {Dobrijevic},
  and {Hersant}}]{Hue2016}
{Hue} V, {Greathouse} TK, {Cavali{\'e}} T, {Dobrijevic} M, {Hersant} F (2016)
  {2D photochemical modeling of Saturn's stratosphere. Part II: Feedback
  between composition and temperature}. \icarus 267:334--343,
  \doi{10.1016/j.icarus.2015.12.007}

\bibitem[{{Hue} et~al.(2018){Hue}, {Hersant}, {Cavali{\'e}}, {Dobrijevic}, and
  {Sinclair}}]{Hue2018}
{Hue} V, {Hersant} F, {Cavali{\'e}} T, {Dobrijevic} M, {Sinclair} JA (2018)
  {Photochemistry, mixing and transport in Jupiter's stratosphere constrained
  by Cassini}. \icarus 307:106--123, \doi{10.1016/j.icarus.2018.02.018}

\bibitem[{{Iess} et~al.(2018){Iess}, {Folkner}, {Durante}, {Parisi}, {Kaspi},
  {Galanti}, {Guillot}, {Hubbard}, {Stevenson}, {Anderson}, {Buccino},
  {Casajus}, {Milani}, {Park}, {Racioppa}, {Serra}, {Tortora}, {Zannoni},
  {Cao}, {Helled}, {Lunine}, {Miguel}, {Militzer}, {Wahl}, {Connerney},
  {Levin}, and {Bolton}}]{Iess2018}
{Iess} L, {Folkner} WM, {Durante} D, {Parisi} M, {Kaspi} Y, {Galanti} E,
  {Guillot} T, {Hubbard} WB, {Stevenson} DJ, {Anderson} JD, {Buccino} DR,
  {Casajus} LG, {Milani} A, {Park} R, {Racioppa} P, {Serra} D, {Tortora} P,
  {Zannoni} M, {Cao} H, {Helled} R, {Lunine} JI, {Miguel} Y, {Militzer} B,
  {Wahl} S, {Connerney} JEP, {Levin} SM, {Bolton} SJ (2018) {Measurement of
  Jupiter{\textquoteright}s asymmetric gravity field}. \nat 555(7695):220--222,
  \doi{10.1038/nature25776}

\bibitem[{{Iess} et~al.(2019){Iess}, {Militzer}, {Kaspi}, {Nicholson},
  {Durante}, {Racioppa}, {Anabtawi}, {Galanti}, {Hubbard}, {Mariani},
  {Tortora}, {Wahl}, and {Zannoni}}]{Iess2019}
{Iess} L, {Militzer} B, {Kaspi} Y, {Nicholson} P, {Durante} D, {Racioppa} P,
  {Anabtawi} A, {Galanti} E, {Hubbard} W, {Mariani} MJ, {Tortora} P, {Wahl} S,
  {Zannoni} M (2019) {Measurement and implications of Saturn's gravity field
  and ring mass}. Science 364(6445):aat2965, \doi{10.1126/science.aat2965}

\bibitem[{{Irwin} et~al.(2018){Irwin}, {Toledo}, {Garland}, {Teanby},
  {Fletcher}, {Orton}, and {B{\'e}zard}}]{Irwin2018}
{Irwin} PGJ, {Toledo} D, {Garland} R, {Teanby} NA, {Fletcher} LN, {Orton} GA,
  {B{\'e}zard} B (2018) {Detection of hydrogen sulfide above the clouds in
  Uranus's atmosphere}. Nature Astronomy 2:420--427,
  \doi{10.1038/s41550-018-0432-1}

\bibitem[{{Irwin} et~al.(2019{\natexlab{a}}){Irwin}, {Toledo}, {Braude},
  {Bacon}, {Weilbacher}, {Teanby}, {Fletcher}, and {Orton}}]{Irwin2019b}
{Irwin} PGJ, {Toledo} D, {Braude} AS, {Bacon} R, {Weilbacher} PM, {Teanby} NA,
  {Fletcher} LN, {Orton} GS (2019{\natexlab{a}}) {Latitudinal variation in the
  abundance of methane (CH$_{4}$) above the clouds in Neptune's atmosphere from
  VLT/MUSE Narrow Field Mode Observations}. \icarus 331:69--82,
  \doi{10.1016/j.icarus.2019.05.011}

\bibitem[{{Irwin} et~al.(2019{\natexlab{b}}){Irwin}, {Toledo}, {Garland},
  {Teanby}, {Fletcher}, {Orton}, and {B{\'e}zard}}]{Irwin2019a}
{Irwin} PGJ, {Toledo} D, {Garland} R, {Teanby} NA, {Fletcher} LN, {Orton} GS,
  {B{\'e}zard} B (2019{\natexlab{b}}) {Probable detection of hydrogen sulphide
  (H$_{2}$S) in Neptune's atmosphere}. \icarus 321:550--563,
  \doi{10.1016/j.icarus.2018.12.014}

\bibitem[{{Jacobson}(2009)}]{Jacobson2009}
{Jacobson} RA (2009) {The Orbits of the Neptunian Satellites and the
  Orientation of the Pole of Neptune}. \aj 137(5):4322--4329,
  \doi{10.1088/0004-6256/137/5/4322}

\bibitem[{{Jacobson}(2014)}]{Jacobson2014}
{Jacobson} RA (2014) {The Orbits of the Uranian Satellites and Rings, the
  Gravity Field of the Uranian System, and the Orientation of the Pole of
  Uranus}. \aj 148(5):76, \doi{10.1088/0004-6256/148/5/76}

\bibitem[{{Jacobson} et~al.(2006){Jacobson}, {Antreasian}, {Bordi}, {Criddle},
  {Ionasescu}, {Jones}, {Mackenzie}, {Meek}, {Parcher}, {Pelletier}, {Owen},
  {Roth}, {Roundhill}, and {Stauch}}]{Jacobson2006}
{Jacobson} RA, {Antreasian} PG, {Bordi} JJ, {Criddle} KE, {Ionasescu} R,
  {Jones} JB, {Mackenzie} RA, {Meek} MC, {Parcher} D, {Pelletier} FJ, {Owen} J
  W~M, {Roth} DC, {Roundhill} IM, {Stauch} JR (2006) {The Gravity Field of the
  Saturnian System from Satellite Observations and Spacecraft Tracking Data}.
  \aj 132(6):2520--2526, \doi{10.1086/508812}

\bibitem[{{Janssen} et~al.(2005){Janssen}, {Hofstadter}, {Gulkis}, {Ingersoll},
  {Allison}, {Bolton}, {Levin}, and {Kamp}}]{Janssen2005}
{Janssen} MA, {Hofstadter} MD, {Gulkis} S, {Ingersoll} AP, {Allison} M,
  {Bolton} SJ, {Levin} SM, {Kamp} LW (2005) {Microwave remote sensing of
  Jupiter's atmosphere from an orbiting spacecraft}. \icarus 173(2):447--453,
  \doi{10.1016/j.icarus.2004.08.012}

\bibitem[{{Janssen} et~al.(2013){Janssen}, {Ingersoll}, {Allison}, {Gulkis},
  {Laraia}, {Baines}, {Edgington}, {Anderson}, {Kelleher}, and
  {Oyafuso}}]{Janssen2013}
{Janssen} MA, {Ingersoll} AP, {Allison} MD, {Gulkis} S, {Laraia} AL, {Baines}
  KH, {Edgington} SG, {Anderson} YZ, {Kelleher} K, {Oyafuso} FA (2013)
  {Saturn{\textquoteright}s thermal emission at 2.2-cm wavelength as imaged by
  the Cassini RADAR radiometer}. \icarus 226(1):522--535,
  \doi{10.1016/j.icarus.2013.06.008}

\bibitem[{{Karkoschka} and {Tomasko}(2009)}]{Karkoschka2009}
{Karkoschka} E, {Tomasko} M (2009) {The haze and methane distributions on
  Uranus from HST-STIS spectroscopy}. \icarus 202:287--309,
  \doi{10.1016/j.icarus.2009.02.010}

\bibitem[{{Karkoschka} and {Tomasko}(2011)}]{Karkoschka2011}
{Karkoschka} E, {Tomasko} MG (2011) {The haze and methane distributions on
  Neptune from HST-STIS spectroscopy}. \icarus 211:780--797,
  \doi{10.1016/j.icarus.2010.08.013}

\bibitem[{{Kaspi} et~al.(2013){Kaspi}, {Showman}, {Hubbard}, {Aharonson}, and
  {Helled}}]{Kaspi2013}
{Kaspi} Y, {Showman} AP, {Hubbard} WB, {Aharonson} O, {Helled} R (2013)
  {Atmospheric confinement of jet streams on Uranus and Neptune}. \nat
  497(7449):344--347, \doi{10.1038/nature12131}

\bibitem[{{Kaspi} et~al.(2018){Kaspi}, {Galanti}, {Hubbard}, {Stevenson},
  {Bolton}, {Iess}, {Guillot}, {Bloxham}, {Connerney}, {Cao}, {Durante},
  {Folkner}, {Helled}, {Ingersoll}, {Levin}, {Lunine}, {Miguel}, {Militzer},
  {Parisi}, and {Wahl}}]{Kaspi2018}
{Kaspi} Y, {Galanti} E, {Hubbard} WB, {Stevenson} DJ, {Bolton} SJ, {Iess} L,
  {Guillot} T, {Bloxham} J, {Connerney} JEP, {Cao} H, {Durante} D, {Folkner}
  WM, {Helled} R, {Ingersoll} AP, {Levin} SM, {Lunine} JI, {Miguel} Y,
  {Militzer} B, {Parisi} M, {Wahl} SM (2018) {Jupiter{\textquoteright}s
  atmospheric jet streams extend thousands of kilometres deep}. \nat
  555(7695):223--226, \doi{10.1038/nature25793}

\bibitem[{{Knacke} et~al.(1982){Knacke}, {Kim}, {Ridgway}, and
  {Tokunaga}}]{Knacke1982}
{Knacke} RF, {Kim} SJ, {Ridgway} ST, {Tokunaga} AT (1982) {The abundances of
  CH4, CH3D, NH3, and PH3 in the troposphere of Jupiter derived from
  high-resolution 1100-1200/cm spectra}. \apj 262:388--395,
  \doi{10.1086/160432}

\bibitem[{{Koskinen} and {Guerlet}(2018)}]{Koskinen2018}
{Koskinen} TT, {Guerlet} S (2018) {Atmospheric structure and helium abundance
  on Saturn from Cassini/UVIS and CIRS observations}. \icarus 307:161--171,
  \doi{10.1016/j.icarus.2018.02.020}

\bibitem[{{Lambrechts} and {Johansen}(2014)}]{Lambrechts2014}
{Lambrechts} M, {Johansen} A (2014) {Forming the cores of giant planets from
  the radial pebble flux in protoplanetary discs}. \aap 572:A107,
  \doi{10.1051/0004-6361/201424343}

\bibitem[{{Landgraf} et~al.(2002){Landgraf}, {Liou}, {Zook}, and
  {Gr{\"u}n}}]{Landgraf2002}
{Landgraf} M, {Liou} JC, {Zook} HA, {Gr{\"u}n} E (2002) {Origins of Solar
  System Dust beyond Jupiter}. \aj 123:2857--2861, \doi{10.1086/339704}

\bibitem[{{Laraia} et~al.(2013){Laraia}, {Ingersoll}, {Janssen}, {Gulkis},
  {Oyafuso}, and {Allison}}]{Laraia2013}
{Laraia} AL, {Ingersoll} AP, {Janssen} MA, {Gulkis} S, {Oyafuso} F, {Allison} M
  (2013) {Analysis of Saturn's thermal emission at 2.2-cm wavelength: Spatial
  distribution of ammonia vapor}. \icarus 226:641--654,
  \doi{10.1016/j.icarus.2013.06.017}

\bibitem[{{Leconte} et~al.(2017){Leconte}, {Selsis}, {Hersant}, and
  {Guillot}}]{Leconte2017}
{Leconte} J, {Selsis} F, {Hersant} F, {Guillot} T (2017)
  {Condensation-inhibited convection in hydrogen-rich atmospheres . Stability
  against double-diffusive processes and thermal profiles for Jupiter, Saturn,
  Uranus, and Neptune}. \aap 598:A98, \doi{10.1051/0004-6361/201629140}

\bibitem[{{Leisawitz} et~al.(2018){Leisawitz}, {Amatucci}, {Carter}, {DiPirro},
  {Flores}, {Staguhn}, {Wu}, {Allen}, {Arenberg}, {Armus}, {Battersby},
  {Bauer}, {Bell}, {Beltran}, {Benford}, {Bergin}, {Bradford}, {Bradley},
  {Burgarella}, {Carey}, {Chi}, {Cooray}, {Corsetti}, {De Beck}, {Denis},
  {Dewell}, {East}, {Edgington}, {Ennico}, {Fantano}, {Feller}, {Folta},
  {Fortney}, {Generie}, {Gerin}, {Granger}, {Harpole}, {Harvey}, {Helmich},
  {Hilliard}, {Howard}, {Jacoby}, {Jamil}, {Kataria}, {Knight}, {Knollenberg},
  {Lightsey}, {Lipscy}, {Mamajek}, {Martins}, {Meixner}, {Melnick}, {Milam},
  {Mooney}, {Moseley}, {Narayanan}, {Neff}, {Nguyen}, {Nordt}, {Olson},
  {Padgett}, {Petach}, {Petro}, {Pohner}, {Pontoppidan}, {Pope}, {Ramspacher},
  {Roellig}, {Sakon}, {Sandin}, {Sandstrom}, {Scott}, {Sheth}, {Steeves},
  {Stevenson}, {Stokowski}, {Stoneking}, {Su}, {Tajdaran}, {Tompkins},
  {Vieira}, {Webster}, {Wiedner}, {Wright}, and {Zmuidzinas}}]{Leisawitz2018}
{Leisawitz} D, {Amatucci} E, {Carter} R, {DiPirro} M, {Flores} A, {Staguhn} J,
  {Wu} C, {Allen} L, {Arenberg} J, {Armus} L, {Battersby} C, {Bauer} J, {Bell}
  R, {Beltran} P, {Benford} D, {Bergin} E, {Bradford} CM, {Bradley} D,
  {Burgarella} D, {Carey} S, {Chi} D, {Cooray} A, {Corsetti} J, {De Beck} E,
  {Denis} K, {Dewell} L, {East} M, {Edgington} S, {Ennico} K, {Fantano} L,
  {Feller} G, {Folta} D, {Fortney} J, {Generie} J, {Gerin} M, {Granger} Z,
  {Harpole} G, {Harvey} K, {Helmich} F, {Hilliard} L, {Howard} J, {Jacoby} M,
  {Jamil} A, {Kataria} T, {Knight} S, {Knollenberg} P, {Lightsey} P, {Lipscy}
  S, {Mamajek} E, {Martins} G, {Meixner} M, {Melnick} G, {Milam} S, {Mooney} T,
  {Moseley} SH, {Narayanan} D, {Neff} S, {Nguyen} T, {Nordt} A, {Olson} J,
  {Padgett} D, {Petach} M, {Petro} S, {Pohner} J, {Pontoppidan} K, {Pope} A,
  {Ramspacher} D, {Roellig} T, {Sakon} I, {Sandin} C, {Sandstrom} K, {Scott} D,
  {Sheth} K, {Steeves} J, {Stevenson} K, {Stokowski} L, {Stoneking} E, {Su} K,
  {Tajdaran} K, {Tompkins} S, {Vieira} J, {Webster} C, {Wiedner} M, {Wright}
  EL, {Zmuidzinas} J (2018) {The Origins Space Telescope: mission concept
  overview}. In: \procspie, Society of Photo-Optical Instrumentation Engineers
  (SPIE) Conference Series, vol 10698, p 1069815, \doi{10.1117/12.2313823}

\bibitem[{{Lellouch} et~al.(1995){Lellouch}, {Paubert}, {Moreno}, {Festou},
  {Bezard}, {Bockelee-Morvan}, {Colom}, {Crovisier}, {Encrenaz}, {Gautier},
  {Marten}, {Despois}, {Strobel}, and {Sievers}}]{Lellouch1995}
{Lellouch} E, {Paubert} G, {Moreno} R, {Festou} MC, {Bezard} B,
  {Bockelee-Morvan} D, {Colom} P, {Crovisier} J, {Encrenaz} T, {Gautier} D,
  {Marten} A, {Despois} D, {Strobel} DF, {Sievers} A (1995) {Chemical and
  Thermal Response of Jupiter's Atmosphere Following the Impact of Comet
  Shoemaker-Levy-9}. \nat 373:592--595, \doi{10.1038/373592a0}

\bibitem[{{Lellouch} et~al.(2001){Lellouch}, {B{\'e}zard}, {Fouchet},
  {Feuchtgruber}, {Encrenaz}, and {de Graauw}}]{Lellouch2001}
{Lellouch} E, {B{\'e}zard} B, {Fouchet} T, {Feuchtgruber} H, {Encrenaz} T, {de
  Graauw} T (2001) {The deuterium abundance in Jupiter and Saturn from ISO-SWS
  observations}. \aap 370:610--622, \doi{10.1051/0004-6361:20010259}

\bibitem[{{Lellouch} et~al.(2005){Lellouch}, {Moreno}, and
  {Paubert}}]{Lellouch2005}
{Lellouch} E, {Moreno} R, {Paubert} G (2005) {A dual origin for Neptune's
  carbon monoxide?} \aap 430:L37--L40, \doi{10.1051/0004-6361:200400127}

\bibitem[{{Lellouch} et~al.(2006){Lellouch}, {B{\'e}zard}, {Strobel},
  {Bjoraker}, {Flasar}, and {Romani}}]{Lellouch2006}
{Lellouch} E, {B{\'e}zard} B, {Strobel} DF, {Bjoraker} GL, {Flasar} FM,
  {Romani} PN (2006) {On the HCN and CO$_{2}$ abundance and distribution in
  Jupiter's stratosphere}. \icarus 184:478--497,
  \doi{10.1016/j.icarus.2006.05.018}

\bibitem[{{Lellouch} et~al.(2015){Lellouch}, {Moreno}, {Orton}, {Feuchtgruber},
  {Cavali{\'e}}, {Moses}, {Hartogh}, {Jarchow}, and {Sagawa}}]{Lellouch2015}
{Lellouch} E, {Moreno} R, {Orton} GS, {Feuchtgruber} H, {Cavali{\'e}} T,
  {Moses} JI, {Hartogh} P, {Jarchow} C, {Sagawa} H (2015) {New constraints on
  the CH$_{4}$ vertical profile in Uranus and Neptune from Herschel
  observations}. \aap 579:A121, \doi{10.1051/0004-6361/201526518}

\bibitem[{{Li} et~al.(2017){Li}, {Ingersoll}, {Janssen}, {Levin}, {Bolton},
  {Adumitroaie}, {Allison}, {Arballo}, {Bellotti}, {Brown}, {Ewald}, {Jewell},
  {Misra}, {Orton}, {Oyafuso}, {Steffes}, and {Williamson}}]{Li2017}
{Li} C, {Ingersoll} A, {Janssen} M, {Levin} S, {Bolton} S, {Adumitroaie} V,
  {Allison} M, {Arballo} J, {Bellotti} A, {Brown} S, {Ewald} S, {Jewell} L,
  {Misra} S, {Orton} G, {Oyafuso} F, {Steffes} P, {Williamson} R (2017) {The
  distribution of ammonia on Jupiter from a preliminary inversion of Juno
  microwave radiometer data}. \grl 44(11):5317--5325,
  \doi{10.1002/2017GL073159}

\bibitem[{{Li} et~al.(2020){Li}, {Ingersoll}, {Bolton}, {Levin}, {Janssen},
  {Atreya}, {Lunine}, {Steffes}, {Brown}, {Guillot}, {Allison}, {Arballo},
  {Bellotti}, {Adumitroaie}, {Gulkis}, {Hodges}, {Li}, {Misra}, {Orton},
  {Oyafuso}, {Santos-Costa}, {Waite}, and {Zhang}}]{Li2020}
{Li} C, {Ingersoll} A, {Bolton} S, {Levin} S, {Janssen} M, {Atreya} S, {Lunine}
  J, {Steffes} P, {Brown} S, {Guillot} T, {Allison} M, {Arballo} J, {Bellotti}
  A, {Adumitroaie} V, {Gulkis} S, {Hodges} A, {Li} L, {Misra} S, {Orton} G,
  {Oyafuso} F, {Santos-Costa} D, {Waite} H, {Zhang} Z (2020) {The water
  abundance in Jupiter's equatorial zone}. Nature Astronomy
  \doi{10.1038/s41550-020-1009-3}

\bibitem[{{Lindal}(1992)}]{Lindal1992}
{Lindal} GF (1992) {The Atmosphere of Neptune: an Analysis of Radio Occultation
  Data Acquired with Voyager 2}. \aj 103:967, \doi{10.1086/116119}

\bibitem[{{Lindal} et~al.(1981){Lindal}, {Wood}, {Levy}, {Anderson},
  {Sweetnam}, {Hotz}, {Buckles}, {Holmes}, {Doms}, {Eshleman}, {Tyler}, and
  {Croft}}]{Lindal1981}
{Lindal} GF, {Wood} GE, {Levy} GS, {Anderson} JD, {Sweetnam} DN, {Hotz} HB,
  {Buckles} BJ, {Holmes} DP, {Doms} PE, {Eshleman} VR, {Tyler} GL, {Croft} TA
  (1981) {The atmosphere of Jupiter: an analysis of the Voyager radio
  occulation measurements}. \jgr 86(A10):8721--8727,
  \doi{10.1029/JA086iA10p08721}

\bibitem[{{Lindal} et~al.(1985){Lindal}, {Sweetnam}, and
  {Eshleman}}]{Lindal1985}
{Lindal} GF, {Sweetnam} DN, {Eshleman} VR (1985) {The atmosphere of Saturn - an
  analysis of the Voyager radio occultation measurements}. \aj 90:1136--1146,
  \doi{10.1086/113820}

\bibitem[{{Lindal} et~al.(1987){Lindal}, {Lyons}, {Sweetnam}, {Eshleman}, and
  {Hinson}}]{Lindal1987}
{Lindal} GF, {Lyons} JR, {Sweetnam} DN, {Eshleman} VR, {Hinson} DP (1987) {The
  atmosphere of Uranus - Results of radio occultation measurements with Voyager
  2}. \jgr 92:14987--15001, \doi{10.1029/JA092iA13p14987}

\bibitem[{{Lindal} et~al.(1990){Lindal}, {Lyons}, {Sweetnam}, {Eshleman}, and
  {Hinson}}]{Lindal1990}
{Lindal} GF, {Lyons} JR, {Sweetnam} DN, {Eshleman} VR, {Hinson} DP (1990) {The
  atmosphere of Neptune - Results of radio occultation measurements with the
  Voyager 2 spacecraft}. \grl 17:1733--1736, \doi{10.1029/GL017i010p01733}

\bibitem[{{Lodders}(2004)}]{Lodders2004}
{Lodders} K (2004) {Jupiter Formed with More Tar than Ice}. \apj 611:587--597,
  \doi{10.1086/421970}

\bibitem[{{Lodders}(2010)}]{Lodders2010}
{Lodders} K (2010) {Solar System Abundances of the Elements}. Astrophysics and
  Space Science Proceedings 16:379, \doi{10.1007/978-3-642-10352-0_8}

\bibitem[{{Lodders} and {Fegley}(1994)}]{Lodders1994}
{Lodders} K, {Fegley} B Jr (1994) {The origin of carbon monoxide in Neptunes's
  atmosphere}. \icarus 112:368--375, \doi{10.1006/icar.1994.1190}

\bibitem[{{Lunine} and {Stevenson}(1985)}]{Lunine1985}
{Lunine} JI, {Stevenson} DJ (1985) {Thermodynamics of clathrate hydrate at low
  and high pressures with application to the outer solar system}. \apjs
  58:493--531, \doi{10.1086/191050}

\bibitem[{{Luszcz-Cook} and {de Pater}(2013)}]{Luszcz-Cook2013}
{Luszcz-Cook} SH, {de Pater} I (2013) {Constraining the origins of Neptune's
  carbon monoxide abundance with CARMA millimeter-wave observations}. \icarus
  222(1):379--400, \doi{10.1016/j.icarus.2012.11.002}

\bibitem[{{Mahaffy} et~al.(2000){Mahaffy}, {Niemann}, {Alpert}, {Atreya},
  {Demick}, {Donahue}, {Harpold}, and {Owen}}]{Mahaffy2000}
{Mahaffy} PR, {Niemann} HB, {Alpert} A, {Atreya} SK, {Demick} J, {Donahue} TM,
  {Harpold} DN, {Owen} TC (2000) {Noble gas abundance and isotope ratios in the
  atmosphere of Jupiter from the Galileo Probe Mass Spectrometer}. \jgr
  105:15061--15072, \doi{10.1029/1999JE001224}

\bibitem[{{Marley} et~al.(1995){Marley}, {G{\'o}mez}, and
  {Podolak}}]{Marley1995}
{Marley} MS, {G{\'o}mez} P, {Podolak} M (1995) {Monte Carlo interior models for
  Uranus and Neptune}. \jgr 100(E11):23349--23354, \doi{10.1029/95JE02362}

\bibitem[{{Marten} et~al.(1993){Marten}, {Gautier}, {Owen}, {Sanders},
  {Matthews}, {Atreya}, {Tilanus}, and {Deane}}]{Marten1993}
{Marten} A, {Gautier} D, {Owen} T, {Sanders} DB, {Matthews} HE, {Atreya} SK,
  {Tilanus} RPJ, {Deane} JR (1993) {First observations of CO and HCN on Neptune
  and Uranus at millimeter wavelengths and the implications for atmospheric
  chemistry}. \apj 406:285--297, \doi{10.1086/172440}

\bibitem[{{Marten} et~al.(2005){Marten}, {Matthews}, {Owen}, {Moreno},
  {Hidayat}, and {Biraud}}]{Marten2005}
{Marten} A, {Matthews} HE, {Owen} T, {Moreno} R, {Hidayat} T, {Biraud} Y (2005)
  {Improved constraints on Neptune's atmosphere from submillimetre-wavelength
  observations}. \aap 429:1097--1105, \doi{10.1051/0004-6361:20041695}

\bibitem[{{Matousek}(2007)}]{Matousek2007}
{Matousek} S (2007) {The Juno New Frontiers mission}. Acta Astronautica
  61:932--939, \doi{10.1016/j.actaastro.2006.12.013}

\bibitem[{{Mazevet} et~al.(2019){Mazevet}, {Licari}, {Chabrier}, and
  {Potekhin}}]{Mazevet2019}
{Mazevet} S, {Licari} A, {Chabrier} G, {Potekhin} AY (2019) {Ab initio based
  equation of state of dense water for planetary and exoplanetary modeling}.
  \aap 621:A128, \doi{10.1051/0004-6361/201833963}

\bibitem[{{Meadows} et~al.(2008){Meadows}, {Orton}, {Line}, {Liang}, {Yung},
  {Van Cleve}, and {Burgdorf}}]{Meadows2008}
{Meadows} VS, {Orton} G, {Line} M, {Liang} MC, {Yung} YL, {Van Cleve} J,
  {Burgdorf} MJ (2008) {First Spitzer observations of Neptune: Detection of new
  hydrocarbons}. \icarus 197(2):585--589, \doi{10.1016/j.icarus.2008.05.023}

\bibitem[{{Miguel} et~al.(2016){Miguel}, {Guillot}, and {Fayon}}]{Miguel2016}
{Miguel} Y, {Guillot} T, {Fayon} L (2016) {Jupiter internal structure: the
  effect of different equations of state}. \aap 596:A114,
  \doi{10.1051/0004-6361/201629732}

\bibitem[{{Militzer} and {Hubbard}(2013)}]{Militzer2013}
{Militzer} B, {Hubbard} WB (2013) {Ab Initio Equation of State for
  Hydrogen-Helium Mixtures with Recalibration of the Giant-planet Mass-Radius
  Relation}. \apj 774(2):148, \doi{10.1088/0004-637X/774/2/148}

\bibitem[{{Molter} et~al.(2019){Molter}, {de Pater}, {Sault}, {Butler},
  {Luszcz-Cook}, {Tollefson}, and {de Boer}}]{Molter2019}
{Molter} E, {de Pater} I, {Sault} RJ, {Butler} B, {Luszcz-Cook} S, {Tollefson}
  J, {de Boer} D (2019) {Uranus's Tropospheric Circulation and Composition
  with ALMA and the VLA}. In: EPSC Abstracts, vol~13, pp EPSC--DPS2019--726--1

\bibitem[{{Moreno} et~al.(2009){Moreno}, {Marten}, and {Lellouch}}]{Moreno2009}
{Moreno} R, {Marten} A, {Lellouch} E (2009) {Search for PH$_3$ in the
  Atmospheres of Uranus and Neptune at Millimeter Wavelength}. In: AAS/Division
  for Planetary Sciences Meeting Abstracts \#41, AAS/Division for Planetary
  Sciences Meeting Abstracts, p 28.02

\bibitem[{{Moreno} et~al.(2011){Moreno}, {Lellouch}, {Courtin}, {Swinyard},
  {Fulton}, {Orton}, {Hartogh}, {Jarchow}, {Cavali\'e}, {Feuchtgruber}, and
  Team}]{Moreno2011}
{Moreno} R, {Lellouch} E, {Courtin} R, {Swinyard} B, {Fulton} T, {Orton} G,
  {Hartogh} P, {Jarchow} C, {Cavali\'e} T, {Feuchtgruber} H, Team TH (2011)
  {Observations of CO and HCN on Neptune with Herschel SPIRE}. In: Geophysical
  Research Abstracts, Geophysical Research Abstracts, vol~13, pp EGU2011--8299

\bibitem[{{Moreno} et~al.(2017){Moreno}, {Lellouch}, {Cavali{\'e}}, and
  {Moullet}}]{Moreno2017}
{Moreno} R, {Lellouch} E, {Cavali{\'e}} T, {Moullet} A (2017) {Detection of CS
  in Neptune's atmosphere from ALMA observations}. \aap 608:L5,
  \doi{10.1051/0004-6361/201731472}

\bibitem[{{Moses}(2014)}]{Moses2014}
{Moses} JI (2014) {Chemical kinetics on extrasolar planets}. Philosophical
  Transactions of the Royal Society of London Series A 372:20130073--20130073,
  \doi{10.1098/rsta.2013.0073}

\bibitem[{{Moses} and {Poppe}(2017)}]{Moses2017}
{Moses} JI, {Poppe} AR (2017) {Dust ablation on the giant planets: Consequences
  for stratospheric photochemistry}. \icarus 297:33--58,
  \doi{10.1016/j.icarus.2017.06.002}

\bibitem[{{Moses} et~al.(2000{\natexlab{a}}){Moses}, {B{\'e}zard}, {Lellouch},
  {Gladstone}, {Feuchtgruber}, and {Allen}}]{Moses2000a}
{Moses} JI, {B{\'e}zard} B, {Lellouch} E, {Gladstone} GR, {Feuchtgruber} H,
  {Allen} M (2000{\natexlab{a}}) {Photochemistry of Saturn's Atmosphere. I.
  Hydrocarbon Chemistry and Comparisons with ISO Observations}. \icarus
  143(2):244--298, \doi{10.1006/icar.1999.6270}

\bibitem[{{Moses} et~al.(2000{\natexlab{b}}){Moses}, {Lellouch}, {B{\'e}zard},
  {Gladstone}, {Feuchtgruber}, and {Allen}}]{Moses2000b}
{Moses} JI, {Lellouch} E, {B{\'e}zard} B, {Gladstone} GR, {Feuchtgruber} H,
  {Allen} M (2000{\natexlab{b}}) {Photochemistry of Saturn's Atmosphere. II.
  Effects of an Influx of External Oxygen}. \icarus 145:166--202,
  \doi{10.1006/icar.1999.6320}

\bibitem[{{Moses} et~al.(2005){Moses}, {Fouchet}, {B{\'e}zard}, {Gladstone},
  {Lellouch}, and {Feuchtgruber}}]{Moses2005}
{Moses} JI, {Fouchet} T, {B{\'e}zard} B, {Gladstone} GR, {Lellouch} E,
  {Feuchtgruber} H (2005) {Photochemistry and diffusion in Jupiter's
  stratosphere: Constraints from ISO observations and comparisons with other
  giant planets}. Journal of Geophysical Research (Planets) 110:E08001,
  \doi{10.1029/2005JE002411}

\bibitem[{{Moses} et~al.(2011){Moses}, {Visscher}, {Fortney}, {Showman},
  {Lewis}, {Griffith}, {Klippenstein}, {Shabram}, {Friedson}, {Marley}, and
  {Freedman}}]{Moses2011}
{Moses} JI, {Visscher} C, {Fortney} JJ, {Showman} AP, {Lewis} NK, {Griffith}
  CA, {Klippenstein} SJ, {Shabram} M, {Friedson} AJ, {Marley} MS, {Freedman} RS
  (2011) {Disequilibrium Carbon, Oxygen, and Nitrogen Chemistry in the
  Atmospheres of HD 189733b and HD 209458b}. \apj 737:15,
  \doi{10.1088/0004-637X/737/1/15}

\bibitem[{{Moses} et~al.(2018){Moses}, {Fletcher}, {Greathouse}, {Orton}, and
  {Hue}}]{Moses2018}
{Moses} JI, {Fletcher} LN, {Greathouse} TK, {Orton} GS, {Hue} V (2018)
  {Seasonal stratospheric photochemistry on Uranus and Neptune}. \icarus
  307:124--145, \doi{10.1016/j.icarus.2018.02.004}

\bibitem[{{Mousis} et~al.(2006){Mousis}, {Alibert}, and {Benz}}]{Mousis2006}
{Mousis} O, {Alibert} Y, {Benz} W (2006) {Saturn's internal structure and
  carbon enrichment}. \aap 449(1):411--415, \doi{10.1051/0004-6361:20054224}

\bibitem[{{Mousis} et~al.(2012){Mousis}, {Lunine}, {Madhusudhan}, and
  {Johnson}}]{Mousis2012}
{Mousis} O, {Lunine} JI, {Madhusudhan} N, {Johnson} TV (2012) {Nebular Water
  Depletion as the Cause of Jupiter's Low Oxygen Abundance}. \apjl 751:L7,
  \doi{10.1088/2041-8205/751/1/L7}

\bibitem[{{Mousis} et~al.(2014{\natexlab{a}}){Mousis}, {Fletcher}, {Lebreton},
  {Wurz}, {Cavali{\'e}}, {Coustenis}, {Courtin}, {Gautier}, {Helled}, {Irwin},
  {Morse}, {Nettelmann}, {Marty}, {Rousselot}, {Venot}, {Atkinson}, {Waite},
  {Reh}, {Simon}, {Atreya}, {Andr{\'e}}, {Blanc}, {Daglis}, {Fischer},
  {Geppert}, {Guillot}, {Hedman}, {Hueso}, {Lellouch}, {Lunine}, {Murray},
  {O`Donoghue}, {Rengel}, {S{\'a}nchez-Lavega}, {Schmider}, {Spiga}, {Spilker},
  {Petit}, {Tiscareno}, {Ali-Dib}, {Altwegg}, {Bolton}, {Bouquet}, {Briois},
  {Fouchet}, {Guerlet}, {Kostiuk}, {Lebleu}, {Moreno}, {Orton}, and
  {Poncy}}]{Mousis2014a}
{Mousis} O, {Fletcher} LN, {Lebreton} JP, {Wurz} P, {Cavali{\'e}} T,
  {Coustenis} A, {Courtin} R, {Gautier} D, {Helled} R, {Irwin} PGJ, {Morse} AD,
  {Nettelmann} N, {Marty} B, {Rousselot} P, {Venot} O, {Atkinson} DH, {Waite}
  JH, {Reh} KR, {Simon} AA, {Atreya} S, {Andr{\'e}} N, {Blanc} M, {Daglis} IA,
  {Fischer} G, {Geppert} WD, {Guillot} T, {Hedman} MM, {Hueso} R, {Lellouch} E,
  {Lunine} JI, {Murray} CD, {O`Donoghue} J, {Rengel} M, {S{\'a}nchez-Lavega} A,
  {Schmider} FX, {Spiga} A, {Spilker} T, {Petit} JM, {Tiscareno} MS, {Ali-Dib}
  M, {Altwegg} K, {Bolton} SJ, {Bouquet} A, {Briois} C, {Fouchet} T, {Guerlet}
  S, {Kostiuk} T, {Lebleu} D, {Moreno} R, {Orton} GS, {Poncy} J
  (2014{\natexlab{a}}) {Scientific rationale for Saturn's in situ exploration}.
  \planss 104:29--47, \doi{10.1016/j.pss.2014.09.014}

\bibitem[{{Mousis} et~al.(2014{\natexlab{b}}){Mousis}, {Lunine}, {Fletcher},
  {Mandt}, {Ali-Dib}, {Gautier}, and {Atreya}}]{Mousis2014b}
{Mousis} O, {Lunine} JI, {Fletcher} LN, {Mandt} KE, {Ali-Dib} M, {Gautier} D,
  {Atreya} S (2014{\natexlab{b}}) {New Insights on Saturn's Formation from its
  Nitrogen Isotopic Composition}. \apjl 796(2):L28,
  \doi{10.1088/2041-8205/796/2/L28}

\bibitem[{{Mousis} et~al.(2016){Mousis}, {Atkinson}, {Spilker}, {Venkatapathy},
  {Poncy}, {Frampton}, {Coustenis}, {Reh}, {Lebreton}, {Fletcher}, {Hueso},
  {Amato}, {Colaprete}, {Ferri}, {Stam}, {Wurz}, {Atreya}, {Aslam}, {Banfield},
  {Calcutt}, {Fischer}, {Holland}, {Keller}, {Kessler}, {Leese}, {Levacher},
  {Morse}, {Mu{\~n}oz}, {Renard}, {Sheridan}, {Schmider}, {Snik}, {Waite},
  {Bird}, {Cavali{\'e}}, {Deleuil}, {Fortney}, {Gautier}, {Guillot}, {Lunine},
  {Marty}, {Nixon}, {Orton}, and {S{\'a}nchez-Lavega}}]{Mousis2016}
{Mousis} O, {Atkinson} DH, {Spilker} T, {Venkatapathy} E, {Poncy} J, {Frampton}
  R, {Coustenis} A, {Reh} K, {Lebreton} JP, {Fletcher} LN, {Hueso} R, {Amato}
  MJ, {Colaprete} A, {Ferri} F, {Stam} D, {Wurz} P, {Atreya} S, {Aslam} S,
  {Banfield} DJ, {Calcutt} S, {Fischer} G, {Holland} A, {Keller} C, {Kessler}
  E, {Leese} M, {Levacher} P, {Morse} A, {Mu{\~n}oz} O, {Renard} JB, {Sheridan}
  S, {Schmider} FX, {Snik} F, {Waite} JH, {Bird} M, {Cavali{\'e}} T, {Deleuil}
  M, {Fortney} J, {Gautier} D, {Guillot} T, {Lunine} JI, {Marty} B, {Nixon} C,
  {Orton} GS, {S{\'a}nchez-Lavega} A (2016) {The Hera Saturn entry probe
  mission}. \planss 130:80--103, \doi{10.1016/j.pss.2015.06.020}

\bibitem[{{Mousis} et~al.(2018){Mousis}, {Atkinson}, {Cavali{\'e}}, {Fletcher},
  {Amato}, {Aslam}, {Ferri}, {Renard}, {Spilker}, {Venkatapathy}, {Wurz},
  {Aplin}, {Coustenis}, {Deleuil}, {Dobrijevic}, {Fouchet}, {Guillot},
  {Hartogh}, {Hewagama}, {Hofstadter}, {Hue}, {Hueso}, {Lebreton}, {Lellouch},
  {Moses}, {Orton}, {Pearl}, {S{\'a}nchez-Lavega}, {Simon}, {Venot}, {Waite},
  {Achterberg}, {Atreya}, {Billebaud}, {Blanc}, {Borget}, {Brugger}, {Charnoz},
  {Chiavassa}, {Cottini}, {d'Hendecourt}, {Danger}, {Encrenaz}, {Gorius},
  {Jorda}, {Marty}, {Moreno}, {Morse}, {Nixon}, {Reh}, {Ronnet}, {Schmider},
  {Sheridan}, {Sotin}, {Vernazza}, and {Villanueva}}]{Mousis2018}
{Mousis} O, {Atkinson} DH, {Cavali{\'e}} T, {Fletcher} LN, {Amato} MJ, {Aslam}
  S, {Ferri} F, {Renard} JB, {Spilker} T, {Venkatapathy} E, {Wurz} P, {Aplin}
  K, {Coustenis} A, {Deleuil} M, {Dobrijevic} M, {Fouchet} T, {Guillot} T,
  {Hartogh} P, {Hewagama} T, {Hofstadter} MD, {Hue} V, {Hueso} R, {Lebreton}
  JP, {Lellouch} E, {Moses} J, {Orton} GS, {Pearl} JC, {S{\'a}nchez-Lavega} A,
  {Simon} A, {Venot} O, {Waite} JH, {Achterberg} RK, {Atreya} S, {Billebaud} F,
  {Blanc} M, {Borget} F, {Brugger} B, {Charnoz} S, {Chiavassa} T, {Cottini} V,
  {d'Hendecourt} L, {Danger} G, {Encrenaz} T, {Gorius} NJP, {Jorda} L, {Marty}
  B, {Moreno} R, {Morse} A, {Nixon} C, {Reh} K, {Ronnet} T, {Schmider} FX,
  {Sheridan} S, {Sotin} C, {Vernazza} P, {Villanueva} GL (2018) {Scientific
  rationale for Uranus and Neptune in situ explorations}. \planss 155:12--40,
  \doi{10.1016/j.pss.2017.10.005}

\bibitem[{{Mousis} et~al.(2019){Mousis}, {Ronnet}, and {Lunine}}]{Mousis2019a}
{Mousis} O, {Ronnet} T, {Lunine} JI (2019) {Jupiter{\textquoteright}s Formation
  in the Vicinity of the Amorphous Ice Snowline}. \apj 875(1):9,
  \doi{10.3847/1538-4357/ab0a72}

\bibitem[{{Mousis} et~al.(2020){Mousis}, {Atreya}, {Lunine}, {Mandt}, {Marty},
  and {Ronnet}}]{Mousis2020}
{Mousis} O, {Atreya} S, {Lunine} JI, {Mandt} KE, {Marty} B, {Ronnet} T (2020)
  {Key atmospheric Signatures for identifying the source reservoirs of
  volatiles in Uranus and Neptune}. \ssr this issue

\bibitem[{{Nettelmann} et~al.(2008){Nettelmann}, {Holst}, {Kietzmann},
  {French}, {Redmer}, and {Blaschke}}]{Nettelmann2008}
{Nettelmann} N, {Holst} B, {Kietzmann} A, {French} M, {Redmer} R, {Blaschke} D
  (2008) {Ab Initio Equation of State Data for Hydrogen, Helium, and Water and
  the Internal Structure of Jupiter}. \apj 683:1217--1228, \doi{10.1086/589806}

\bibitem[{{Nettelmann} et~al.(2013){Nettelmann}, {Helled}, {Fortney}, and
  {Redmer}}]{Nettelmann2013}
{Nettelmann} N, {Helled} R, {Fortney} JJ, {Redmer} R (2013) {New indication for
  a dichotomy in the interior structure of Uranus and Neptune from the
  application of modified shape and rotation data}. \planss 77:143--151,
  \doi{10.1016/j.pss.2012.06.019}

\bibitem[{{Nettelmann} et~al.(2016){Nettelmann}, {Wang}, {Fortney}, {Hamel},
  {Yellamilli}, {Bethkenhagen}, and {Redmer}}]{Nettelmann2016}
{Nettelmann} N, {Wang} K, {Fortney} JJ, {Hamel} S, {Yellamilli} S,
  {Bethkenhagen} M, {Redmer} R (2016) {Uranus evolution models with simple
  thermal boundary layers}. \icarus 275:107--116,
  \doi{10.1016/j.icarus.2016.04.008}

\bibitem[{{Niemann} et~al.(1998){Niemann}, {Atreya}, {Carignan}, {Donahue},
  {Haberman}, {Harpold}, {Hartle}, {Hunten}, {Kasprzak}, {Mahaffy}, {Owen}, and
  {Way}}]{Niemann1998}
{Niemann} HB, {Atreya} SK, {Carignan} GR, {Donahue} TM, {Haberman} JA,
  {Harpold} DN, {Hartle} RE, {Hunten} DM, {Kasprzak} WT, {Mahaffy} PR, {Owen}
  TC, {Way} SH (1998) {The composition of the Jovian atmosphere as determined
  by the Galileo probe mass spectrometer}. \jgr 103:22831--22846,
  \doi{10.1029/98JE01050}

\bibitem[{{Noll} and {Larson}(1991)}]{Noll1991}
{Noll} KS, {Larson} HP (1991) {The spectrum of Saturn from 1990 to 2230 cm
  $^{-1}$: Abundances of AsH $_{3}$, CH $_{3}$D, CO, GeH $_{4}$, NH $_{3}$, and
  PH $_{3}$}. \icarus 89(1):168--189, \doi{10.1016/0019-1035(91)90096-C}

\bibitem[{{Noll} et~al.(1986){Noll}, {Knacke}, {Geballe}, and
  {Tokunaga}}]{Noll1986}
{Noll} KS, {Knacke} RF, {Geballe} TR, {Tokunaga} AT (1986) {Detection of carbon
  monoxide in Saturn}. \apjl 309:L91--L94, \doi{10.1086/184768}

\bibitem[{{Norwood} et~al.(2016{\natexlab{a}}){Norwood}, {Hammel}, {Milam},
  {Stansberry}, {Lunine}, {Chanover}, {Hines}, {Sonneborn}, {Tiscareno},
  {Brown}, and {Ferruit}}]{Norwood2016b}
{Norwood} J, {Hammel} H, {Milam} S, {Stansberry} J, {Lunine} J, {Chanover} N,
  {Hines} D, {Sonneborn} G, {Tiscareno} M, {Brown} M, {Ferruit} P
  (2016{\natexlab{a}}) {Solar System Observations with the James Webb Space
  Telescope}. \pasp 128(2):025004, \doi{10.1088/1538-3873/128/960/025004}

\bibitem[{{Norwood} et~al.(2016{\natexlab{b}}){Norwood}, {Moses}, {Fletcher},
  {Orton}, {Irwin}, {Atreya}, {Rages}, {Cavali{\'e}}, {S{\'a}nchez-Lavega},
  {Hueso}, and {Chanover}}]{Norwood2016a}
{Norwood} J, {Moses} J, {Fletcher} LN, {Orton} G, {Irwin} PGJ, {Atreya} S,
  {Rages} K, {Cavali{\'e}} T, {S{\'a}nchez-Lavega} A, {Hueso} R, {Chanover} N
  (2016{\natexlab{b}}) {Giant Planet Observations with the James Webb Space
  Telescope}. \pasp 128(1):018005, \doi{10.1088/1538-3873/128/959/018005}

\bibitem[{{{\"O}berg} and {Bergin}(2016)}]{Oberg2016}
{{\"O}berg} KI, {Bergin} EA (2016) {Excess C/O and C/H in Outer Protoplanetary
  Disk Gas}. \apjl 831(2):L19, \doi{10.3847/2041-8205/831/2/L19}

\bibitem[{{{\"O}berg} et~al.(2011){{\"O}berg}, {Murray-Clay}, and
  {Bergin}}]{Oberg2011}
{{\"O}berg} KI, {Murray-Clay} R, {Bergin} EA (2011) {The Effects of Snowlines
  on C/O in Planetary Atmospheres}. \apjl 743(1):L16,
  \doi{10.1088/2041-8205/743/1/L16}

\bibitem[{{Ollivier} et~al.(2000){Ollivier}, {Dobrij{\'e}vic}, and
  {Parisot}}]{Ollivier2000}
{Ollivier} JL, {Dobrij{\'e}vic} M, {Parisot} JP (2000) {New photochemical model
  of Saturn's atmosphere}. \planss 48:699--716,
  \doi{10.1016/S0032-0633(00)00035-0}

\bibitem[{{Orton} et~al.(1998){Orton}, {Fisher}, {Baines}, {Stewart},
  {Friedson}, {Ortiz}, {Marinova}, {Ressler}, {Dayal}, {Hoffmann}, {Hora},
  {Hinkley}, {Krishnan}, {Masanovic}, {Tesic}, {Tziolas}, and
  {Parija}}]{Orton1998}
{Orton} GS, {Fisher} BM, {Baines} KH, {Stewart} ST, {Friedson} AJ, {Ortiz} JL,
  {Marinova} M, {Ressler} M, {Dayal} A, {Hoffmann} W, {Hora} J, {Hinkley} S,
  {Krishnan} V, {Masanovic} M, {Tesic} J, {Tziolas} A, {Parija} KC (1998)
  {Characteristics of the Galileo probe entry site from Earth-based remote
  sensing observations}. \jgr 103(E10):22791--22814, \doi{10.1029/98JE02380}

\bibitem[{{Orton} et~al.(2014){Orton}, {Moses}, {Fletcher}, {Mainzer}, {Hines},
  {Hammel}, {Martin-Torres}, {Burgdorf}, {Merlet}, and {Line}}]{Orton2014b}
{Orton} GS, {Moses} JI, {Fletcher} LN, {Mainzer} AK, {Hines} D, {Hammel} HB,
  {Martin-Torres} J, {Burgdorf} M, {Merlet} C, {Line} MR (2014) {Mid-infrared
  spectroscopy of Uranus from the Spitzer infrared spectrometer: 2.
  Determination of the mean composition of the upper troposphere and
  stratosphere}. \icarus 243:471--493, \doi{10.1016/j.icarus.2014.07.012}

\bibitem[{{Owen} and {Encrenaz}(2003)}]{Owen2003}
{Owen} T, {Encrenaz} T (2003) {Element Abundances and Isotope Ratios in the
  Giant Planets and Titan}. \ssr 106:121--138, \doi{10.1023/A:1024633603624}

\bibitem[{{Owen} and {Encrenaz}(2006)}]{Owen2006}
{Owen} T, {Encrenaz} T (2006) {Compositional constraints on giant planet
  formation}. \planss 54:1188--1196, \doi{10.1016/j.pss.2006.05.030}

\bibitem[{{Owen} et~al.(1999){Owen}, {Mahaffy}, {Niemann}, {Atreya}, {Donahue},
  {Bar-Nun}, and {de Pater}}]{Owen1999}
{Owen} T, {Mahaffy} P, {Niemann} HB, {Atreya} S, {Donahue} T, {Bar-Nun} A, {de
  Pater} I (1999) {A low-temperature origin for the planetesimals that formed
  Jupiter}. \nat 402:269--270, \doi{10.1038/46232}

\bibitem[{{Pearl} and {Conrath}(1991)}]{Pearl1991}
{Pearl} JC, {Conrath} BJ (1991) {The albedo, effective temperature, and energy
  balance of Neptune, as determined from Voyager data}. \jgr 96:18921

\bibitem[{{Pearl} et~al.(1990){Pearl}, {Conrath}, {Hanel}, and
  {Pirraglia}}]{Pearl1990}
{Pearl} JC, {Conrath} BJ, {Hanel} RA, {Pirraglia} JA (1990) {The albedo,
  effective temperature, and energy balance of Uranus, as determined from
  Voyager IRIS data}. \icarus 84:12--28, \doi{10.1016/0019-1035(90)90155-3}

\bibitem[{Perry et~al.(2018)Perry, Waite~Jr., Mitchell, Miller, Cravens,
  Perryman, Moore, Yelle, Hsu, Hedman, Cuzzi, Strobel, Hamil, Glein, Paxton,
  Teolis, and McNutt~Jr.}]{Perry2018}
Perry ME, Waite~Jr JH, Mitchell DG, Miller KE, Cravens TE, Perryman RS, Moore
  L, Yelle RV, Hsu HW, Hedman MM, Cuzzi JN, Strobel DF, Hamil OQ, Glein CR,
  Paxton LJ, Teolis BD, McNutt~Jr RL (2018) Material flux from the rings of
  saturn into its atmosphere. \grl 45(19):10,093--10,100,
  \doi{10.1029/2018GL078575},
  \urlprefix\url{https://agupubs.onlinelibrary.wiley.com/doi/abs/10.1029/2018GL078575}

\bibitem[{{Podolak} and {Helled}(2012)}]{Podolak2012}
{Podolak} M, {Helled} R (2012) {What Do We Really Know about Uranus and
  Neptune?} \apjl 759(2):L32, \doi{10.1088/2041-8205/759/2/L32}

\bibitem[{{Podolak} et~al.(1991){Podolak}, {Hubbard}, and
  {Stevenson}}]{Podolak1991}
{Podolak} M, {Hubbard} WB, {Stevenson} DJ (1991) {Models of Uranus' interior
  and magnetic field.}, pp 29--61

\bibitem[{{Pollack} et~al.(1996){Pollack}, {Hubickyj}, {Bodenheimer},
  {Lissauer}, {Podolak}, and {Greenzweig}}]{Pollack1996}
{Pollack} JB, {Hubickyj} O, {Bodenheimer} P, {Lissauer} JJ, {Podolak} M,
  {Greenzweig} Y (1996) {Formation of the Giant Planets by Concurrent Accretion
  of Solids and Gas}. \icarus 124:62--85, \doi{10.1006/icar.1996.0190}

\bibitem[{{Prang{\'e}} et~al.(2006){Prang{\'e}}, {Fouchet}, {Courtin},
  {Connerney}, and {McConnell}}]{Prange2006}
{Prang{\'e}} R, {Fouchet} T, {Courtin} R, {Connerney} JEP, {McConnell} JC
  (2006) {Latitudinal variation of Saturn photochemistry deduced from
  spatially-resolved ultraviolet spectra}. \icarus 180:379--392,
  \doi{10.1016/j.icarus.2005.11.005}

\bibitem[{{Prinn} and {Barshay}(1977)}]{Prinn1977}
{Prinn} RG, {Barshay} SS (1977) {Carbon monoxide on Jupiter and implications
  for atmospheric convection}. Science 198:1031--1034,
  \doi{10.1126/science.198.4321.1031}

\bibitem[{{Redmer} et~al.(2011){Redmer}, {Mattsson}, {Nettelmann}, and
  {French}}]{Redmer2011}
{Redmer} R, {Mattsson} TR, {Nettelmann} N, {French} M (2011) {The phase diagram
  of water and the magnetic fields of Uranus and Neptune}. \icarus
  211(1):798--803, \doi{10.1016/j.icarus.2010.08.008}

\bibitem[{{Roelfsema} et~al.(2018){Roelfsema}, {Shibai}, {Armus}, {Arrazola},
  {Audard}, {Audley}, {Bradford}, {Charles}, {Dieleman}, {Doi}, {Duband},
  {Eggens}, {Evers}, {Funaki}, {Gao}, {Giard}, {di Giorgio}, {Gonz{\'a}lez
  Fern{\'a}ndez}, {Griffin}, {Helmich}, {Hijmering}, {Huisman}, {Ishihara},
  {Isobe}, {Jackson}, {Jacobs}, {Jellema}, {Kamp}, {Kaneda}, {Kawada},
  {Kemper}, {Kerschbaum}, {Khosropanah}, {Kohno}, {Kooijman}, {Krause}, {van
  der Kuur}, {Kwon}, {Laauwen}, {de Lange}, {Larsson}, {van Loon}, {Madden},
  {Matsuhara}, {Najarro}, {Nakagawa}, {Naylor}, {Ogawa}, {Onaka}, {Oyabu},
  {Poglitsch}, {Reveret}, {Rodriguez}, {Spinoglio}, {Sakon}, {Sato},
  {Shinozaki}, {Shipman}, {Sugita}, {Suzuki}, {van der Tak}, {Torres Redondo},
  {Wada}, {Wang}, {Wafelbakker}, {van Weers}, {Withington}, {Vandenbussche},
  {Yamada}, and {Yamamura}}]{Roelfsema2018}
{Roelfsema} PR, {Shibai} H, {Armus} L, {Arrazola} D, {Audard} M, {Audley} MD,
  {Bradford} CM, {Charles} I, {Dieleman} P, {Doi} Y, {Duband} L, {Eggens} M,
  {Evers} J, {Funaki} I, {Gao} JR, {Giard} M, {di Giorgio} A, {Gonz{\'a}lez
  Fern{\'a}ndez} LM, {Griffin} M, {Helmich} FP, {Hijmering} R, {Huisman} R,
  {Ishihara} D, {Isobe} N, {Jackson} B, {Jacobs} H, {Jellema} W, {Kamp} I,
  {Kaneda} H, {Kawada} M, {Kemper} F, {Kerschbaum} F, {Khosropanah} P, {Kohno}
  K, {Kooijman} PP, {Krause} O, {van der Kuur} J, {Kwon} J, {Laauwen} WM, {de
  Lange} G, {Larsson} B, {van Loon} D, {Madden} SC, {Matsuhara} H, {Najarro} F,
  {Nakagawa} T, {Naylor} D, {Ogawa} H, {Onaka} T, {Oyabu} S, {Poglitsch} A,
  {Reveret} V, {Rodriguez} L, {Spinoglio} L, {Sakon} I, {Sato} Y, {Shinozaki}
  K, {Shipman} R, {Sugita} H, {Suzuki} T, {van der Tak} FFS, {Torres Redondo}
  J, {Wada} T, {Wang} SY, {Wafelbakker} CK, {van Weers} H, {Withington} S,
  {Vandenbussche} B, {Yamada} T, {Yamamura} I (2018) {SPICA-A Large Cryogenic
  Infrared Space Telescope: Unveiling the Obscured Universe}. \pasa 35:e030,
  \doi{10.1017/pasa.2018.15}

\bibitem[{{R{\"o}ttgering}(2003)}]{Rottgering2003}
{R{\"o}ttgering} H (2003) {LOFAR, a new low frequency radio telescope}. \nar
  47(4-5):405--409, \doi{10.1016/S1387-6473(03)00057-5}

\bibitem[{{Roulston} and {Stevenson}(1995)}]{Roulston1995}
{Roulston} MS, {Stevenson} DJ (1995) Prediction of neon depletion in
  jupiter's atmosphere. Eos, Transactions American Geophysical Union 76:343

\bibitem[{{Safronov}(1966)}]{Safronov1966}
{Safronov} VS (1966) {Sizes of the largest bodies falling onto the planets
  during their formation}. \sovast 9:987--991

\bibitem[{{Scott} et~al.(2015){Scott}, {Grevesse}, {Asplund}, {Sauval}, {Lind},
  {Takeda}, {Collet}, {Trampedach}, and {Hayek}}]{Scott2015}
{Scott} P, {Grevesse} N, {Asplund} M, {Sauval} AJ, {Lind} K, {Takeda} Y,
  {Collet} R, {Trampedach} R, {Hayek} W (2015) {The elemental composition of
  the Sun. I. The intermediate mass elements Na to Ca}. \aap 573:A25,
  \doi{10.1051/0004-6361/201424109}

\bibitem[{{Seiff} et~al.(1998){Seiff}, {Kirk}, {Knight}, {Young}, {Mihalov},
  {Young}, {Milos}, {Schubert}, {Blanchard}, and {Atkinson}}]{Seiff1998}
{Seiff} A, {Kirk} DB, {Knight} TCD, {Young} RE, {Mihalov} JD, {Young} LA,
  {Milos} FS, {Schubert} G, {Blanchard} RC, {Atkinson} D (1998) {Thermal
  structure of Jupiter's atmosphere near the edge of a 5-{$\mu$}m hot spot in
  the north equatorial belt}. \jgr 103:22857--22890, \doi{10.1029/98JE01766}

\bibitem[{{Simon} et~al.(2018){Simon}, {Stern}, and {Hofstadter}}]{Simon2018}
{Simon} AA, {Stern} SA, {Hofstadter} M (2018) {Outer Solar System Exploration:
  A Compelling and Unified Dual Mission Decadal Strategy for Exploring Uranus,
  Neptune, Triton, Dwarf Planets, and Small KBOs and Centaurs}. arXiv e-prints
  arXiv:1807.08769

\bibitem[{{Simon} et~al.(2020){Simon}, {Fletcher}, {Arridge}, {Atkinson},
  {Coustenis}, {Ferri}, {Hofstadter}, {Masters}, {Mousis}, {Reh}, {Turrini},
  and {Witasse}}]{Simon2020}
{Simon} AA, {Fletcher} LN, {Arridge} C, {Atkinson} D, {Coustenis} A, {Ferri} F,
  {Hofstadter} M, {Masters} A, {Mousis} O, {Reh} K, {Turrini} D, {Witasse} O
  (2020) {A Review of the in Situ Probe Designs from Recent Ice Giant Mission
  Concept Studies}. \ssr 216(1):17, \doi{10.1007/s11214-020-0639-1}

\bibitem[{{Smith}(1998)}]{Smith1998}
{Smith} MD (1998) {Estimation of a Length Scale to Use with the Quench Level
  Approximation for Obtaining Chemical Abundances}. \icarus 132:176--184,
  \doi{10.1006/icar.1997.5886}

\bibitem[{{Sromovsky} and {Fry}(2008)}]{Sromovsky2008}
{Sromovsky} LA, {Fry} PM (2008) {The methane abundance and structure of Uranus'
  cloud bands inferred from spatially resolved 2006 Keck grism spectra}.
  \icarus 193:252--266, \doi{10.1016/j.icarus.2007.08.037}

\bibitem[{{Sromovsky} et~al.(2011){Sromovsky}, {Fry}, and
  {Kim}}]{Sromovsky2011}
{Sromovsky} LA, {Fry} PM, {Kim} JH (2011) {Methane on Uranus: The case for a
  compact CH $_{4}$ cloud layer at low latitudes and a severe CH $_{4}$
  depletion at high-latitudes based on re-analysis of Voyager occultation
  measurements and STIS spectroscopy}. \icarus 215:292--312,
  \doi{10.1016/j.icarus.2011.06.024}

\bibitem[{{Sromovsky} et~al.(2014){Sromovsky}, {Karkoschka}, {Fry}, {Hammel},
  {de Pater}, and {Rages}}]{Sromovsky2014}
{Sromovsky} LA, {Karkoschka} E, {Fry} PM, {Hammel} HB, {de Pater} I, {Rages} K
  (2014) {Methane depletion in both polar regions of Uranus inferred from
  HST/STIS and Keck/NIRC2 observations}. \icarus 238:137--155,
  \doi{10.1016/j.icarus.2014.05.016}

\bibitem[{{Stanley} and {Bloxham}(2004)}]{Stanley2004}
{Stanley} S, {Bloxham} J (2004) {Convective-region geometry as the cause of
  Uranus' and Neptune's unusual magnetic fields}. \nat 428(6979):151--153,
  \doi{10.1038/nature02376}

\bibitem[{{Stanley} and {Bloxham}(2006)}]{Stanley2006}
{Stanley} S, {Bloxham} J (2006) {Numerical dynamo models of Uranus' and
  Neptune's magnetic fields}. \icarus 184(2):556--572,
  \doi{10.1016/j.icarus.2006.05.005}

\bibitem[{{Stone}(1976)}]{Stone1976}
{Stone} PH (1976) {The meteorology of the Jovian atmosphere}. In: {Gehrels} T
  (ed) IAU Colloq. 30: Jupiter: Studies of the Interior, Atmosp here,
  Magnetosphere and Satellites, pp 586--618

\bibitem[{Teanby and Irwin(2013)}]{Teanby2013}
Teanby NA, Irwin PGJ (2013) An external origin for carbon monoxide on uranus
  from herschel/spire? \apjl 775(2):L49

\bibitem[{{Teanby} et~al.(2019){Teanby}, {Irwin}, and {Moses}}]{Teanby2019}
{Teanby} NA, {Irwin} PGJ, {Moses} JI (2019) {Neptune's carbon monoxide profile
  and phosphine upper limits from Herschel/SPIRE: Implications for interior
  structure and formation}. \icarus 319:86--98,
  \doi{10.1016/j.icarus.2018.09.014}

\bibitem[{{Thorngren} and {Fortney}(2019)}]{Thorngren2019}
{Thorngren} D, {Fortney} JJ (2019) {Connecting Giant Planet Atmosphere and
  Interior Modeling: Constraints on Atmospheric Metal Enrichment}. \apjl
  874(2):L31, \doi{10.3847/2041-8213/ab1137}

\bibitem[{{Tollefson} et~al.(2019{\natexlab{a}}){Tollefson}, {de Pater},
  {Luszcz-Cook}, and {DeBoer}}]{Tollefson2019a}
{Tollefson} J, {de Pater} I, {Luszcz-Cook} S, {DeBoer} D (2019{\natexlab{a}})
  {Neptune's Latitudinal Variations as Viewed with ALMA}. \aj 157(6):251,
  \doi{10.3847/1538-3881/ab1fdf}

\bibitem[{{Tollefson} et~al.(2019{\natexlab{b}}){Tollefson}, {de Pater},
  {Sault}, {Butler}, {Luszcz-Cook}, and {DeBoer}}]{Tollefson2019b}
{Tollefson} J, {de Pater} I, {Sault} S, {Butler} B, {Luszcz-Cook} S, {DeBoer} D
  (2019{\natexlab{b}}) {Spatial variations on Neptune in the radio}. In: EPSC
  Abstracts, pp EPSC--DPS2019--728--1

\bibitem[{{Tsai} et~al.(2017){Tsai}, {Lyons}, {Grosheintz}, {Rimmer},
  {Kitzmann}, and {Heng}}]{Tsai2017}
{Tsai} SM, {Lyons} JR, {Grosheintz} L, {Rimmer} PB, {Kitzmann} D, {Heng} K
  (2017) {VULCAN: An Open-source, Validated Chemical Kinetics Python Code for
  Exoplanetary Atmospheres}. \apjs 228(2):20, \doi{10.3847/1538-4365/228/2/20}

\bibitem[{{Valletta} and {Helled}(2019)}]{Valletta2019}
{Valletta} C, {Helled} R (2019) {The Deposition of Heavy Elements in Giant
  Protoplanetary Atmospheres: The Importance of Planetesimal-Envelope
  Interactions}. \apj 871(1):127, \doi{10.3847/1538-4357/aaf427}

\bibitem[{{Vazan} and {Helled}(2020)}]{Vazan2020}
{Vazan} A, {Helled} R (2020) {Explaining the low luminosity of Uranus: a
  self-consistent thermal and structural evolution}. \aap 633:A50,
  \doi{10.1051/0004-6361/201936588}

\bibitem[{{Venot} et~al.(2012){Venot}, {H{\'e}brard}, {Ag{\'u}ndez},
  {Dobrijevic}, {Selsis}, {Hersant}, {Iro}, and {Bounaceur}}]{Venot2012}
{Venot} O, {H{\'e}brard} E, {Ag{\'u}ndez} M, {Dobrijevic} M, {Selsis} F,
  {Hersant} F, {Iro} N, {Bounaceur} R (2012) {A chemical model for the
  atmosphere of hot Jupiters}. \aap 546:A43, \doi{10.1051/0004-6361/201219310}

\bibitem[{{Venot} et~al.(2019){Venot}, {Bounaceur}, {Dobrijevic},
  {H{\'e}brard}, {Cavali{\'e}}, {Tremblin}, {Drummond}, and
  {Charnay}}]{Venot2019}
{Venot} O, {Bounaceur} R, {Dobrijevic} M, {H{\'e}brard} E, {Cavali{\'e}} T,
  {Tremblin} P, {Drummond} B, {Charnay} B (2019) {Reduced chemical scheme for
  modelling warm to hot hydrogen-dominated atmospheres}. \aap 624:A58,
  \doi{10.1051/0004-6361/201834861}

\bibitem[{{Venot} et~al.(2020){Venot}, {Cavali{\'e}}, {Bounaceur}, {Tremblin},
  {Brouillard}, and {Lhoussaine Ben Brahim}}]{Venot2020}
{Venot} O, {Cavali{\'e}} T, {Bounaceur} R, {Tremblin} P, {Brouillard} L,
  {Lhoussaine Ben Brahim} R (2020) {New chemical scheme for giant planet
  thermochemistry. Update of the methanol chemistry and new reduced chemical
  scheme}. \aap 634:A78, \doi{10.1051/0004-6361/201936697}

\bibitem[{{Visscher} and {Fegley}(2005)}]{Visscher2005}
{Visscher} C, {Fegley} B Jr (2005) {Chemical Constraints on the Water and Total
  Oxygen Abundances in the Deep Atmosphere of Saturn}. \apj 623:1221--1227,
  \doi{10.1086/428493}

\bibitem[{{Visscher} and {Moses}(2011)}]{Visscher2011}
{Visscher} C, {Moses} JI (2011) {Quenching of Carbon Monoxide and Methane in
  the Atmospheres of Cool Brown Dwarfs and Hot Jupiters}. \apj 738:72,
  \doi{10.1088/0004-637X/738/1/72}

\bibitem[{{Visscher} et~al.(2010){Visscher}, {Moses}, and
  {Saslow}}]{Visscher2010}
{Visscher} C, {Moses} JI, {Saslow} SA (2010) {The deep water abundance on
  Jupiter: New constraints from thermochemical kinetics and diffusion
  modeling}. \icarus 209:602--615, \doi{10.1016/j.icarus.2010.03.029}

\bibitem[{{von Zahn} et~al.(1998){von Zahn}, {Hunten}, and
  {Lehmacher}}]{vonZahn1998}
{von Zahn} U, {Hunten} DM, {Lehmacher} G (1998) {Helium in Jupiter's
  atmosphere: Results from the Galileo probe helium interferometer experiment}.
  \jgr 103:22815--22830, \doi{10.1029/98JE00695}

\bibitem[{{Vorburger} et~al.(2020){Vorburger}, {Wurz}, and
  {Waite}}]{Vorburger2020}
{Vorburger} A, {Wurz} P, {Waite} JH (2020) {Chemical and isotopic composition
  measurements on atmospheric probes}. \ssr this issue

\bibitem[{{Waite} et~al.(2018){Waite}, {Perryman}, {Perry}, {Miller}, {Bell},
  {Glein}, {Grimes}, {Hedman}, {Cuzzi}, {Brockwell}, {Teolis}, {Moore},
  {Mitchell}, {Persoon}, {Kurth}, {Wahlund}, {Morooka}, {Hadid}, {Walker},
  {Nagy}, {Yelle}, {Ledvina}, {Johnson}, {Tseng}, {Tucker}, and
  {Ip}}]{Waite2018}
{Waite} JH, {Perryman} RS, {Perry} ME, {Miller} KE, {Bell} J, {Glein} CR,
  {Grimes} J, {Hedman} M, {Cuzzi} J, {Brockwell} T, {Teolis} B, {Moore} L,
  {Mitchell} DG, {Persoon} A, {Kurth} WS, {Wahlund} JE, {Morooka} M, {Hadid}
  LZ, {Walker} J, {Nagy} A, {Yelle} R, {Ledvina} S, {Johnson} R, {Tseng} W,
  {Tucker} OJ, {Ip} WH (2018) {Chemical interactions between Saturn's
  atmosphere and its rings}. \science 362:51--

\bibitem[{{Wang} et~al.(2015){Wang}, {Gierasch}, {Lunine}, and
  {Mousis}}]{Wang2015}
{Wang} D, {Gierasch} PJ, {Lunine} JI, {Mousis} O (2015) {New insights on
  Jupiter's deep water abundance from disequilibrium species}. \icarus
  250:154--164, \doi{10.1016/j.icarus.2014.11.026}

\bibitem[{{Wang} et~al.(2016){Wang}, {Lunine}, and {Mousis}}]{Wang2016}
{Wang} D, {Lunine} JI, {Mousis} O (2016) {Modeling the disequilibrium species
  for Jupiter and Saturn: Implications for Juno and Saturn entry probe}.
  \icarus 276:21--38, \doi{10.1016/j.icarus.2016.04.027}

\bibitem[{{Wilson} and {Militzer}(2010)}]{Wilson2010}
{Wilson} HF, {Militzer} B (2010) {Sequestration of Noble Gases in Giant Planet
  Interiors}. \prl 104(12):121101, \doi{10.1103/PhysRevLett.104.121101}

\bibitem[{{Wong} et~al.(2004){Wong}, {Mahaffy}, {Atreya}, {Niemann}, and
  {Owen}}]{Wong2004}
{Wong} MH, {Mahaffy} PR, {Atreya} SK, {Niemann} HB, {Owen} TC (2004) {Updated
  Galileo probe mass spectrometer measurements of carbon, oxygen, nitrogen, and
  sulfur on Jupiter}. \icarus 171:153--170, \doi{10.1016/j.icarus.2004.04.010}

\bibitem[{{Wurz} et~al.(2012){Wurz}, {Abplanalp}, {Tulej}, and
  {Lammer}}]{Wurz2012}
{Wurz} P, {Abplanalp} D, {Tulej} M, {Lammer} H (2012) {A neutral gas mass
  spectrometer for the investigation of lunar volatiles}. \planss
  74(1):264--269, \doi{10.1016/j.pss.2012.05.016}

\bibitem[{{Yung} et~al.(1988){Yung}, {Drew}, {Pinto}, and {Friedl}}]{Yung1988}
{Yung} YL, {Drew} WA, {Pinto} JP, {Friedl} RR (1988) {Estimation of the
  reaction rate for the formation of CH3O from H + H2CO - Implications for
  chemistry in the solar system}. \icarus 73:516--526,
  \doi{10.1016/0019-1035(88)90061-9}

\end{thebibliography}

%
%

\end{document}